\documentclass[twocolumn,twocolappendix]{aastex631}

\newcommand{\hst}{{\it HST}}
\newcommand{\jwst}{{\it JWST}}

\newcommand{\beagle}{\textsc{Beagle}}
\newcommand{\prospector}{\textsc{Prospector}}
\newcommand{\cloudy}{\textsc{Cloudy}}

\newcommand{\muv}{$M_{\rm UV}$}

\newcommand{\lya}{\hbox{Ly$\alpha$}}
\newcommand{\ha}{\hbox{H$\alpha$}}
\newcommand{\hb}{\hbox{H$\beta$}}
\newcommand{\hg}{\hbox{H$\gamma$}}
\newcommand{\hd}{\hbox{H$\delta$}}
\newcommand{\he}{\hbox{H$\epsilon$}}
\newcommand{\oiii}{\hbox{O\,{\sc iii}}}
\newcommand{\oii}{\hbox{O\,{\sc ii}}}
\newcommand{\oi}{\hbox{O\,{\sc i}}}
\newcommand{\hi}{\hbox{H\,{\sc i}}}
\newcommand{\hii}{\hbox{H\,{\sc ii}}}

\newcommand{\niii}{\hbox{N\,{\sc iii}}}
\newcommand{\niv}{\hbox{N\,{\sc iv}}}
\newcommand{\nv}{\hbox{N\,{\sc v}}}

\newcommand{\neiii}{\hbox{Ne\,{\sc iii}}}
\newcommand{\cii}{\hbox{C\,{\sc ii}}}
\newcommand{\ciii}{\hbox{C\,{\sc iii}}}

\newcommand{\civ}{\hbox{C\,{\sc iv}}}
\newcommand{\heii}{\hbox{He\,{\sc ii}}}

\newcommand{\siii}{\hbox{Si\,{\sc ii}}}
\newcommand{\siiv}{\hbox{Si\,{\sc iv}}}
\newcommand{\alii}{\hbox{Al\,{\sc ii}}}

\newcommand{\feii}{\hbox{Fe\,{\sc ii}}}
\newcommand{\feiii}{\hbox{Fe\,{\sc iii}}}
\newcommand{\feiv}{\hbox{Fe\,{\sc iv}}}

\newcommand{\ergscm}{\,erg~s$^{-1}$~cm$^{-2}$}
\newcommand{\ergscmA}{\,erg~s$^{-1}$~cm$^{-2}$~\AA$^{-1}$}
\newcommand{\kms}{\,km~s$^{-1}$}
\newcommand{\um}{\,$\mu$m}

\newcommand{\srcname}{SPURS-A2744-7}

\graphicspath{{./}{figures/}}

\begin{document}

\title{SPURS: Bursty Star Formation in an Extremely Luminous Weak Emission Line Galaxy at $z=9.3$ }

\shorttitle{SPURS-A2744-7}
\shortauthors{Chen et al.}

\correspondingauthor{Zuyi Chen}
 \email{zuyi.chen@nbi.ku.dk}

 \author[0000-0002-2178-5471]{Zuyi Chen}
 \affiliation{Cosmic Dawn Center (DAWN)}
 \affiliation{Niels Bohr Institute, University of Copenhagen, Jagtvej 128, 2200 Copenhagen N, Denmark}

 \author[0000-0001-6106-5172]{Daniel P. Stark}
 \affiliation{Department of Astronomy, University of California, Berkeley, Berkeley, CA 94720, USA}

 \author[0000-0002-3407-1785]{Charlotte A. Mason}
 \affiliation{Cosmic Dawn Center (DAWN)}
 \affiliation{Niels Bohr Institute, University of Copenhagen, Jagtvej 128, 2200 Copenhagen N, Denmark}

\author[0000-0003-0390-0656]{Adele Plat}
\affiliation{Institute of Physics, GalSpec Laboratory, Ecole Polytechnique Federale de Lausanne, Observatoire de Sauverny, Chemin Pegasi 51, 1290
Versoix, Switzerland}

\author[0000-0001-5487-0392]{Viola Gelli}
\affiliation{Cosmic Dawn Center (DAWN)}
\affiliation{Niels Bohr Institute, University of Copenhagen, Jagtvej 128, 2200 Copenhagen N, Denmark}

\author[0000-0002-9132-6561]{Peter Senchyna}
\affiliation{The Observatories of the Carnegie Institution for Science, 813 Santa Barbara Street, Pasadena, CA 91101, USA}

\author[0000-0002-2645-679X]{Keerthi Vasan G. C.}
\affiliation{The Observatories of the Carnegie Institution for Science, 813 Santa Barbara Street, Pasadena, CA 91101, USA}

\author[0000-0003-4564-2771]{Ryan Endsley}
\affiliation{Department of Astronomy, The University of Texas at Austin, Austin, TX 78712, USA}
\affiliation{Cosmic Frontier Center, The University of Texas at Austin, Austin, TX 78712, USA}

\author[0000-0001-5940-338X]{Mengtao Tang}
\affiliation{Tsung-Dao Lee Institute, Shanghai Jiao Tong University, 1 Lisuo Road, Shanghai 201210, People’s Republic of China}
\affiliation{School of Physics and Astronomy, Shanghai Jiao Tong University, 800 Dongchuan Road, Shanghai 200240, People’s Republic of China}
\affiliation{State Key Laboratory of Dark Matter Physics, Shanghai Jiao Tong University, 1 Lisuo Road, Shanghai 201210, People’s Republic of China}

\author[0000-0001-8426-1141]{Michael W. Topping}
\affiliation{Steward Observatory, University of Arizona, 933 N Cherry Avenue, Tucson, AZ 85721, USA}

\author[0000-0003-1432-7744]{Lily Whitler}
\affiliation{Kavli Institute for Cosmology, University of Cambridge, Madingley Road, Cambridge, CB3 0HA, UK}
\affiliation{Cavendish Laboratory, University of Cambridge, JJ Thomson Avenue, Cambridge, CB3 0US, UK}



\begin{abstract}
{\it JWST} has revealed a population of super-luminous early galaxies with a volume density in excess of most expectations. The spectra reveal diverse properties: while some reveal strong emission lines characteristic of galaxies in the midst of strong bursts, others show weak emission lines that could reflect old stellar populations, large escape fractions, or post-burst star formation histories.  Through the {\it JWST} Cycle 4 large program SPURS, we have obtained ultra-deep (29 hr) rest-frame UV spectroscopy of a $z=9.3$ super-luminous ($M_{\rm UV}=-21.66$) galaxy with large assembled stellar mass (1.6$\times$10$^9$ $M_\odot$) and extremely weak emission lines (H$\beta$ EW $\approx25$~\AA).  The strong stellar wind features and rest-optical line ratios suggest the galaxy is already significantly enriched, with a metallicity of 0.4--0.7~Z$_\odot$.  The interstellar absorption lines reveal outflows ($v\simeq -161$~km~s$^{-1}$) with a large neutral gas covering fraction, suggesting that the weak emission lines are not due to large escape fractions.  The combination of the Balmer break, weak emission lines, and stellar wind features constrains the star formation history, indicating a recent burst of star formation lasting 10--20 Myr followed by a downturn over the last 10~Myr. The observations suggest that  $z\gtrsim 9$ weak emission line galaxies such as this source can be explained by stochastic star formation, provided that the downturns in star formation are recent (i.e., $<10$ Myr prior to observation).  The ultra-deep grating spectrum enables the IGM damping wing to be characterized, decoupling the effects of local absorption. The smooth Ly$\alpha$ break indicates that this source, one of the most massive galaxies known at $z>9$, is likely situated in a small ionized bubble ($0.29_{-0.09}^{+0.11}$~pMpc), as is common at large neutral hydrogen fractions ($\bar{x}_{\rm HI}=0.81_{-0.21}^{+0.14}$).

\end{abstract}



\section{Introduction} \label{sec:intro}

The first deep NIRCam datasets revealed the presence of super luminous galaxies at $z\gtrsim 10$ with volume density well in excess of what was expected prior to {\it JWST} (see reviews by \citealt{Ellis2025,Matthee2025,Stark2026}).
Over the last two years, new surveys have continued to identify luminous galaxies at the redshift frontier 
\citep{Carniani2024_z14,Castellano2024,Naidu2026_MoMz14,Donnan2026}. Meanwhile, as larger areas have been imaged with NIRCam, the luminosity functions at $z>10$ have firmed up, confirming that the integrated luminosity density is in excess of many pre-{\it JWST} expectations \citep[e.g.,][]{Castellano2023,McLeod2024,Finkelstein2024,Donnan2024,Harikane2025,Weibel2025a,Whitler2025,Kreilgaard2026,McLeod2026}. 

The physical processes that create such luminous galaxies in the early universe remain disputed. Some have argued that stochasticity is the primary driver, allowing a subset of low-mass halos to appear UV-luminous  \citep[e.g.,][]{Mason2023,Mirocha2023,Shen2023,Sun2023,Kravtsov2024,Gelli2024}. 
Others suggest star formation efficiencies may be higher in the low-mass halos that dominate at early epochs, leading to larger UV luminosities in galaxies at $z>10$ \citep[e.g.,][]{Dekel2023,Li2024,Nikopoulos2024,Ceverino2024,Kar2026}. 
Furthermore, several of the proposed solutions do not require significant redshift evolution in physics, relying instead on the mass-dependence of either stochasticity \citep{Gelli2024,Munoz2026} or star formation efficiency \citep{Feldmann2025}, with the lowest mass halos coming into view at the highest redshifts owing to the redshift-dependence of the baryon accretion rates. In addition, the emergent luminosity of the earliest galaxies could be boosted by variations in the ionizing properties (i.e., initial mass function; e.g., \citealt{Yung2024}), presence of active galactic nuclei (AGN; e.g., \citealt{Hegde2024}), and reduced dust attenuation (e.g., \citealt{Ferrara2023,Fiore2023,Narayanan2025}).

Isolating the primary explanation for the excess luminosity seen at $z>10$ is critical if we are to understand the growth of the first generations of galaxies. Attention is now turning to other observables.
Spectroscopy is one of the most powerful tools for distinguishing between the various scenarios listed above. The first spectra of UV-bright galaxies at $z>10$  revealed prominent rest-frame UV emission lines  \citep[e.g.,][]{Bunker2023,Castellano2024,Naidu2026_MoMz14}, as expected for galaxies that are in the midst of strong upturns in star formation. The physical conditions implied by these spectra are markedly different from those typical at later cosmic epochs. In particular, several of the most luminous galaxies at  $z>10$ (GNz11, GHZ2; \citealt{Bunker2023,Castellano2024}) have spectra that suggest some combination of extremely high gas densities, strong high-ionization emission lines, and nitrogen-enhanced abundance patterns that may be linked to the formation of globular clusters \citep[e.g.,][]{Senchyna2024,Ji2026}. The origin of the high-ionization emission has been actively debated, with both low-metallicity massive stars and accreting black holes proposed as viable power sources \citep[e.g.,][]{Maiolino2024_GNz11,AlvarezMarquez2024,Castellano2025,Ortiz2025,Zhu2026}. The discovery of such extreme spectra may hint at a link between episodic bursts of star formation, possibly accompanied by AGN activity, and the presence of luminous galaxies at $z>10$.

However, it has recently been shown that not all super-luminous galaxies have spectra that indicate a bursting mode of star formation. This was first demonstrated with the discovery of GS-z14-0, a $z=14.32$ galaxy with M$_{\rm{UV}}=-20.8$ \citep{Carniani2024_z14}. Both rest-UV and rest-optical spectra reveal weak emission lines, with MIRI spectroscopic observations indicating an H$\beta$ EW $\approx 80$~\AA\ \citep{Helton2025}.  More recently, several more super-luminous early galaxies have been shown to have strikingly weak emission line spectra (e.g., \citealt{Donnan2026,Harikane2026,MarquesChaves2026}), generally with more extended morphological structures \citep{Donnan2026,Harikane2026}. Recent compilations of the spectroscopic database at $z>9$ reveal these are a significant subset of the early galaxy population \citep{Tang2025,Roberts-Borsani2025}.

The nature of the weak line sources remains somewhat enigmatic. On one hand, it is possible these are more massive galaxies that are luminous because they have built up an old stellar population, producing a stronger optical continuum which leads to lower EW rest-optical emission lines. Alternatively, the weak lines could arise if ionizing photon escape fractions are very large, or if there is a 
dearth of O stars following a recent downturn in star formation.  If we are to understand what is driving the formation of luminous galaxies at $z>9$, it is very important that we get to the bottom of the physics responsible for weak emission lines. The ultraviolet provides our best path to progress. For example, if escape fractions are large and there remains a dominant population of recently-formed stars, spectra should reveal weak interstellar absorption lines and direct wind and photospheric signatures of a dominant massive star population. Characterizing faint absorption lines requires deep spectroscopy with at least moderate spectral resolution. Unfortunately, none of the super-luminous galaxies with weak emission lines have been observed with sufficient depth and resolution in the rest-frame ultraviolet.

\begin{figure}[t]
    \centering
    \includegraphics[width=\columnwidth]{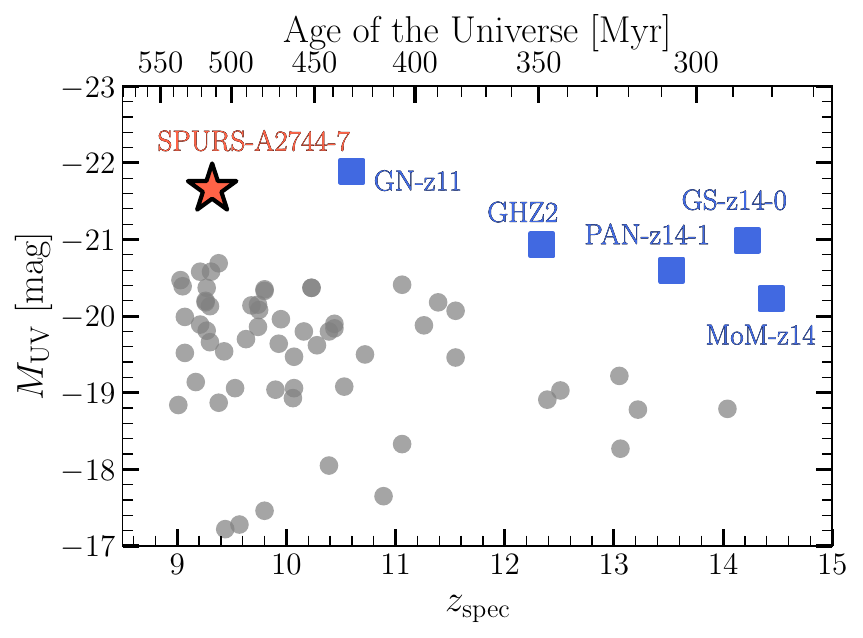}
    \caption{UV absolute magnitude of \srcname{} compared to other spectroscopically confirmed galaxies at $z>9$.
    We plot the NIRSpec sample compiled in \cite{Tang2025} as gray dots and highlight other notably luminous galaxies at these redshifts, including GN-z11 (\citealt{Oesch2015,Bunker2023,Tacchella2023_GNz11}), GHZ2 (\citealt{Castellano2024}), PAN-z14-1 (\citealt{Donnan2026}), GS-z14-0 (\citealt{Carniani2024_z14}),  and MoM-z14 (\citealt{Naidu2026_MoMz14}).}
    \label{fig:Muv_z}
\end{figure}

Motivated by this shortcoming, we have initiated the SPectroscopic Ultra-deep Reionization-era Survey (SPURS, GO 9214, PIs Mason and Stark), providing the first ultra-deep grating observations of the rest-frame UV in the earliest star forming galaxies. The SPURS observations are being conducted in four pointings across four fields. Here we utilize our first observations in Abell 2744 to investigate the origin of weak emission lines in one of the most UV-luminous systems at $z>9$ (\muv{}~= $-21.66\pm0.03$; Figure~\ref{fig:Muv_z}), known previously as both Gz9p3 \citep{Boyett2024_z9p3}, UNCOVER-3686 \citep{Fujimoto2024_uncover} and DHZ1 \citep{Castellano2023}.
Early \jwst{}/NIRSpec spectroscopic observations  confirmed its redshift at $z=9.3127$ \citep{Boyett2024_z9p3,Fujimoto2024_uncover}, while also revealing extremely weak rest-optical emission lines, with [\oiii{}]~$\lambda5008$ EW $\approx 210$~\AA{} and H$\beta$ EW $\approx 25$~\AA{}. Subsequent analysis has revealed it as one of the weakest line emitters known among super-luminous galaxies at $z>9$.\citep[e.g.,][]{Tang2025}. Here we present the new SPURS observations of this source, revealing the stellar populations and gas properties of the system with the detail afforded by ultra-deep  grating spectroscopy.

The organization of this paper is as follows.
In \S~\ref{sec:data}, we present the new SPURS observations and summarize what was known about the source prior to our program. 
We characterize its rest-frame UV and optical spectroscopic properties using the SPURS spectrum in \S~\ref{sec:results}.
We perform photoionization modeling of the full SPURS spectrum to infer the ISM and stellar population properties in \S~\ref{sec:modeling}, and discuss the implications in \S~\ref{sec:discussion}.
Finally, we summarize our conclusions in \S~\ref{sec:conclusion}. 
Throughout this paper, we adopt a flat $\Lambda$CDM cosmology with $H_0 = 70$ km~s$^{-1}$~Mpc$^{-1}$, $\Omega_{\rm m} = 0.3$, and $\Omega_\Lambda = 0.7$.
All magnitudes are measured in the AB system \citep{Oke1983}.
The line equivalent widths are calculated in the rest frame, with positive values corresponding to emission lines and negative values to absorption lines.

\section{Observations and Reduction}\label{sec:data}

\begin{figure}[t]
    \centering
    \includegraphics[width=\columnwidth]{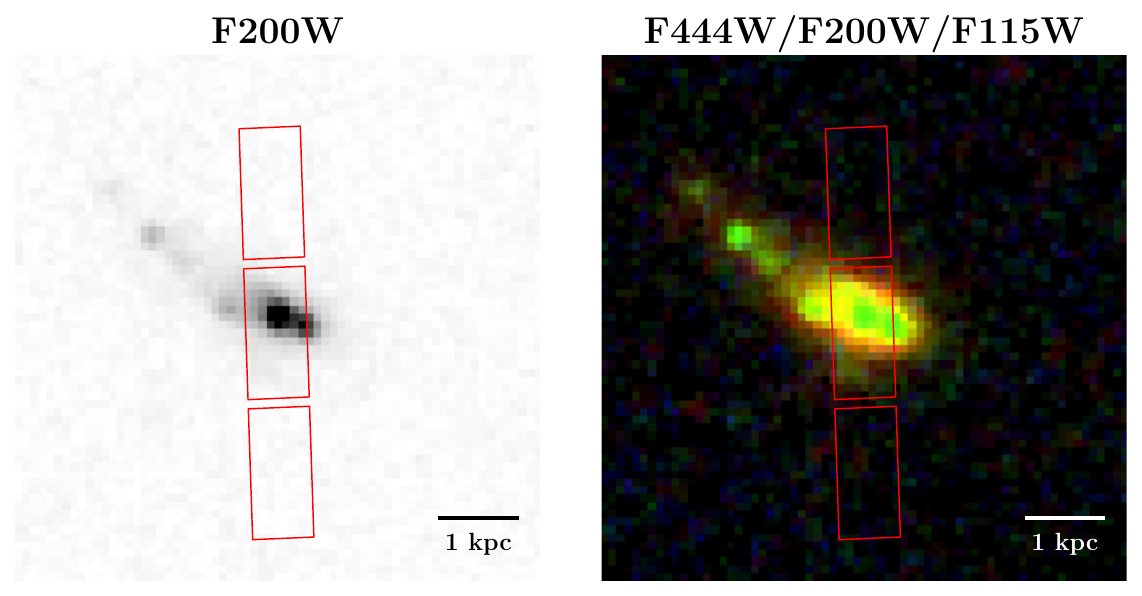}
    \caption{NIRCam F200W and RGB images ($2\arcsec{}\times2\arcsec{}$) of \srcname{}.
    A scale bar of 1~kpc (lensing magnification corrected) is also shown.
    Our NIRSpec shutter (red) is centered around the brightest clump in the rest-UV.}
    \label{fig:image}
\end{figure}

\begin{figure*}
    \centering
    \includegraphics[width=1\linewidth]{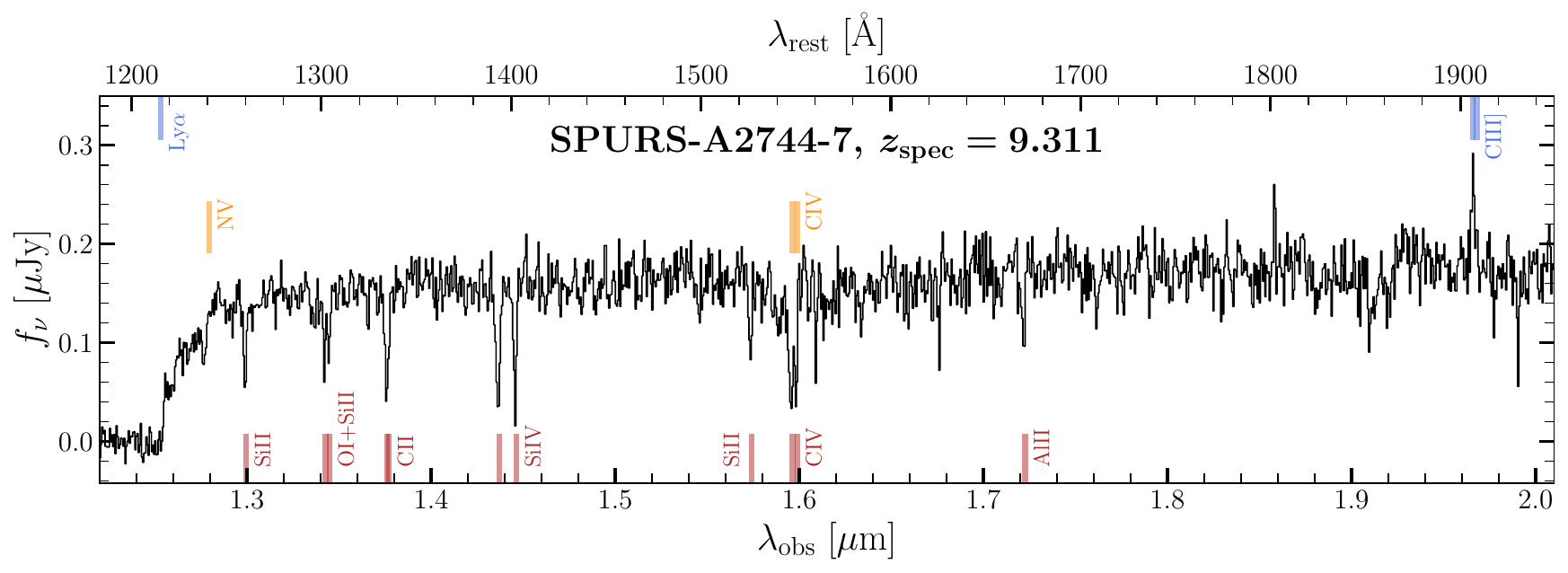}
    \caption{Rest-UV spectrum for \srcname{} taken with G140M.
    With 30-hour depth, we detect the continuum along with a wealth of interstellar absorption lines (red).
    The spectrum is characterized by weak rest-UV emission lines, and we only detect the \ciii{}]~$\lambda1907,1909$ doublet, in addition to a very tentative \lya{} emission (blue).
    We also indicate the detection of stellar wind features in \nv{} and \civ{} in orange.
    }
    \label{fig:restUV_spec}
\end{figure*}

The $z=9.3$ galaxy that forms the basis of this paper  was initially selected as a luminous $z>8$ photometric candidate from \hst{} imaging \citep{Castellano2016a} and later also with \jwst{} NIRCam data \citep{Atek2023,Castellano2023} in the Abell 2744 field (with coordinates RA = $3.6171694$, Dec = $-30.4255494$).
It was subsequently observed with \jwst{}/NIRSpec high-resolution ($R\sim2700$) grating by the GLASS program (exposure time 4.9~hr in each of F100LP/G140H, F170LP/G235H, and F290LP/G395H; \citealt{Treu2022,Boyett2024_z9p3}) and with low-resolution ($R\sim100$) prism by the UNCOVER program (exposure time 2.6~hr; \citealt{Bezanson2024,Fujimoto2024_uncover}), but very deep spectroscopy detecting the rest-UV and optical continuum at grating ($R\gtrsim1000$) resolution has not been obtained.
It is also known as Gz9p3 \citep{Boyett2024_z9p3}, UNCOVER-3686 \citep{Fujimoto2024_uncover}, and DHZ1 \citep{Castellano2023}, but for consistency with our survey source catalog terminology, 
in what follows, we will refer to it as  \srcname{}, using the identification number from the SPURS Abell 2744 MSA.

\srcname{} is mildly gravitationally lensed, with a magnification factor of $\mu=1.62$ using the recent lensing maps in this field \citep{Furtak2023_lesning_model,Price2025}.
It was resolved by the NIRCam imaging into a bright compact clump that dominates the light in rest-UV ($\simeq 140$ pc in size) and a fainter nearby clump with a  tail extending  $\approx3\mbox{--}4$~kpc in length (also see Figure~\ref{fig:image}), which has been suggested to be undergoing interaction \citep{Boyett2024_z9p3}.
The NIRCam SED of the entire galaxy reveals a blue UV slope ($\beta=-1.94_{-0.06}^{+0.05}$; \citealt{Boyett2024_z9p3}), likely indicating relatively low dust attenuation.
Stellar population modeling of this full NIRCam SED also suggests a large integrated stellar mass ($1.6_{-0.6}^{+0.4}\times10^9~M_\odot$; \citealt{Boyett2024_z9p3}), placing it as one of the most massive galaxies known above $z\simeq 9$.
The NIRCam SED modeling also suggests ongoing star formation activity with a star formation rate (SFR) of  $19_{-4}^{+5}$~$M_\odot$~yr$^{-1}$.
As already mentioned in \S~\ref{sec:intro}, the previous NIRSpec observations have demonstrated \srcname{} has weak rest-UV and optical emission lines ([\oiii{}]~$\lambda5008$ EW $\approx210$~\AA{} and H$\beta$ EW $\approx25$~\AA{}), which we will discuss in more detail in \S~\ref{sec:results} using the newly obtained  SPURS spectrum.

More recently, ALMA observations provide additional insights into the ISM properties of \srcname{} \citep{Algera2025}.
These data reveal  detection of the [\oiii{}]~$88\mu$m line, identifying it as the most intrinsically luminous [\oiii{}]~88 $\mu$m emitter ($L_{\rm [OIII]88} = 1.13_{-0.32}^{+0.37}\times10^9\,L_\odot$)  at $z>9$.
Intriguingly, the centroid of [\oiii{}]~$88\mu$m appears  offset from the dominant rest-UV component of this galaxy, peaking around the extended tail, suggesting the presence of an extended ionized gas reservoir and a deficit of [\oiii{}]~$88\mu$m emission in the main UV-emitting region.
In contrast to its luminous far-IR line emission, \srcname{} was not detected in the thermal dust continuum ($<54.6$~$\mu$Jy at 335.5 GHz), with the upper limit suggesting a low dust mass ($M_d < 1.9\times10^6~M_\odot$;  assuming a fixed dust temperature of 50~K; \citealt{Algera2025}).
\citet{Algera2025} demonstrate that this suggests one of the lowest 
 dust-to-stellar mass ratios known at $z\gtrsim 9$ ($\lesssim0.001$).

\subsection{Deep Spectroscopy with SPURS}
\label{sec:spurs_obs}

We have obtained new ultra-deep \jwst{}/NIRSpec medium resolution ($R\sim1000$) spectroscopy of \srcname{}.
The full description of the SPURS dataset will be presented in a future paper, and we briefly summarize the first observation targeting the Abell 2744 field used in this work.
Data were acquired in Multi-Object Spectroscopy (MOS) mode in two separate visits between 2025 November~6 and November~9.
We use three grating/filter pairs at medium resolution ($R\sim1000$): G140M/F100LP, G235M/F170LP, and G395M/F290LP, with integration times of 29.18, 7.90, and 2.92 hours in each grating, respectively. 
\srcname{} was placed on the NIRSpec Micro-Shutter Assembly with the open shutter centered around the bright clump in the rest-frame UV (Figure~\ref{fig:image}).
Its SPURS spectra were reduced using the \texttt{msaexp} package (v0.9.13), based on the official \jwst{} pipeline version 1.16.1 and reference file mapping \texttt{jwst\_1303.pmap}, following the procedures in previous works (\citealt{deGraaff2024_kinematics,Heintz2025_PRIMAL,Valentino2025}).
During the reduction, we have assumed a point source pathloss correction, consistent with the compact morphology of the bright UV clump of this galaxy (see Figure~\ref{fig:image}).

The resulting spectrum covers 1.05--2.34~\um{} in G140M, 1.76--3.78~\um{} in G235M, and 2.87--5.50~\um{} in G395M in the observed frame.
We achieve wavelength coverage to longer wavelengths than the nominal coverage of each grating through the extended wavelength extraction available in \texttt{msaexp} \citep{Valentino2025}, where we pick up the wavelength range covered by the detector but may be contaminated by second-order spectra.
However, for $z>9$ galaxies, where the \lya{} break shifts to $>1.2$~\um{}, higher-order contamination begins at wavelengths longer than the detector cutoff ($2.4$~\um{}) in G140M.
We also do not expect it to affect G395M, given the wavelength range covered.
For G235M, we exclude wavelengths at $>3.52$~\um{} where the second-order overlap begins, and use only the contamination-free region at 1.76--3.52~\um{}.
Using the systemic redshift determined below, this corresponds to rest-frame wavelengths of 1018--1765~\AA{} in G140M, 1706--3414~\AA{} in G235M, and 2812--5333~\AA{} in G395M, providing continuous coverage from \lya{} to [\oiii{}]$\lambda5008$.
In our analysis below, we will adopt the combined spectrum from all three gratings.
Here, we first inverse-variance stack the G140M spectrum taken in both visits after normalizing them to the same continuum flux over 1.4--1.8~\um{}.
We then inverse-variance stack the other two gratings, both from the second visit, using normalization factors determined from the overlapping regions between gratings (1.8--2.3~\um{} between G140M and G235M, and 2.9--3.5~\um{} between G235M and G395M).

We detect the continuum trace in the deepest G140M grating, with SNR $=10.1$ per pixel at the observed wavelength 1.4~$\mu$m, which is presented in Figure~\ref{fig:restUV_spec}.
With the systemic redshift derived below, we find that the continuum SNR exceeds 6 per pixel across the wavelength range from \lya{} to \ciii{}] (observed frame 1.26--1.97~$\mu$m).
The typical $3\sigma$ flux limit is $4.2\times10^{-20}$~\ergscmA{}, assuming an integration window of 1000~km~s$^{-1}$ ($\sim2$ resolution elements).
At the redshift of this galaxy, this translates to a $3\sigma$ rest-frame EW limit of 0.2~\AA{}.
The continuum is also detected in the two redder gratings, albeit at lower SNRs given the shallower integration time (SNR $=3.2$ per pixel at 2.4~$\mu$m in G235M, and SNR $=2.3$ per pixel at 4.0~$\mu$m in G395M).
We estimate $3\sigma$ flux limits of $7.9\times10^{-20}$~\ergscmA{} in G235M, and $7.0\times10^{-20}$~\ergscmA{} in G395M, with corresponding $3\sigma$ EW limits of 0.9 and 1.9~\AA{}, respectively.

We determine the systemic redshift from the strongest rest-frame optical emission lines (i.e., [\oiii{}]+\hb{}).
We fit Gaussian profiles to each line, fixing their line widths and relative wavelengths, along with a linear function for the continuum, and find $z=9.311$.
We derive consistent redshifts from other emission lines in different gratings (\ciii{}] in G140M and [\oii{}] in the second-order contaminated region of G235M), and they match the values previously reported in \cite{Boyett2024_z9p3}.
For each detected emission line, we measure the line flux by fitting a Gaussian profile plus a linear continuum, fixing the systemic redshift.
For closely separated doublets (e.g., \ciii{}]), we fit all components simultaneously while fixing their line width and relative wavelength.
We repeat the process 1000 times and report 3$\sigma$ upper limits from the measured line flux uncertainties.
We also place 3$\sigma$ upper limits on individual emission lines not detected in our deep spectrum.
To do so, we integrate the continuum-subtracted spectrum over a 5~\AA\ window for single lines (e.g., \heii{}) and an 8~\AA{} window for doublets or multiplets (e.g., \civ{}, \niv{}], \niii{}]), where the window sizes are set by the measured widths of the individual \ciii{}] components and take into account the slightly lower resolution at the blue end.
Here, the continuum is determined from the median flux level in adjacent line-free regions.
The line fluxes and uncertainties (or upper limits) reported in this paper are not corrected for gravitational magnification.
For emission line FWHMs, we correct for instrumental broadening, assuming 1.3$\times$ the resolution curve from the \jwst{} User Documentation~\footnote{\url{https://jwst-docs.stsci.edu/jwst-near-infrared-spectrograph/nirspec-instrumentation/nirspec-dispersers-and-filters}}. The factor of 1.3 accounts for the fact that the actual spectral resolution is $\approx1.3\times$ better than the values tabulated in the \jwst{} User Documentation, which are based on pre-launch estimates \citep[e.g.,][]{deGraaff2024_kinematics,Shajib2025_NRSresol}.
We report the measurements for rest-UV and optical emission lines in Table~\ref{tab:restUVOptLines}.

\section{Results}\label{sec:results}
In this section, we characterize the SPURS spectra of \srcname{}, starting with emission lines (\S~\ref{subsec:uvlines} and \S~\ref{subsec:optspec}), then characterizing absorption from the ISM and CGM (\S~\ref{subsec:ism_abs}), IGM (\S~\ref{subsec:lya_abs}), and stars (\S~\ref{subsec:wind}). We summarize the picture that emerges in \S~\ref{subsec:res_summary}.

\begin{figure*}[t]
    \centering
    \includegraphics[width=1\linewidth]{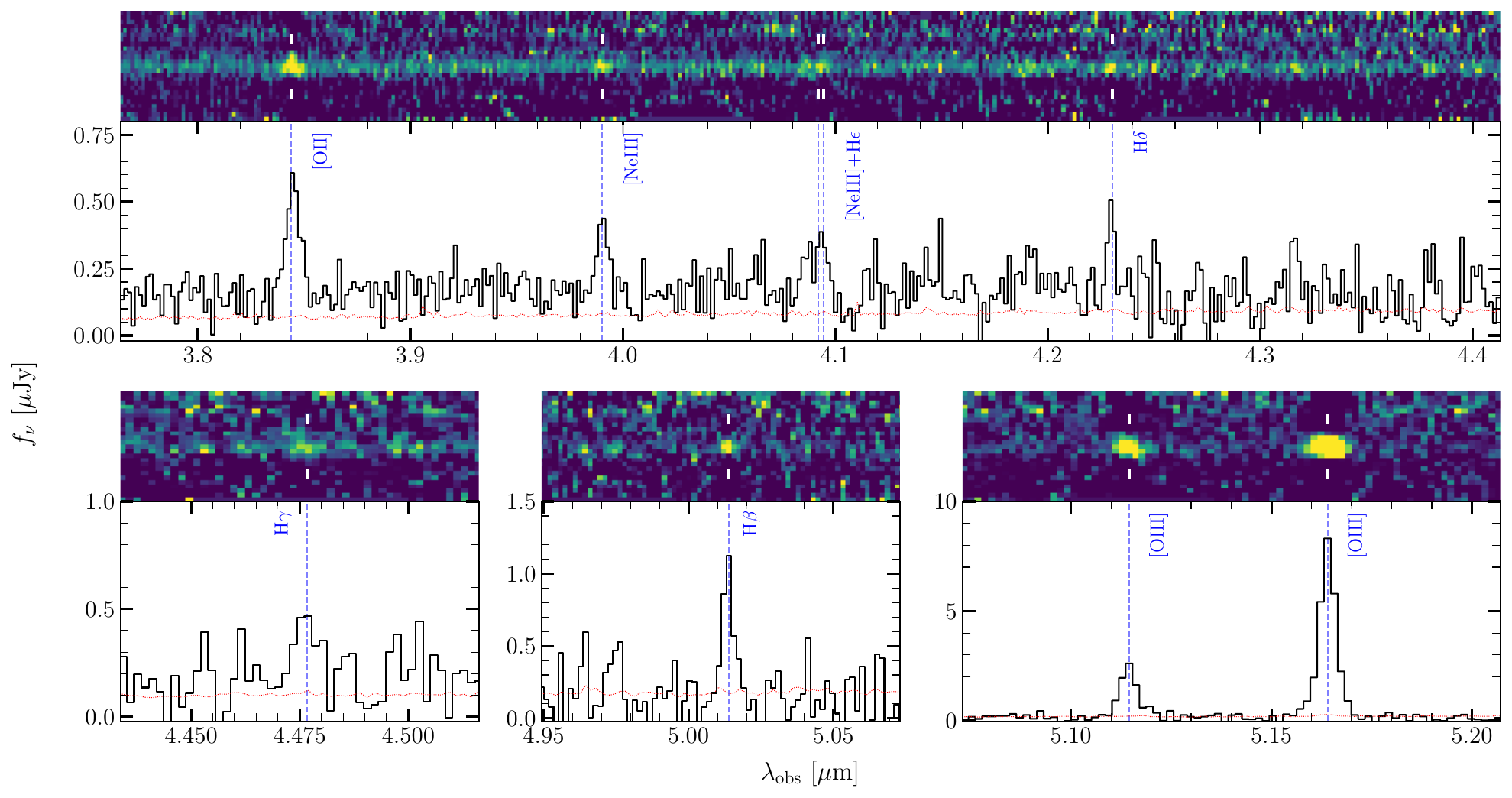}
    \caption{The rest-frame optical spectrum from G395M. 
    We mark the detected (SNR$>3$) emission lines with blue dashed lines on the 1D spectrum and white bars on the 2D spectrum.
    The red dotted line corresponds to the 1D error spectrum.}
    \label{fig:restOpt_spec}
\end{figure*}

\begin{table}[t!]
    \centering
    \begin{tabular}{cccc}
    \hline \hline
        Line                              & Flux                       & EW                       & FWHM \\
                                          & [$10^{-20}$ergs/s/cm$^2$]  & [\AA]                  & [km\,s$^{-1}$] \\
    \hline \hline
    \multicolumn{4}{c}{Rest-UV Lines} \\
    \hline 
        \niv{}]~$\lambda\lambda1483,1486$  & $            < 17.4$       & $            <  0.8$   &  -- \\
        \civ{}~$\lambda\lambda1548,1550$  & $            <  5.6$       & $            <  0.3$   &  -- \\
        \heii{}~$\lambda1640$              & $            < 13.9$       & $            <  0.8$   &  -- \\
        \oiii{}]~$\lambda\lambda1660,1666$ & $            < 34.7$       & $            <  2.0$   &  -- \\
        \niii{}]~$\lambda1750$             & $            < 28.2$       & $            <  1.8$   &  -- \\
        \ciii{}]~$\lambda1907$             & $15.8_{-3.6}^{+3.2}$       & $ 1.2_{-0.3}^{+0.3}$   &  $259_{-78}^{+41}$\\
        \ciii{}]~$\lambda1909$             & $ 5.3_{-3.4}^{+3.2}$       & $ 0.4_{-0.2}^{+0.2}$   &  $259_{-78}^{+41}$\\
        
    \hline \hline
    \multicolumn{4}{c}{Rest-optical Lines} \\
    \hline
       $[\oii{}]$~$\lambda3728$      & $      73_{-    9}^{+    9} $ & $     24_{-   4}^{+   5} $ &                        -- \\
       $[\neiii{}]$~$\lambda3869$                & $      35_{-    9}^{+    9} $ & $     12_{-   3}^{+   4} $ & $   369_{- 99}^{+ 126} $ \\
       $[\neiii{}]$~$\lambda3968$+H$\epsilon$    & $      20_{-    7}^{+    6} $ & $    6.6_{- 2.3}^{+ 2.1} $ &                        -- \\
       H$\delta$                                 & $      22_{-    6}^{+    6} $ & $    7.3_{- 1.9}^{+ 2.4} $ &                        -- \\
       H$\gamma$                                 & $      32_{-    8}^{+    8} $ & $     13_{-   4}^{+   4} $ & $   206_{- 130}^{+ 185} $ \\
       H$\beta$                                  & $      53_{-    7}^{+    7} $ & $     27_{-   4}^{+   4} $ & $   233_{-   9}^{+  10} $ \\
       $[\oiii]$~$\lambda4960$                   & $     144_{-    9}^{+    9} $ & $     82_{-  12}^{+  14} $ & $   229_{-   9}^{+   9} $ \\
       $[\oiii]$~$\lambda5008$                   & $     463_{-   14}^{+   12} $ & $    257_{-  22}^{+  23} $ & $   227_{-   9}^{+   9} $ \\

    \hline
    \end{tabular}
    \caption{Measurements of emission lines in the rest-UV and optical spectrum. 
    We list each detected emission line, reporting its line flux (not corrected for gravitational magnification), equivalent width, and FWHM.
    We also report the upper limits of flux and equivalent width for other rest-UV emission lines not detected in the SPURS spectrum.}
    \label{tab:restUVOptLines}
\end{table}

\subsection{Rest-UV Emission Lines}\label{subsec:uvlines}

The deep rest-UV spectrum of \srcname{}  reveals only a single emission line, the 
 \ciii{}]~$\lambda\lambda$1907,1909 doublet (see Figure~\ref{fig:restUV_spec}). 
Using the method described in \S~\ref{sec:data}, we measure a flux of $15.8_{-3.6}^{+3.2}\times10^{-20}$~\ergscmA{} for \ciii{}]~$\lambda$1907 and $5.3_{-3.4}^{+3.2}\times10^{-20}$~\ergscmA{} for \ciii{}]~$\lambda$1909.
The total EWs for both components are $1.6_{-0.3}^{+0.3}$~\AA{}, which is a factor of 8 lower than the values in $z>9$ composite spectra \citep[e.g.,][]{Tang2025,Roberts-Borsani2025}. 
Such weak \ciii{}] emission has been commonly observed at lower redshifts (e.g., \citealt{Du2017,LeFevre2019,Llerena2022,Mainali2023}) and is also seen in UV-luminous galaxies at $z\simeq 6\mbox{--}7$ \citep[e.g.,][]{Endsley2025,Shapley2025_z7}. 
In these cases, the weaker \ciii{}] EWs likely reflect a combination of older stellar population ages and larger gas-phase metallicities.

The \ciii{}] profile of \srcname{} is moderately resolved, revealing a dominant blue peak at 1907~\AA{} and a flux ratio of $f_{\rm 1907}/f_{\rm 1909} = 2.9_{-1.2}^{+3.8}$. 
The \ciii{}] doublet ratio is sensitive to the electron density, approaching unity at very high densities ($>2\times10^4$~cm$^{-3}$). 
Converting the \ciii{}] flux ratio to an ionized gas density using the {\tt pyneb} package \citep{Luridiana2015}, we find that the observed limit on the doublet ratio ($f_{\rm 1907}/f_{\rm 1909}>1.5$) rules out such very high densities. However, it is consistent with a range of moderate-to-lower densities ($<10^{3}$~cm$^{-3}$). We note that \cite{Boyett2024_z9p3} has also used the [\oii{}] doublet resolved in the GLASS high-resolution spectrum to derive an electron density of $590_{-250}^{+570}$~cm$^{-3}$. 
The densities of \srcname{} appear  lower than the super-luminous galaxies at $z>9$ with stronger emission lines \citep[e.g.,][]{Senchyna2024,Maiolino2024_GNz11,Zavala2024}.

The absence of high ionization line emission (\civ{}, \heii{}) indicates that hard radiation fields are not present in \srcname{}. 
The implied \civ{}/\ciii{}] ratio ($< 0.27$) is at least $10\times$ lower than that seen in GHZ2 \citep{Castellano2024}. 
The upper limits on the \civ{} and \heii{} EWs ($<0.29$ and $<0.77$~\AA{}, respectively) similarly suggest a weaker radiation field. 
We place $3\sigma$ upper limits on \niv{}] and \niii{}], with equivalent widths EW \niv{}]~$<0.8$~\AA{} and EW \niii{}]~$<$ 1.8~\AA{}.
As the \ciii{}] is also weak, we obtain only an upper limit on the \niii{}]/\ciii{}] ratio of 1.38.

The Ly$\alpha$ profile is dominated by absorption, although we find evidence for a faint, tentative emission line superimposed on the absorption. 
We will characterize this portion of the spectrum in detail in \S~\ref{subsec:lya_abs}.  
The weakness of Ly$\alpha$ is consistent with expectations given the strong IGM damping wing that is expected to be typical at $z\simeq 9$ \citep[e.g.,][]{Tang2024_nirspec,Jones2025,Kageura2025,Mason2026,Napolitano2025_uds}.

\subsection{Rest-Optical Emission Lines and Balmer Break}\label{subsec:optspec}

\begin{figure*}
    \centering
    \includegraphics[width=1\linewidth]{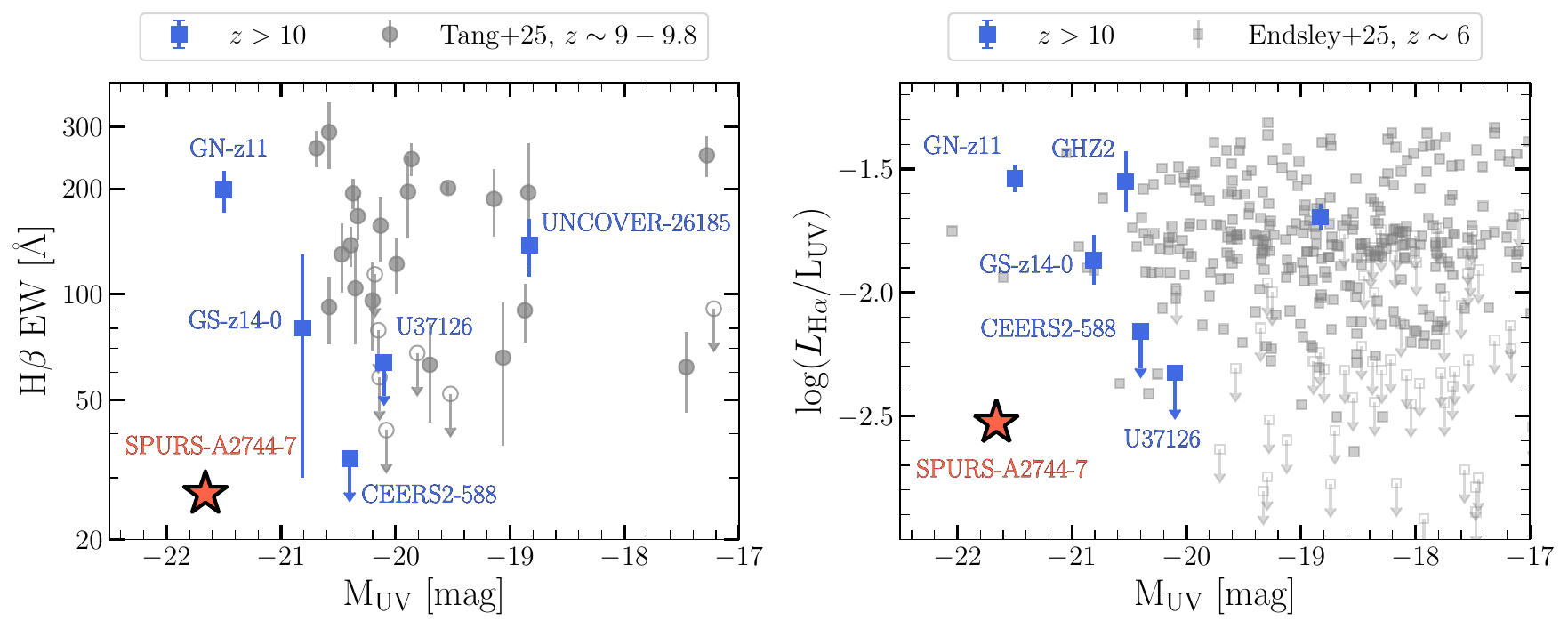}
    \caption{
    Left: Equivalent width (EW) of H$\beta$ as a function of absolute UV magnitude ($M_{\rm UV}$). 
    For comparison, we include the $z \simeq 9\mbox{--}9.8$ NIRSpec sample from \cite{Tang2025}, alongside $z > 10$ galaxies with MIRI observations \citep{Helton2025, MarquesChaves2026, AlvarezMarquez2026, Harikane2026} or NIRSpec H$\gamma$ measurements \citep{Bunker2023}. 
    Here, when \hb{} EW is not directly available, values are scaled from H$\alpha$ or H$\gamma$ assuming intrinsic flux ratios and a flat continuum in $f_\nu$. 
    Despite being among the most intrinsically UV-luminous galaxies yet discovered at $z>9$, \srcname{} exhibits one of the lowest \hb{} equivalent widths observed in this epoch.
    Right: \ha{}-to-UV luminosity ratio ($L_{\rm H\alpha}/L_{\rm UV}$) as a function of absolute UV magnitude. 
    For \srcname{}, we use \ha{} luminosity scaled from that of \hb{} (assuming no dust correction as expected from other Balmer line ratios) and adopt the UV luminosity directly measured from the spectrum to avoid aperture issues.
    The derived $L_{\rm H\alpha}/L_{\rm UV}$ ratio is significantly lower than that of extreme sources like GHZ2 and GN-z11, and is smaller than the lowest upper limits at $z>10$ with MIRI observations (\citealt{AlvarezMarquez2024,Zavala2024}, in addition to the aforementioned references).
    For comparison, we also plot the $z\sim6$ Lyman break galaxies from \cite{Endsley2025_bursty}.}
    \label{fig:weakline}
\end{figure*}

\begin{figure*}
    \centering
    \includegraphics[width=1\linewidth]{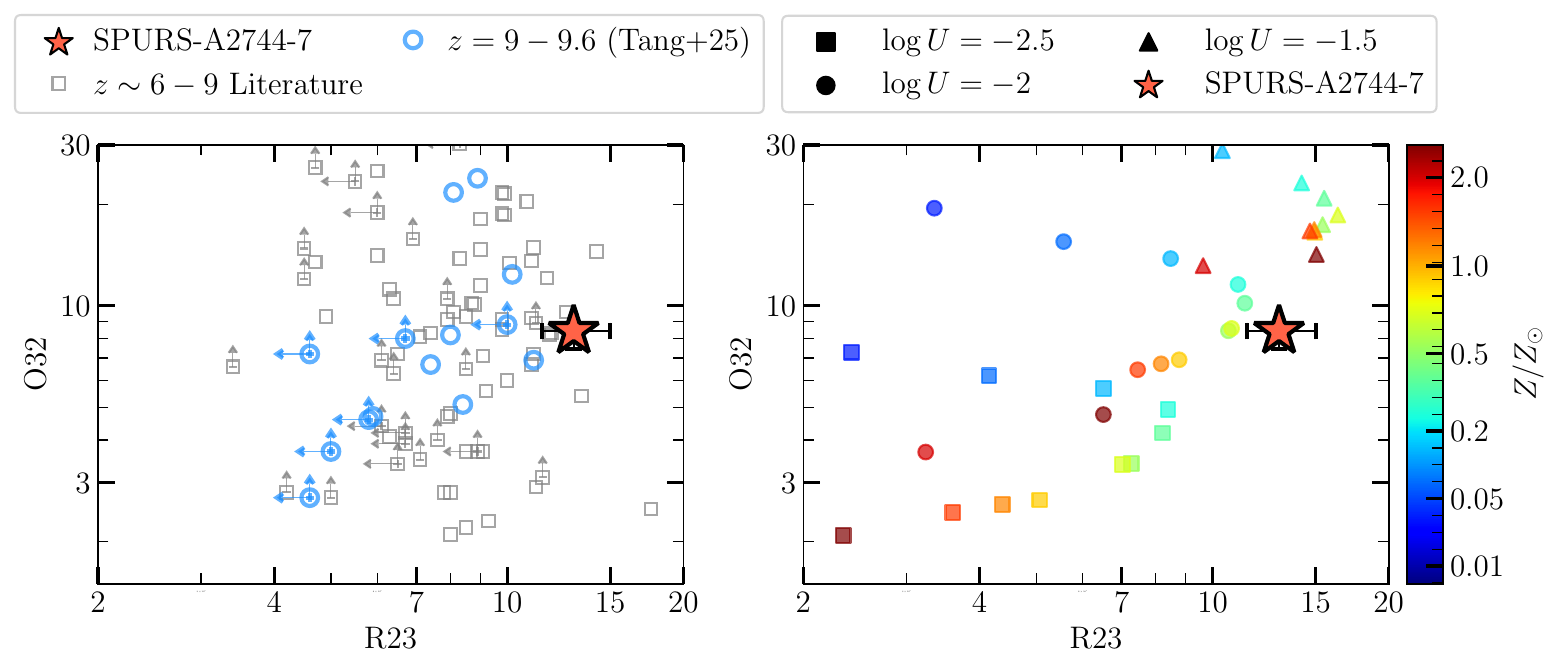}
    \caption{Left: Observed O32 ([\oiii{}]$\lambda\lambda$4960,5008/[\oii{}]) versus R23  (([\oiii{}]$\lambda\lambda$4960,5008+[\oii{}])/\hb{}) in \srcname{}.
    Compared to literature samples at $z=9.0\mbox{--}9.6$ \citep{Tang2025} and $z\sim6\mbox{--}9$ \citep{Cameron2023,Nakajima2023,Tang2023}, this source has a comparably large O32 ratio, yet its R23 ratio still lies at the high end.
    Here, we only plot the literature samples with [\oii{}] detections, but including the non-detections does not alter the fact that \srcname{} has relatively large R23 given its observed O32 value.
    Right: Observed O32 and R23 ratios in the context of photoionization models of star-forming galaxies \citep{Gutkin2016} for a grid of ionization parameters and metallicities.
    Reproducing the observed line ratio requires models with both a large ionization parameter (log~$U\sim-2$) and moderately enriched gas ($\sim 0.4\mbox{--}0.6\,Z_\odot$).}
    \label{fig:O32R23}
\end{figure*}

The rest-optical spectrum of \srcname{} reveals weak emission lines, which are shown in Figure~\ref{fig:restOpt_spec}.
We measure a [\oiii{}]$\lambda$5007 ([\oiii{}]$\lambda$4959) flux of $4.63_{-0.14}^{+0.12}\times10^{-18}$~\ergscm{} ($1.43_{-0.09}^{+0.09}\times10^{-18}$~\ergscm{}).
This corresponds to a [\oiii{}]$\lambda$5007 EW of $257_{-22}^{+23}$~\AA{}, which is $3\times$ lower than the average values derived from the composite prism spectra of $z>9$ galaxies (e.g., \citealt{RobertsBorsani2024,Tang2025}).
The [\oiii{}]$\lambda$5007 emission line can be fit by a single Gaussian profile (FWHM = $225_{-9}^{+9}$~\kms{} after correction for instrumental resolution as described in \S~\ref{sec:data}).

We also detect a series of Balmer emission lines in the SPURS spectrum.
We measure an \hb{} line flux of $0.53_{-0.07}^{+0.07}\times10^{-18}$~\ergscm{},  which corresponds to an \hb{} EW of $27_{-4}^{+4}$~\AA{}.
As shown in the left panel of Figure~\ref{fig:weakline}, this EW is among the smallest values known at these redshifts.
We  detect individual higher-order Balmer lines  with EW~\hg{} $= 13_{-4}^{+4}$~\AA{} and EW~\hd{} $= 7_{-3}^{+3}$~\AA{}, as well as the unresolved [\neiii{}]$\lambda3968$+\he{} blend with EW~$= 7_{-2}^{+2}$~\AA{}. 
All of these lines are consistently weaker (by a factor of $\sim5$) than the typical values derived from the composite prism spectra at $z>9$ (e.g., \citealt{RobertsBorsani2024,Tang2025}).

The \ha{}-to-UV luminosity ratio, $L_{\rm H\alpha}/L_{\rm UV}$, is sensitive to variations in recent star formation activity \citep[e.g.][]{Emami2019,Asada2023,Clarke2024,Endsley2025_bursty}.
For galaxies undergoing steady star formation (at a constant rate over $>100$~Myr), we expect an equilibrium value of $\log(L_{\rm H\alpha}/L_{\rm UV})\approx-1.8$ (based on the \citealt{Gutkin2016} photoionization models assuming no dust).
Because \ha{} shifts out of the G395M wavelength coverage, we estimate its luminosity by scaling from \hb{}, assuming no dust correction (as expected from the Balmer emission line ratios discussed below).
For UV luminosity, we also adopt the value directly measured from the spectrum (median flux over a 100~\AA{} window centered at 1500~\AA{} in the rest frame), so that both luminosities are measured with the same aperture.
We plot the resulting \ha{}-to-UV luminosity ratio in the right panel of Figure~\ref{fig:weakline}.
We find an extremely small luminosity ratio, $\log(L_{\rm H\alpha}/L_{\rm UV}) = -2.53_{-0.06}^{+0.05}$, which is a factor of 10 lower than what has been found in the extreme sources such as GN-z11 and GHZ2 \citep{AlvarezMarquez2024,Zavala2024}.
Such small values have been observed only in a small fraction of $z\sim 6$ Lyman break galaxies identified with strong recent declines in SFR \citep[e.g.][]{Endsley2025_bursty}.

The Balmer emission line ratios provide additional constraints on the dust attenuation.
We measure \hg{}/\hb{} $= 0.59_{-0.15}^{+0.19}$, consistent with the theoretical ratio expected from case B recombination (0.468 assuming an electron temperature of $T_e=10^4$~K; \citealt{Osterbrock2006}). 
The weaker \hd{} line also implies flux ratios consistent with, or exceeding, the theoretical values, with the measured \hd{}/\hb{} $= 0.41_{-0.21}^{+0.27}$ (intrinsic 0.259). 
These measurements suggest the dust attenuation is unlikely to be very significant, though a small amount of dust may be allowed given the uncertainties.
While our measurements here are not corrected for stellar absorption, we note that the stellar \hg{} absorption is comparable or stronger than \hb{} such that the intrinsic \hg{}/\hb{} ratio of nebular emission could be higher and thus more consistent with little dust.
The UV continuum slope is also blue ($\beta=-1.94$), albeit somewhat redder than typical at $z\simeq 9$ (e.g., \citealt{Austin2025,Cullen2024,Saxena2024_slope,Topping2024}). 
We will also show in \S~\ref{sec:modeling} that the grating continuum is also consistent with only modest attenuation, similar to what might be expected based on the absence of significant dust continuum emission in the far-infrared \citep{Algera2025}.

The emission line flux ratios allow us to additionally investigate the physical conditions of the nebular gas.
The O32 index ($\equiv$[\oiii{}]$\lambda\lambda4960,5008$/[\oii{}]$\lambda\lambda3728$) is sensitive to the ionization state and thus probes the ionization parameter (log~$U$) of the nebular gas.
The Ne3O2 index ($\equiv$[\neiii{}]$\lambda3869$/[\oii{}]$\lambda\lambda3728$) is similarly sensitive to the gas ionization conditions as O32, as both neon and oxygen are $\alpha$-elements and Ne$^{++}$ and O$^{++}$ have similar ionization potentials.
We find large values for ionization-sensitive ratios (O32 $= 8.4_{-0.9}^{+1.2}$ and Ne3O2 $= 0.47_{-0.12}^{+0.13}$),  comparable to those found in typical $z\gtrsim9$ galaxies (e.g., \citealt{RobertsBorsani2024,Tang2025,Cameron2023}). 
While we will perform full emission line modeling in \S~\ref{sec:modeling}, here we empirically characterize the gas ionization parameter using literature calibrations between these indices and log~$U$.
We find that SPURS-A2744-7 has ionized gas with a large ionization parameter  (log~$U=-1.8$ to $-2.2$), despite its weak emission line spectrum. 
These line ratios and inferred gas ionization parameter are considerably higher than those typically found at lower redshifts (O32 $\simeq1\mbox{--}2$ at $z\sim2$ and 3--5 at $z\sim3\mbox{--}5$; e.g., \citealt{Sanders2016,Steidel2016,Shapley2023}).
It has been shown that very large O32 and Ne3O2 ratios scale with the rest-frame optical emission line EWs, such that the line ratios we measure are usually associated with galaxies with very high [\oiii{}] EWs ($\sim$1000~\AA{};  \citealt{Tang2019,Tang2023,Sanders2020,Boyett2024}).
The combination of large O32 and large Ne3O2 in a galaxy with relatively weak rest-optical lines may hint at differences in physical conditions of the \hii{} regions in SPURS-A2744-7 relative to galaxies at lower redshifts with weak emission line spectra.

The O3 ($\equiv$[\oiii{}]$\lambda$5008/\hb{}) and R23 ($\equiv$([\oiii{}]+[\oii{}])/\hb{}) ratios probe the gas excitation conditions.
We measure very large values for both ratios, with O3 $= 8.7_{-1.0}^{+1.3}$ and R23 $= 13.0_{-1.5}^{+2.0}$, comparable to the largest values seen in other $z\gtrsim9$ galaxies (see the left panel of Figure~\ref{fig:O32R23}; \citealt{AlvarezMarquez2024,Hsiao2024a,Zavala2024,Helton2025,Tang2025}).
Both are also significantly greater than values typically measured at lower redshifts (O3 $=3.4$--$4.5$, R23 $=8.5$--$9.1$; \citealt{Steidel2016,Sanders2021}), indicating gas with extreme excitation conditions.
In fact, as shown in the left panel of Figure~\ref{fig:O32R23}, \srcname{} has one of the largest R23 values among  $z>6$ galaxies with similarly large O32 ratios.
Given the large ionization parameter (log~$U\sim-2$) implied by O32 and Ne3O2, the large R23 ratio may indicate a moderately large gas-phase metallicity. To illustrate this, 
we compare the line ratios directly to those expected from the \citet{Gutkin2016} photoionization models in the right panel of Figure~\ref{fig:O32R23}. To reproduce the line ratios, we see that models with gas-phase metallicity in the range
12+log~O/H $\sim8.3\mbox{--}8.5$ (corresponding to 0.4--0.6~$Z_\odot$) provide the best fit to the line ratios.  
We note that individual strong line ratios give a varied picture of the metallicity. 
The large observed R23 and R3 values are both larger than the mean relations between these ratios and metallicities (e.g., \citealt{Sanders2025}). If we consider the 
 scatter of these relations, we find solutions in the range 12+log~O/H $=7.8\mbox{--}8.3$ considering the intrinsic(\citealt{Sanders2025}).
As we noted above, the ionization parameter is large, so not surprisingly, the O32 strong line calibrations suggest a lower metallicity (12+log O/H = 7.9 from \citealt{Sanders2025}).  These values are somewhat lower than those implied by the photoionization models described above. A deeper optical spectrum capable of detecting [OIII]$\lambda4363$ will ultimately be required for a more robust gas-phase metallicity measurement.

The continuum shape also constrains the recent star formation history. 
Here, we focus on empirical measurements of the Balmer break, which is sensitive to the stellar population age.
We empirically quantify the strength of the Balmer break using the flux ratio (in $f_\nu$) between rest-frame 4050~\AA{} and 3500~\AA{}, $f_{\nu, 4050}/f_{\nu, 3500}$.
For simple stellar populations (SSPs) with a moderate metallicity of $0.5\,Z_\odot$ (as suggested by the photoionization models described above), we expect a flat Balmer break or a Balmer jump ($f_{\nu, 4050}/f_{\nu, 3500}\leq1$) at very young ages ($\leq 5$~Myr). The break rapidly strengthens to $f_{\nu, 4050}/f_{\nu, 3500}\simeq1.5$ after 20~Myr following the burst of star formation.
We find a weak Balmer break in \srcname{}, $f_{\nu, 4050}/f_{\nu, 3500} = 1.14_{-0.09}^{+0.07}$. In the context of an SSP, this implies the source is observed shortly after the burst of star formation, with an age of only $\sim5\mbox{--}10$~Myr.
Alternatively, similar Balmer break amplitudes can also arise in galaxies undergoing constant star formation for a more prolonged period ($\sim30$~Myr, assuming $0.5\,Z_\odot$ as above).
These two scenarios predict very different rest-optical emission line strengths, with the EW \hb{} in the constant star formation case $\gtrsim3\times$ larger than that of the SSP \citep{Endsley2025_bursty}.
The observed EW \hb{} is better reproduced by the post burst SSP solution (with age of $\simeq 5$ Myr).  We note that a large ionizing photon escape fraction could also contribute to weaker emission lines, but we will show that this is unlikely to be the case in \S~\ref{sec:modeling}.

\subsection{Interstellar Absorption Lines }\label{subsec:ism_abs}

\begin{figure*}[t!]
    \centering
    \includegraphics[width=0.68\linewidth]{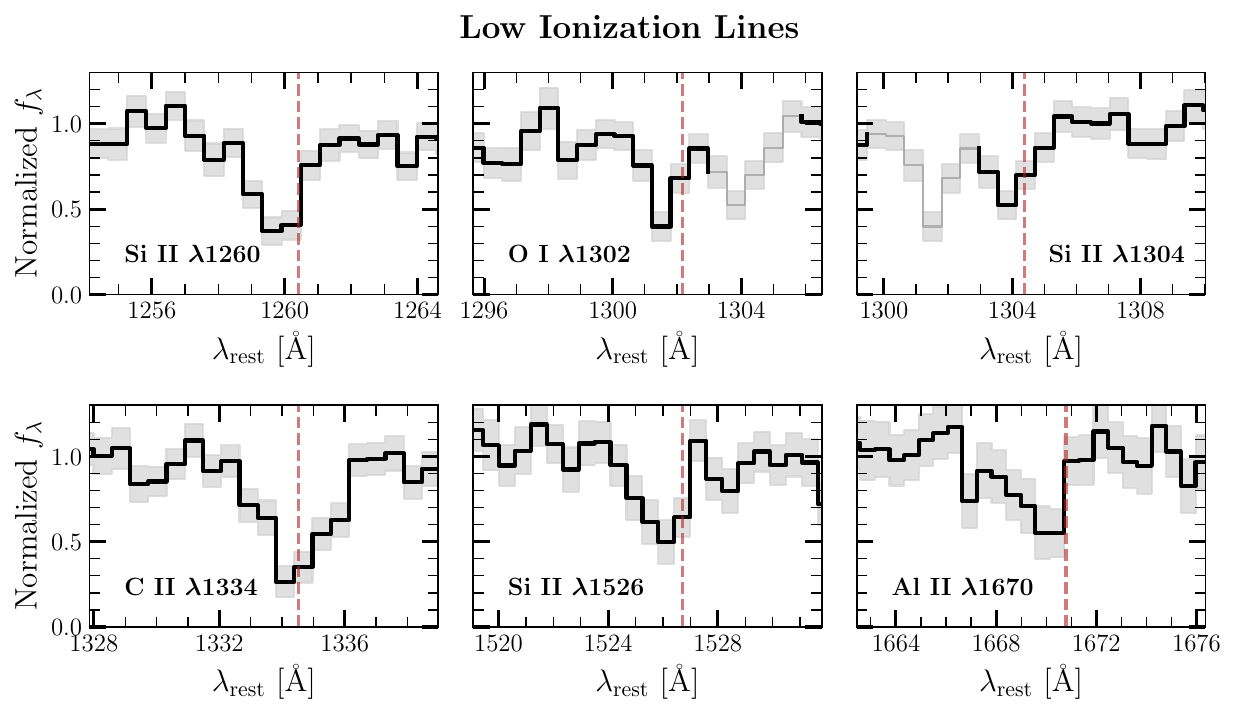}
    \includegraphics[width=0.2735\linewidth]{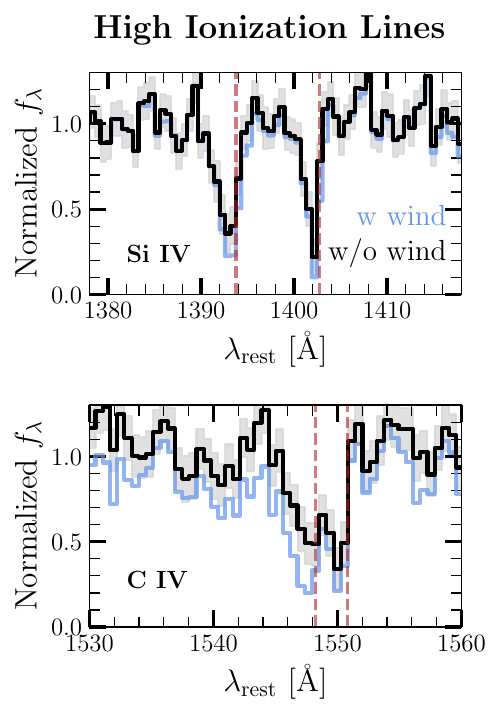}
    \caption{Rest-frame UV interstellar absorption lines detected in the G140M spectrum of \srcname{}. 
    The left three columns show the low ionization absorption lines, while the rightmost column shows the high ionization absorption lines. 
    For the high ionization \siiv{} and \civ{} lines, the ISM absorption could potentially be contaminated by broad absorption by the stellar wind.
    Therefore, we show both the original absorption profiles (light blue lines) as well as the profiles after we subtract the contribution from the stellar wind as modeled by \beagle{} in \S~\ref{subsubsec:fit_continuum} (black lines). 
    In each panel, the systemic wavelength of each line is indicated by a red dashed line. The error spectrum is shown in shaded gray.}
    \label{fig:absorption_lines}
\end{figure*}

\begin{figure}
    \centering
    \includegraphics[width=\linewidth]{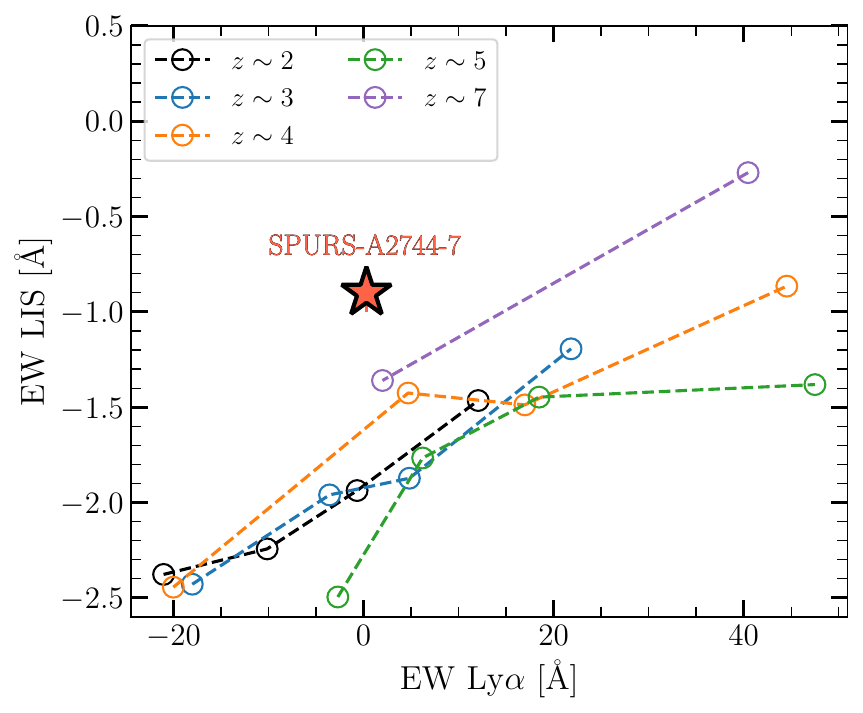}
    \caption{Equivalent width (EW) of low-ionization state (LIS) absorption lines compared to the \lya{} EW.
    The LIS absorption line is known to correlate with the \lya{} EW, such that weak LIS absorption is primarily seen in strong \lya{} emitters as found in the samples of Lyman Break Galaxies at $z\sim2--7$ \citep{Shapley2003,Jones2012,Pahl2020,Glazer2025}.
    \srcname{} appears to be an outlier with very weak \lya{}, despite the fact that the average LIS EW is relatively smaller than that of the lower-redshift galaxies.}
    \label{fig:LIS_EW_z}
\end{figure}

\begin{table}[t!]
    \centering
    \begin{tabular}{cccccc}
    \hline
        Ion & $\lambda_{\rm rest}$ & f & EW & FWHM & $\Delta v$ \\
             & [\AA] &  & [\AA] & [km\,s$^{-1}$] & [km\,s$^{-1}$] \\
    \hline \hline
        \siii{} & 1260.422 & 1.007               & $-1.2_{-0.1}^{+0.1}$ & $408_{- 56}^{+ 48}$ & $-161_{- 23}^{+ 24}$  \\
        \oi{}   & 1302.169 & 0.04887             & $-0.7_{-0.1}^{+0.2}$ & $240_{- 45}^{+ 79}$ & $-132_{- 22}^{+ 36}$  \\
        \siii{} & 1304.370 & 0.094               & $-0.7_{-0.1}^{+0.1}$ & $283_{- 56}^{+ 75}$ & $-114_{- 28}^{+ 26}$  \\
        \cii{}  & 1334.532 & 0.1278              & $-1.8_{-0.2}^{+0.2}$ & $481_{- 85}^{+ 91}$ & $-50_{- 47}^{+ 70}$   \\
        \siiv{} & 1393.760 & 0.5140              & $-1.8_{-0.2}^{+0.2}$ & $530_{- 89}^{+ 86}$ & $-137_{- 57}^{+ 58}$  \\
        \siiv{} & 1402.773 & 0.2553              & $-1.2_{-0.2}^{+0.2}$ & $314_{- 53}^{+ 44}$ & $-160_{- 26}^{+ 22}$  \\
        \siii{} & 1526.707 & 0.130               & $-0.9_{-0.2}^{+0.2}$ & $310_{- 68}^{+ 84}$ & $-158_{- 36}^{+ 46}$  \\
        \civ{}  & 1548.204 & 0.1908              & $-1.4_{-0.2}^{+0.2}$ & $440_{- 100}^{+114}$ & $-64_{- 178}^{+ 61}$  \\
        \civ{}  & 1550.781 & 0.09522             & $-1.0_{-0.1}^{+0.1}$ & $340_{- 51}^{+ 42}$ & $-177_{- 25}^{+ 12}$  \\
        \alii{} & 1670.789 & 1.833               & $-1.1_{-0.3}^{+0.4}$ & $347_{-118}^{+229}$ & $-187_{- 83}^{+ 44}$  \\
    \hline
    \end{tabular}
    \caption{Measurements of rest-UV interstellar absorption lines.
    For each absorption line, we report the associated ion, rest-frame wavelength, oscillator strength, observed rest-frame equivalent width, line width, and the velocity offset ($\Delta v$) of the maximum absorption relative to the systemic redshift.}
    \label{tab:absorption_lines}
\end{table}

We detect a suite of metal absorption lines tracing an enriched interstellar medium and circumgalactic medium (ISM and CGM) surrounding the young stellar population powering the far-UV continuum. 
We find absorption from both low ionization species (LIS; \siii{}~$\lambda1260$, \oi{}~$\lambda1302$, \siii{}~$\lambda1304$, \ciii{}~$\lambda1334$, \siii{}~$\lambda1526$, \alii{}~$\lambda1670$) and higher ionization species (HIS; \siiv{}~$\lambda\lambda1393,1402$, \civ{}~$\lambda\lambda1548,1550$).
We present the absorption profile of each line in Figure~\ref{fig:absorption_lines}.
We characterize their basic properties (EWs, velocities) and describe constraints on the gas covering fraction and kinematics below.

Before measuring the absorption line properties, we must normalize the absorption profile by the continuum.
We fit the continuum with a power-law function ($f_\lambda\propto\lambda^{\beta}$) over rest-frame 1300--1700~\AA{}, where we mask each absorption feature detected above with a conservative window of 20~\AA{}. 
For each line, we assume a Gaussian optical depth profile, but we note that assuming a direct Gaussian fit to the absorption profile does not affect our measurement.
As we will show below, some of the absorption lines appear asymmetric with blueshifted wings, which can arise from the presence of more than one absorption component at different velocities. 
In these cases, we fit with two Gaussian functions to describe both the slower- and faster-moving velocity components \citep[e.g.,][]{Xu2022,G.C.2025}. 
We additionally include a linear component to model the local continuum.
We convolve the models with the NIRSpec line-spread function at the corresponding wavelength before comparing them to the observed absorption profile.
We measure the equivalent widths, velocity at peak absorption ($v_{\rm peak}$), and velocity width (FWHM) from the Gaussian fits.
We report the measurements and uncertainties from 1000 perturbations of the observed spectrum according to the errors (Table~\ref{tab:absorption_lines}).

The physical interpretation of LIS absorption lines depends on their optical depth: while their EWs are sensitive to gas column densities in the optically thin regime, these lines are often saturated in Lyman-break galaxies, meaning their residual intensity provides an important diagnostic of the covering fraction of neutral gas \citep[e.g.,][]{Shapley2003,Jones2013}.
We investigate whether the LIS rest-UV absorption lines in \srcname{} are saturated by comparing their relative equivalent widths.
We first consider the three \siii{} transitions \siii{}~$\lambda1260$, \siii{}~$\lambda1304$, and \siii{}~$\lambda1526$.
In the optically thin regime, we expect the absorption line EW to scale as EW~$\propto Nf\lambda^2$, where $N$ is the ion column density, $f$ is the oscillator strength, and $\lambda$ is the rest-frame wavelength.
Considering the ratio of oscillator strengths and wavelengths of the \siii{} transitions, we expect EW~(1260)/EW~(1304) = 10.00 and EW~(1260)/EW~(1526) = 5.24 when the lines are optically thin.
We measure EW~(1260) = $-1.2_{-0.1}^{+0.1}$~\AA{}, EW~(1304) = $-0.7_{-0.1}^{+0.2}$~\AA{}, and EW~(1526) = $-0.9_{-0.2}^{+0.2}$~\AA{}.
This results in ratios of EW~(1260)/EW~(1304) = $1.7_{-0.4}^{+0.4}$ and EW~(1260)/EW~(1526) = $1.4_{-0.3}^{+0.4}$, both of which likely indicate saturation in at least the stronger \siii{}~$\lambda1260$ line. Thus, we expect that the absorption depth is more sensitive to the gas covering fraction rather than column density.

The fact that the absorption does not reach zero flux is consistent with a non-uniform covering fraction of absorbing gas along the line of sight. The observed depth of the stronger \siii{}~$\lambda1260$ line suggests a covering fraction of $f_c=0.67_{-0.03}^{+0.02}$, which is comparable to what we infer from the other lines ($f_c=0.61_{-0.09}^{+0.03}$ from \siii{}~$\lambda1304$ and $f_c=0.78_{-0.08}^{+0.07}$ from \cii{}~$\lambda1334$) assuming saturation. 
The covering fraction of LIS absorption lines is a known proxy for that of \hi{}, with \hi{} systematically having a larger covering fraction than the LIS lines.
Adopting empirical calibrations between the two \citep[e.g.,][]{Reddy2016,Reddy2022,Gazagnes2018,Saldana-Lopez2022}, we find a large covering fraction (0.82--0.90, depending on the calibration used) for the \hi{} gas in \srcname{}.
We note that at a spectral resolution of $R \sim 1000$, the inferred LIS covering fractions may only represent lower limits if there are unresolved, very narrow absorption components, in which case the \hi{} covering fraction may be even larger.
Consequently, it appears unlikely that SPURS-A2744-7 has a large ionizing photon escape fraction along the line of sight (i.e., $>$0.5), with an inferred escape fraction of only $0.1\mbox{--}0.2$ assuming no dust in the uncovered sightlines.

The average LIS absorption EW measured from the median stacked profile is $-0.9_{-0.1}^{+0.1}$~\AA{}, somewhat weaker than typical values seen in $z\sim2\mbox{--}3$ UV-continuum selected galaxies ($\approx -1.8$~\AA{}; e.g., \citealt{Shapley2003,Steidel2010,Du2018}) and at $z\sim4\mbox{--}5$ ($\approx -1.5$~\AA{}, \citealt{Jones2012,Pahl2020}).
It is also slightly weaker than the values measured from the stacked NIRSpec grating spectra at $z\sim7$ ($\approx -1.2$~\AA{}, \citealt{Glazer2025}). 
This is consistent with previous findings that the average low ionization metal lines become weaker towards higher redshifts, likely reflecting reduced gas covering fractions at $z>6$ \citep{Glazer2025}. 
In Figure~\ref{fig:LIS_EW_z}, we plot the LIS absorption EW as a function of \lya{} EW.
It has been known that the EW of LIS absorption lines correlates with that of \lya{}, with weaker LIS lines more commonly seen in strong \lya{} emitters \citep[e.g.,][]{Shapley2003,Jones2012,Du2018,Pahl2020,Glazer2025}.
In contrast, while \srcname{} has  LIS line EW smaller than the lower-redshift galaxies, the \lya{} we observe is also very weak (EW = $0.46_{-0.14}^{+0.13}$~\AA{}, adopting the value derived in \S~\ref{subsec:lya_abs}).
This likely reflects significant IGM attenuation to \lya{} photons at $z\sim9$, which we will show to be the case through full modeling of the \lya{} (absorption and emission) profile in \S~\ref{subsec:ism_abs}.

We also detect high-ionization absorption lines from \siiv{} and \civ{} which trace the warm ionized gas.
We measure absorption EWs of $-1.8_{-0.2}^{+0.2}$~\AA{} for \siiv{}~$\lambda1393$ and $-1.2_{-0.2}^{+0.2}$~\AA{} for \siiv{}~$\lambda1402$, both significantly larger than the low ionization lines.
The EW ratio of the \siiv{} doublet is EW~(1393)/EW~(1402) = $1.5_{-0.3}^{+0.3}$, lower than the optically thin expectation of 1.97, suggesting that the stronger \siiv{}~$\lambda1393$ line is likely saturated. 
The \civ{} transitions suggest a similar picture, with stronger absorption ($-1.4_{-0.2}^{+0.2}$~\AA{} for the line at 1548~\AA{} and $-1.0_{-0.1}^{+0.1}$~\AA{} at 1550~\AA{}) than the low-ionization transitions.
As we expect EW~(1548)/EW~(1550) = 2.0 at the optically thin limit, our measurements are also consistent with saturation at least for the stronger line (\civ{}~$\lambda1548$).

The absorption line centroids and profiles allow us to probe the kinematics of the multiphase ISM and CGM at $z\sim9$. 
We find the LIS absorption is  blueshifted relative to the systemic redshift (Figure~\ref{fig:absorption_lines}), with peak velocity offsets of $-161_{-23}^{+24}$~\kms{} from \siii{}~$\lambda1260$ and $-187_{-83}^{+44}$~\kms{} from \alii{}~$\lambda1670$ (Figure~\ref{fig:absorption_lines}).
Their velocity profiles appear asymmetric, with a broad tail at the blue side of the absorption center.
To quantify the maximum outflow velocity, we measure the velocity at 90\% of the continuum level in the fitted profile, $v_{90}$ \citep{Chisholm2015}.
We find outflow velocities reaching up to $v_{90}\sim500$~\kms{} from the \siii{}~$\lambda1260$ absorption profile.
The high ionization lines also appear blueshifted from systemic.
We measure velocity offsets at peak absorption of $137\mbox{--}177$~\kms{} for both \civ{} (excluding \civ{}$\lambda1548$ that has a large uncertainty) and \siiv{}, suggesting outflows in warm ionized gas.
A broader analysis of the absorption lines in \srcname{} and other SPURS spectra will be presented in a future work.

Additionally, we detect a narrow absorption feature at 1.277~\um{} (see Figure~\ref{fig:restUV_spec}), consistent with \nv{}$\lambda$1238, which is rarely seen at lower redshifts \citep[e.g.,][]{Shapley2003,Steidel2016,Berg2022}.
The absorption has an EW of $-0.46_{-0.14}^{+0.16}$~\AA{}, with a small FWHM of $230_{-67}^{+129}$~\kms{}. The feature is distinct from the broad stellar wind component discussed later in the section (\S~\ref{subsec:wind}).
Assuming it is \nv{}~$\lambda$1238, we find the centroid is blueshifted by $-192_{-63}^{+65}$~\kms{}, which is similar to what we measure for \civ{} and \siiv{} narrow absorption.
This absorption may correspond to \nv{} arising in shock-heated gas (driven by the recent starburst) at very high temperature and expanding into the CGM \citep[e.g.,][]{Powell2011}, but we defer a more detailed investigation of its physical nature to future work.

\subsection{Damped Ly$\alpha$ Absorption Profile}\label{subsec:lya_abs}

\begin{figure}
    \centering
    \includegraphics[width=\linewidth]{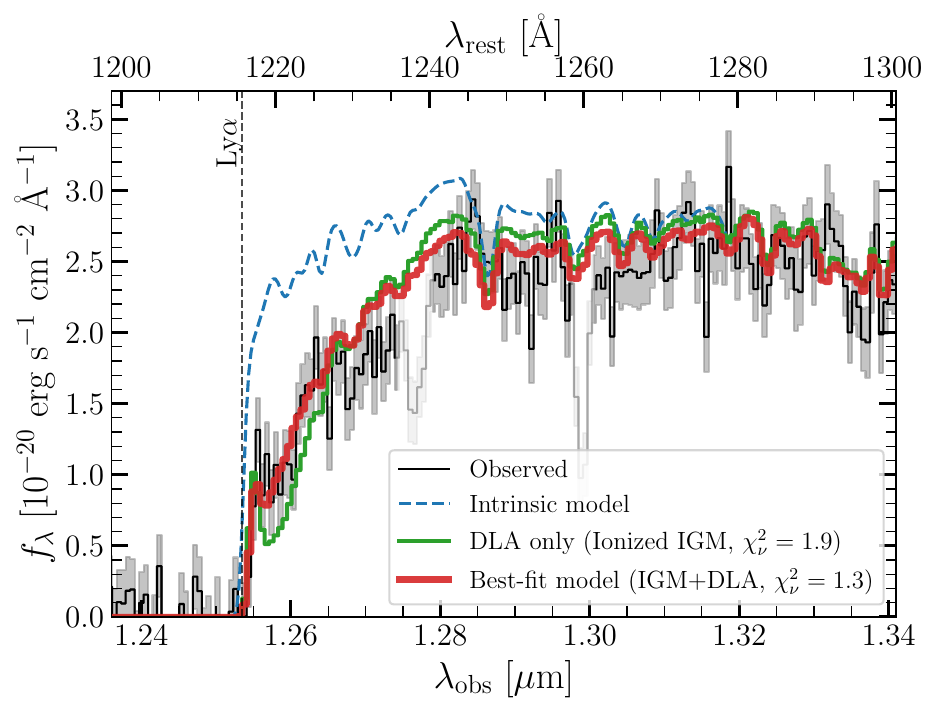}
    \caption{Fit to the observed Ly$\alpha$ profile of \srcname{}. We show the observed spectrum in black, with the error spectrum shaded in gray. The mask region around the interstellar absorption. We show the intrinsic continuum spectrum (without \lya{}) from \beagle{} in blue, with our best-fit model including both the IGM damping wing and DLA absorption in red. For comparison, we also show the median model assuming only DLA absorption (i.e., with an ionized IGM) in green.}
    \label{fig:damping_wing}
\end{figure}

The detection of UV continuum enables the measurement of the \lya{} damping wing due to scattering by neutral hydrogen in the IGM. This can be used to estimate the distance of the source to the neutral IGM and thus to infer the intergalactic neutral fraction and the timeline of reionization at $z>9$.
Previous prism and shallow grating observations of \srcname{} revealed a lack of strong \lya{} emission (EW $<5$--8~\AA{}; \citealt{Boyett2024_z9p3,Tang2024_nirspec}), likely suggesting significant damping wing attenuation.
The presence of damping wing absorption was initially suggested by the GLASS grating spectrum binned over wide rest-frame wavelength windows (yielding SNRs of 5.7 and 6.5 for 4~\AA{} and 24~\AA{} bins, respectively; \citealt{Boyett2024_z9p3}).
Low SNR detections of the LIS absorption lines in the GLASS spectrum suggested that damped \lya{} absorption (DLA) by neutral gas in or surrounding the galaxy could also contribute to the observed break, a finding later supported by modeling of the \lya{} break profile in the prism spectrum  \citep[][]{Umeda2024,Mason2026}.
However, due to the low spectral resolution of the prism and the limited depth of existing grating data, it has been challenging to disentangle the absorption due to the IGM and potential DLA from any possible \lya{} emission \citep[e.g.,][]{Keating2024,Huberty2025,Mason2026}.

The new SPURS spectrum reveals a \lya{} break profile characterized by strong absorption, likely driven by a combination of a DLA and IGM damping wing, as well as a tentative ($2.4\sigma$) emission feature consistent with being \lya{} as discussed in \S~\ref{subsec:uvlines}.
The presence of LIS absorption lines ($\S$\ref{subsec:ism_abs}) implies a high covering fraction of enriched neutral gas in the ISM and CGM. 
Based on trends in rest-UV spectra at lower redshifts \citep[see Figure~\ref{fig:absorption_lines}; e.g.,][]{Shapley2003,Du2018,Reddy2016,Reddy2022}, LIS absorption is typically associated with the presence of a DLA, allowing us to place physically-motivated priors on the DLA component in the damping wing absorption.

To constrain the contribution of each component to the observed damping wing profile, we forward model the spectrum following the methodology detailed in \cite{Mason2026}.
Briefly, we start by generating an intrinsic continuum model by fitting the observed spectrum at $>$1260~\AA{} with the photoionization code \beagle{} (see extended description in \S~\ref{sec:modeling}; \citealt{Chevallard2016}), and subsequently add DLA attenuation, emergent \lya{} emission, resonant scattering due to neutral gas infalling onto the halo \citep[e.g.,][]{Santos2004,Mason2020_bubble}, and IGM damping wing attenuation to posterior samples of the \beagle{} spectrum.
We convolve the model spectrum with the NIRSpec line spread function and sample it to the observed wavelength grid.
We fit the observed SPURS spectrum at rest-frame wavelength $<1300$~\AA{}, masking \siii{}$\lambda$1260, using a Bayesian framework. 
Since the metal LIS lines suggest a large covering fraction of neutral gas which would likely produce a DLA, as expected from both lower redshifts \citep[e.g.,][]{Reddy2016,Reddy2022,Hu2023} and cosmological hydrodynamic simulations \citep{Gelli2025}, we use the LIS absorption lines to provide empirical priors for the DLA properties.
The free parameters are the IGM neutral fraction $\bar{x}_{\rm HI}$ (uniform prior over [0,1]), the distance to the neutral IGM along the line of sight $D_{b}$ (conditional prior based on $\bar{x}_{\rm HI}$ and \muv{}, sampling sightlines from inhomogeneous reionization simulations \citep{Lu2024}; see \citealt{Mason2018,Mason2026}), the emergent \lya{} EW (prior based on the empirical distribution at $z\sim5-6$ by \citealt{Tang2024_nirspec}) and \lya{} velocity offset (prior based on the empirical distribution by \citealt{Mason2018}), the DLA column density $N_{\rm HI}$ (uniform prior on $\log N_{\rm HI}/{\rm cm^{-2}} \in [19,23]$, with the lower limit conservatively motivated from the observed \cii{}$\lambda$1334 EW; \citealt{Pettini1999}, assuming solar abundance), as well as the DLA covering fraction $f_{c, {\rm DLA}}$ (uniform prior from [0.8--1] as expected from the LIS absorption covering fraction; see~\ref{subsec:ism_abs}) and velocity offset $\Delta v_{\rm DLA}$ (uniform prior over [-250,0]~\kms{}; encompassing the range of observed velocities of LIS absorption lines).
We fix the systemic redshift to $z=9.311$ and sample the posterior distribution using the Markov Chain Monte Carlo (MCMC) package \texttt{emcee} \citep{Foreman-Mackey2013}.
In Figure~\ref{fig:damping_wing}, we show the fit to the observed \lya{} profile, as well as a comparison model assuming an ionized IGM (i.e., with only DLA absorption).

To reproduce the observed profile, our fit suggests substantial damping wing absorption from the IGM.
The inferred distance to the neutral IGM is $D_{b} = 0.29_{-0.09}^{+0.11}$~pMpc, indicating a small ionized bubble 
in a largely neutral IGM ($\bar{x}_{\rm HI}=0.81_{-0.21}^{+0.14}$), consistent with other estimates at this redshift \citep[e.g.,][]{Mason2026,Tang2024_nirspec,Nakane2024,Kageura2025}.
While some DLA contribution is also inferred ($\log (N_{\rm HI} / {\rm cm}^{-2}) = 19.5_{-0.3}^{+0.4}$), as expected from the LIS absorption lines, a DLA alone cannot match the gradual \lya{} absorption profile, because its optical depth drops more sharply with wavelength ($\tau_{\lambda, \rm DLA} \propto 1/(\Delta \lambda)^2$) than the IGM damping wing ($\tau_{\lambda, \rm IGM} \propto 1/\Delta \lambda$; \citealt{McQuinn2008,Lidz2021}, due to the contribution of gas at $>10$\,cMpc distances). This is particularly clear at $\sim1222-1230$\,\AA\ where the DLA-only fit ($\log (N_{\rm HI} / {\rm cm}^{-2}) \approx 21$) produces steep absorption, and at $>1230$\,\AA\ where the DLA-only fit underpredicts the absorption compared to the IGM damping wing. A higher column density DLA would increase the optical depth in the wings, but would produce an absorption trough extending to $\gtrsim 1220$\,\AA, which we do not detect.
These conclusions are unchanged if we relax the DLA priors.
The posterior distributions also indicate a weak \lya{} emission line with EW $= 0.46_{-0.14}^{+0.13}$~\AA{}, offset by $327_{-1}^{+9}$\,km/s from the systemic velocity, which is around $2\times$ the median Ly$\alpha$ velocity offset in $z\sim5-6$ galaxies \citep{Tang2024_z56}. 
The line is consistent with the expectation that most \lya{} emission visible during the early stages of reionization is weak and preferentially observed with higher velocity offsets than in the ionized IGM, due to the IGM damping wing \citep{Dijkstra2011,Mason2018_transmission,Yuan2025}.

\subsection{Stellar Wind and Photospheric Features}\label{subsec:wind}

\begin{figure*}[ht!]
    \centering
    \includegraphics[width=1\linewidth]{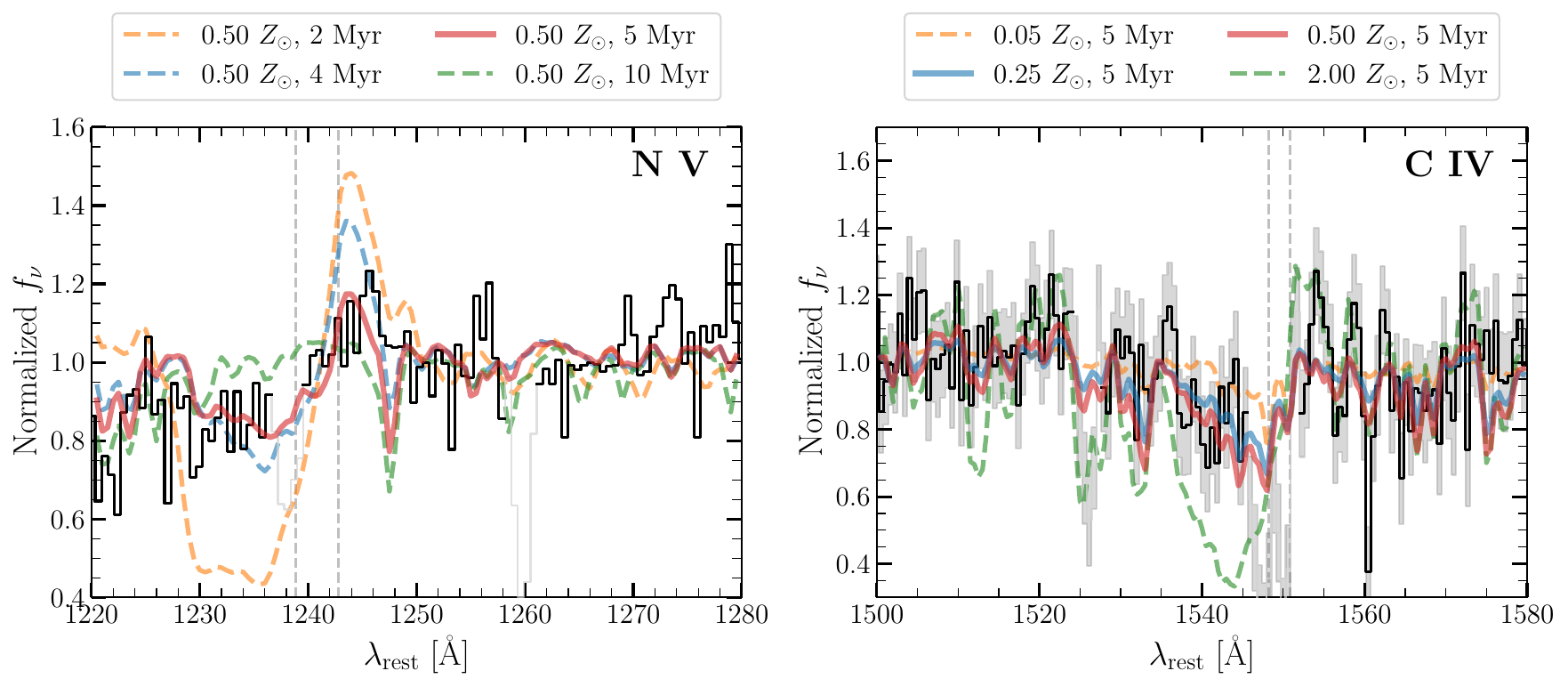}
    \caption{Broad stellar wind absorption features in \nv{} (left panel) and \civ{} (right panel) presented in the SPURS rest-UV spectrum.
    Comparison to the stellar wind profiles from Charlot \& Bruzual SSP models (colored lines) suggests that moderately large metallicities ($\sim$0.25-0.5~$Z_\odot$) in stars are required to reproduce the observed spectrum.
    The vertical dashed lines mark the rest-frame wavelengths of each doublet.
    We also mask the narrow ISM absorption component in light gray.
    }
    \label{fig:wind}
\end{figure*}

\begin{figure*}
    \centering
    \includegraphics[width=1\linewidth]{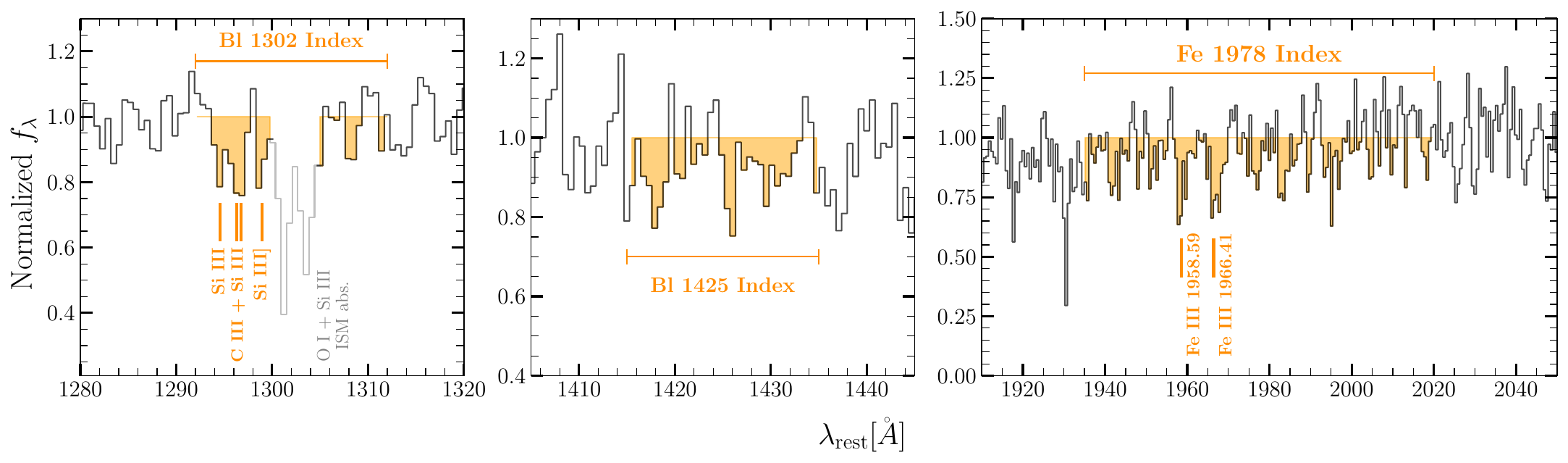}
    \caption{Photospheric features presented in the deep SPURS rest-UV spectrum.
    We present the spectrum at several commonly used indices: Bl 1302 (left), Bl 1425 (middle), and Fe 1978 (right).
    We also mark several strong absorption features likely corresponding to individual photospheric lines with orange bars.
    }
    \label{fig:photospheric}
\end{figure*}

The deep rest-UV spectrum characterizes the strength of P-Cygni stellar wind features in both \nv{}~$\lambda1240$ and the \civ{}~$\lambda1549$ (Figure~\ref{fig:wind}). 
These features are usually powered by luminous O (B as well for \civ{}) stars, with \nv{} providing a sensitive clock on recent star formation and the strength of \civ{} absorption additionally sensitive to the stellar metallicity and IMF \citep[e.g.,][]{Steidel2016,Chisholm2019}.
We estimate the P-Cygni absorption EWs by integrating the continuum normalized absorption profile over a velocity window of $[-2500, -500]$~\kms{}, where the window size is chosen to capture the full extent of wind absorption while excluding the narrow interstellar component. 
Here, the continuum is estimated from the median flux over nearby line-free regions: 1248--1258~\AA{} for \nv{}, and 1500--1510~\AA{} and 1570--1580~\AA{} for \civ{}. 
As \nv{} is additionally impacted by the IGM damping wing, we correct the spectrum for this absorption using the median transmission curve derived in \S~\ref{subsec:lya_abs}.
We measure an EW of $-1.2_{-0.2}^{+0.2}$~\AA{} for the broad \nv{} absorption and $-1.6_{-0.2}^{+0.3}$~\AA{} for \civ{}. We note that the NV profile additionally shows evidence of the P-Cygni emission component.

We empirically characterize the stellar population properties implied by the observed \nv{} and \civ{} profiles through comparison to predictions from simple stellar population models (adopting the \citealt{Bruzual2003} models) in Figure~\ref{fig:wind}.
We show that the \nv{} P-Cygni feature in both absorption and emission only appears at very young stellar population ages for simple stellar population models ($\lesssim 5$~Myr), suggesting some very recent star formation activity.
The broad \civ{} absorption can also be reproduced by young ($\sim 5$ Myr) simple stellar population models with moderate metallicities ($\approx0.25\mbox{--}0.5~Z_\odot$), indicative of the presence of already metal-enriched young stars.
At lower stellar metallicities ($Z_*<0.25~Z_\odot$), the stellar population models underpredict the strength of \civ{} absorption (assuming a SSP age of 5 Myr as required to match the \nv{} profile).
In this case, one explanation is that an overabundance of very massive stars ($M>150~M_\odot$) close to the Eddington limit will drive stronger winds to match the broad \civ{} absorption (e.g. \citealt{Vink2018,Senchyna2021}), which we will also discuss in Appendix \S~\ref{appendix:low_stellar_metallicity}.

In addition to stellar winds, the deep rest-UV continuum spectrum provides a view of the photospheric absorption features imprinted by the atmospheres of massive O and B stars.
We quantify their strength adopting several commonly used indices  \citep[e.g.,][]{Fanelli1992,Rix2004,Vidal-Garcia2017}\,: Bl~1302, Bl~1425, and Fe~1978, which are dominated by Si~{\sc ii}/Si~{\sc iii}/C~{\sc iii}/O~{\sc i}, Si~{\sc iii}/Fe~{\sc v}/C~{\sc ii}/C~{\sc iii}, and Fe~{\sc iii} photospheric lines, respectively.
We illustrate these features in Figure~\ref{fig:photospheric}, where individual photospheric lines constituting these indices are also seen for Bl~1302 (\siii{}~$\lambda1295$, \ciii{}~$\lambda1296$+\ciii{}~$\lambda1297$ blend, and \siii{}]~$\lambda1299$) and Fe~1978 (\feiii{}]~$\lambda1959$ and \feiii{}]~$\lambda1966$).
For the Bl 1302 and Bl 1425 indices, we use the identical central and continuum bandpasses defined in Table 1 of \cite{Vidal-Garcia2017} to calculate the EWs.
We also measure Fe~1978 index closely following the definition in \cite{Rix2004} by integrating the continuum-normalized spectrum over rest-frame 1935--2020~\AA{}, where the continuum is determined by spline fitting to the line-free regions defined in their Table~3.
We measure EW~(Bl~1302) = $-1.6_{-0.5}^{+0.5}$~\AA{},  EW~(Bl~1425) = $-1.2_{-0.4}^{+0.4}$~\AA{}, and EW~(Fe~1978) = $-4.2_{-1.2}^{+1.1}$~\AA{}.
Comparing these EW measurements with calibrations from stellar population synthesis models \cite{Rix2004,Sommariva2011,Vidal-Garcia2017}, we find they are consistent with a slightly enriched metallicity (0.3--0.5~$Z_\odot$).

\subsection{Summary of Spectroscopic Properties from SPURS}\label{subsec:res_summary}

We have presented new observations of the dominant UV-emitting component in \srcname{}.
In summary, we update constraints on the weakness of the emission lines in both rest-UV (total \ciii{}] EW = $1.6_{-0.3}^{+0.3}$~\AA{}) and optical (\hb{} EW= $27_{-4}^{+4}$~\AA{}), confirming it as one of the weakest line emitters at $z>9$.
The rest-frame optical spectrum reveals elevated ionization and excitation in the ionized gas, with O32 = $8.4_{-0.9}^{+1.2}$ and R23 = $13.0_{-1.5}^{+2.0}$.
We find a weak Balmer break ($f_{\nu, 4050}/f_{\nu, 3500} = 1.14_{-0.09}^{+0.07}$), suggesting that the stellar population is dominated by relatively young stars ($<20$~Myr).
The rest-UV continuum also shows  the \civ{} and \nv{} stellar wind features, with the strength consistent with moderately enriched stellar populations ($\sim0.25\mbox{--}0.5$~$Z_\odot$).
The deep rest-UV spectrum additionally provides insight into the gas properties in the ISM and CGM through absorption lines, and the IGM through the Ly$\alpha$ damping wing.
We detect a large number of low- and high-ionization metal absorption lines by the ISM and CGM, suggesting the presence of outflows (with peak velocity at $-114$ to $-187$~\kms{} for the majority of the lines) in both neutral and ionized phases.
In particular, the strong low-ionization absorption lines suggest a large covering fraction of the neutral gas ($f_c\approx0.67$), suggesting that the ionizing photon escape fraction is unlikely to be large, which has implications for the origin of the weak emission lines.
The strong Ly$\alpha$ damping wing suggests \srcname{}, one of the brightest and most massive galaxies at $z>9$, resides close to neutral IGM ($D_b = 0.29_{-0.09}^{+0.11}$\,pMpc).
In what follows, we will synthesize these findings and infer the physical nature of \srcname{} through photoionization modeling.

\section{Photoionization Modeling and Star Formation History of SPURS-A2744-7}\label{sec:modeling}

Our main goal in this section is to use photoionization models to investigate the range of stellar and gas-phase properties that are consistent with the spectrum of SPURS-A2744-7.
We begin by describing the modeling procedures in \S~\ref{subsec:modeling_method} and the results in \S~\ref{subsec:fit_res}.
We synthesize the implications for SFH in \S~\ref{subsec:sfh}.

\subsection{Methodology}\label{subsec:modeling_method}

To infer the physical properties and star formation history of \srcname{}, we fit the full SPURS grating spectrum using the BayEsian Analysis of GaLaxy sEds (\beagle{}) code (v0.29.2; \citealt{Chevallard2016}).
As described in \cite{Gutkin2016}, \beagle{} self-consistently combines the updated \cite{Bruzual2003} stellar population models, underpinned by the isochrones computed with the PAdova and TRieste Stellar Evolution Code (PARSEC; \citealt{Bressan2012,Chen2015}), with the nebular emission modeled using the photoionization code \cloudy{} (\citealt{Ferland2013}).

We fit the combined SPURS spectrum from all three medium-resolution gratings (G140M, G235M, and G395M; see \S~\ref{sec:data} for details in the construction of the combined spectrum) at the native NIRSpec grating pixel sampling (1--6~\AA{} per pixel in the observed frame), including both the continuum and emission lines across the rest-frame UV and optical.
We restrict our fitting window to 1.3--5.2~\um{} (1260--5050~\AA{} in the rest frame), thereby avoiding wavelengths affected by the IGM damping wing at the blue end and regions dominated by low SNR pixels at the red end.
We also mask each low- and high-ionization rest-UV absorption line with a window of [$-500$, 300]~\kms{} around the line center,  where the window size is chosen to cover the full extent of the narrow interstellar component (see~\S~\ref{subsec:ism_abs}). 
We additionally mask a window of [$-10$, 10]~\AA{} around [\neiii{}]~$\lambda3968$, which is detected but not included in default \beagle{} emission line list, but we do not expect it to impact the results as the other stronger Balmer lines and [\neiii{}]~$\lambda3869$ are included in the fit.
The NIRSpec spectral resolution is wavelength-dependent and varies between gratings, with higher resolution toward longer wavelengths in each grating ($R=400\mbox{--}2000$ over the fitted range). This affects the modeling of the rest-UV continuum and the detected emission lines (i.e., \ciii{}], [\oii{}], [\neiii{}], \hb{}, and [\oiii{}]).
As these features are primarily covered by the G140M and G395M gratings, we fit a single eighth-order polynomial that adequately describes the resolution of both gratings as a function of wavelength, which is then used as input for \beagle{} to model the wavelength-dependent resolution.
To account for the internal broadening of emission lines in the modeling, we additionally allow the intrinsic velocity dispersion (same for all emission lines) to vary with a uniform prior of 50--150~\kms, encompassing the range of individual line widths measured in \S~\ref{subsec:optspec}.

To infer the star formation history, we assume a non-parametric form following that described in \cite{Endsley2025_bursty}, in which the SFH is modeled as a piecewise series of tophat functions.
We adopt 8 age bins, with dense binning at recent lookback times to sample the evolution of emission line EWs and continuum shape after a burst of star formation.
Specifically, the first three age bins cover 0--3 Myr, 3--5 Myr, and 5--10~Myr, capturing the recent star formation episodes that dominate the ionizing photon output and, therefore, the emission line fluxes. 
The subsequent two bins cover 10--20 and 20--30~Myr, which, together with the younger bins, track changes in the rest-UV continuum from young massive stars (and in the rest-optical continuum as well when the entire system is dominated by very young stellar populations).
The following two bins isolate the aging populations (30--50, 50--100~Myr) that strengthen the Balmer break.
The final bin spans from 100 Myr to a lookback time corresponding to $z_{\rm f}=20$, allowing for a possible contribution from an old stellar component.
We assume that the star formation rate in each bin is completely independent, with a log-uniform prior over [0.001, 300]~$M_\odot$~yr$^{-1}$, to give the models full flexibility to explore SFHs that reproduce the observations.
We will also explore cases in which SFRs in adjacent bins are correlated, thereby imposing a prior on how rapidly the SFR can vary with time, as well as simpler constant SFHs in Appendix~\ref{appendix:other_models}.

Throughout our modeling, we adopt a \cite{Chabrier2003} stellar initial mass function with mass limits of 0.1--300~$M_\odot$ and the \citealt{Pei1992} SMC dust attenuation curve.
We assume log-uniform priors on the total (gas+dust) interstellar metallicity ($-2 \leq {\rm log} (Z_{\rm ISM}/Z_\odot) \leq 0$) and the gas ionization parameter ($-4 \leq {\rm log} U \leq -1$, defined as the ratio of H-ionizing-photon to gas density at the Strömgren radius), as well as a uniform prior on the V-band dust optical depth ($0.001\leq \tau_{\rm \scriptscriptstyle{V}} \leq 5$).
The gas-phase abundance is tied to the total interstellar metallicity value via the dust-to-metal mass ratio, $\xi_{\rm d}$, which we vary from 0.1 to 0.5.
These models assume the stellar metallicity to be the same as the ISM metallicity, but we will also explore models allowing a lower stellar metallicity in \S~\ref{appendix:low_stellar_metallicity}.
Motivated by the large LIS covering fraction, we assume fitting with a zero ionizing photon escape fraction, but we will also comment on its impact below and in the appendix (\S~\ref{appendix:csfh_fesc}).

When fitting directly to the full spectrum, we find that the fit is often driven by the continuum, given the large number of data points, rather than by the emission lines.
To overcome this issue, we separately fit the emission line fluxes along with pseudo-photometry (integrating the full spectrum through a subset of NIRCam broadband filters), and then use the results as priors when modeling the full spectrum.
Specifically, we fit the rest-frame optical emission line fluxes ([\oii{}]~$\lambda3728$, [\neiii{}]~$\lambda3869$, \hg{}, \hb{}, and [\oiii{}]~$\lambda\lambda4960,5008$) using \beagle{}.
Along with the lines, we include a pseudo-NIRCam SED, obtained by integrating the full spectrum in commonly used broadband filters (F150W, F200W, F277W, F356W, and F444W), which allows us to also self-consistently model the stellar populations.
We maintain the same assumptions and priors on the ISM parameters (metallicity, ionization parameter, dust-to-metal ratio, and dust attenuation), and adopt the same non-parametric SFH as described above.
Next, we perform the full spectral modeling, using Gaussian priors for each ISM parameter as well as the SFR in each SFH bin from the posterior distributions obtained from the initial emission line + pseudo-photometry fits.

In the following, we first present the constraints on ionized gas properties from the fit to emission line fluxes and pseudo-photometry, focusing on the range of parameters consistent with the observed line ratios.
We then discuss the stellar population properties and star formation history from the full spectrum modeling, adopting the median of the marginalized posterior probability distribution for each parameter.

\begin{figure*}[ht!]
    \centering
    \includegraphics[width=1\linewidth]{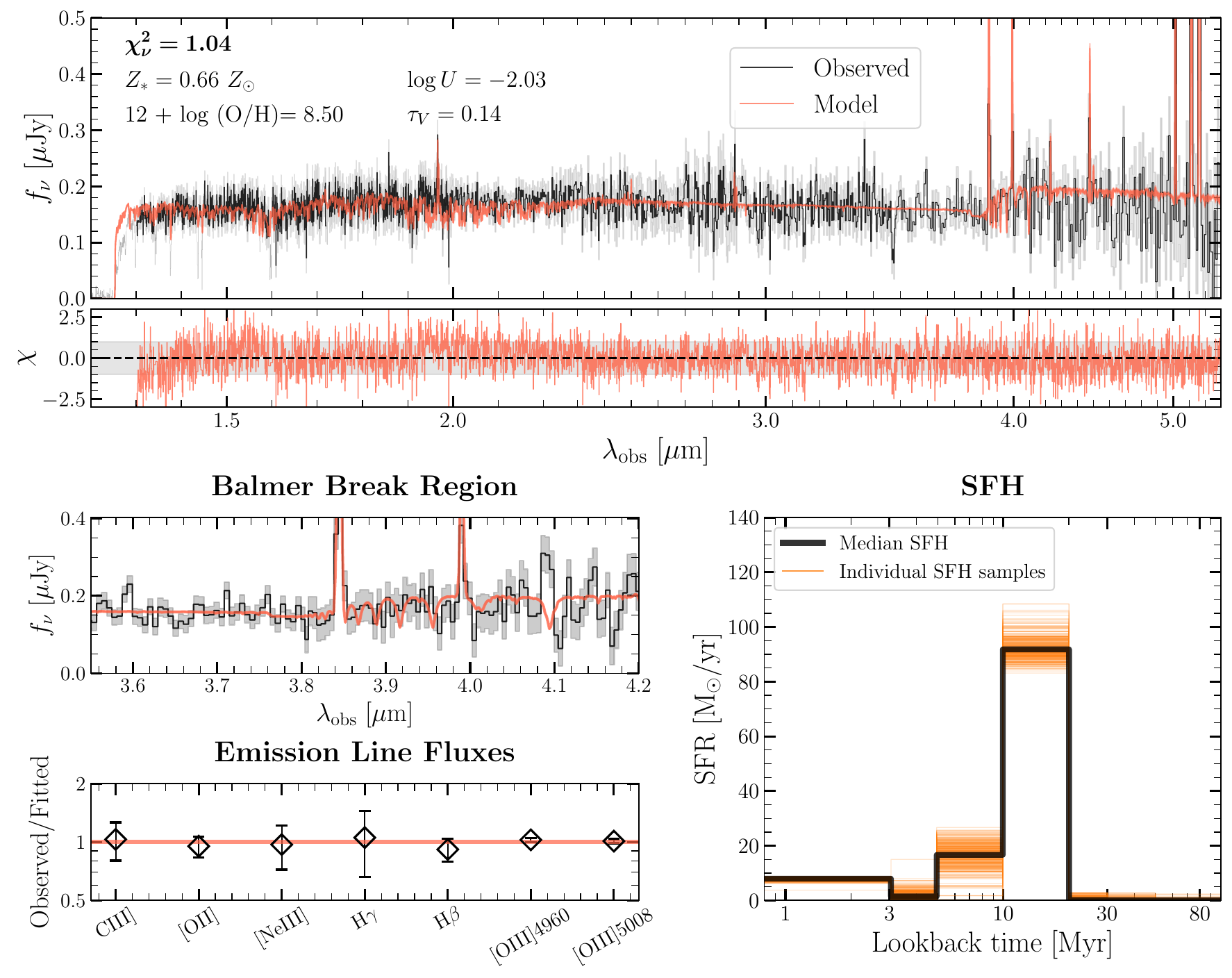}
    \caption{Top panel: Fit to the observed grating spectrum, assuming a non-parametric SFH. 
    We plot the grating spectrum as a solid black line, with errors shaded in gray.
    We do not fit the spectrum at wavelengths shorter than 1260~\AA{} or regions impacted by interstellar absorption features, which are masked in gray.
    We show the model SED in red, with the shaded region representing the 16th-to-84th percentile range of the posterior.
    While our fit is performed on the grating spectrum at the native resolution and pixel sampling, we plot the observed spectrum binned by 3--6 pixels at wavelengths longer than 2~\um{} (regions covered by G235M+G395M gratings) for clarity (no binning for model spectrum).
    We additionally show $\chi = (f_{\rm \nu, obs} -  f_{\rm \nu, m} ) / \sigma_{\rm obs}$ as a function of the observed wavelength, where $f_{\rm \nu, obs}$, $ \sigma_{\rm obs}$, and $f_{\rm \nu, obs}$ are the observed flux, uncertainty, and the model flux, respectively.
    Bottom left panels: zoom in on the continuum fits to the \civ{} wind profile and the Balmer break region, as well as comparison of \ciii{}] and optical emission line fluxes between the model and observation.
    Bottom right panel: star formation history inferred from the \beagle{} fits to the spectrum. 
    We plot the median SFR within each age bin in black, with individual SFH samples from the posterior shown in orange lines.
    }
    \label{fig:spectral_fit_fiducial}
\end{figure*}

\subsection{Results}\label{subsec:fit_res}

\subsubsection{Emission line and pseudo-SED fits }\label{subsubsec:fit_line}

We first explore the range of ionized gas properties inferred by the rest-optical emission-line fluxes and the pseudo-photometry (from the full spectrum integrated in NIRCam broadband filters), which serve as priors for the following full grating fits.
Using \beagle{}, we find that the \cite{Gutkin2016} models assuming $Z_* = Z_{\rm ISM}$ successfully reproduce all input rest-optical line fluxes (within $1\sigma$ uncertainties) and thus their corresponding flux ratios.
The large O32 and Ne3O2 values are best fit with a high ionization parameter ($\log U = -2.08_{-0.11}^{+0.10}$).
To reproduce the full set of emission line ratios, the models permit a spread of metallicities spanning from $0.47\,Z_\odot$ to $0.75\,Z_\odot$.
This spread is expected from the degeneracy with the ionization parameter in fitting both O32 and R23 ratios (as shown in Figure~\ref{fig:O32R23}), where we find that models at a higher metallicity can also fit the observed line ratios with a larger ionization parameter to compensate for the weakening of the [\oiii{}] flux caused by enhanced metal cooling.
These ISM metallicities are large, with a median of $Z=0.60\,Z_\odot$.
At the fitted dust-to-metal ratio ($\xi_d = 0.34_{-0.08}^{+0.08}$), this ISM metallicity also translates to a large gas-phase oxygen abundance of $12 + \log(\mathrm{O/H}) = 8.48_{-0.11}^{+0.09}$. 

The spectrum fit also suggests modest dust attenuation, yielding a V-band optical depth of $\tau_{\rm V}=0.12_{-0.02}^{+0.02}$.
Dust attenuation affects the observed line ratios, for example increasing the O32 ratio at fixed ionization parameter.  As a result, a large dust optical depth allows a lower ionization parameter to fit a given O32 ratio. In the case of \srcname{}, this in turn results in model fits returning lower metallicities to match the 
observed R23 ratio. So obtaining reliable inferences of ionization parameter and metallicity requires the effects of dust reddening to be known. 
If we were to fit the observations without reddening-sensitive information (i.e., without a Balmer decrement or UV continuum slope), somewhat lower metallicities would be possible. However, 
we find that to simultaneously fit the flat UV continuum (in the pseudo-SED) and the \hb{}/\hg{} flux ratio, strong dust attenuation is disfavored (also consistent with the non-detection of dust continuum in recent ALMA observations, \citealt{Algera2025}).  The combination of a significant metal reservoir without detectable dust continuum or significant reddening may suggest dust removal via outflows  \citep[e.g.,][]{Ferrara2023,Fiore2023,Ziparo2023,Ferrara2025} if dust is preferentially removed, or it may suggest that dust is present but not as clearly detectable owing to the grain size distributions \citep{Markov2023,McKinney2025,Narayanan2025,Shivaei2025}.

\subsubsection{Full grating spectrum fits}\label{subsubsec:fit_continuum}

\begin{table}[t]
\centering
\caption{Derived median physical properties from \beagle{} photoionization modeling of the full SPURS grating spectrum. 
We report the median inferred values for the ionized gas and stellar population properties, as well as the star formation history.
We note that the listed SFRs (averaged values over the age bin) and sSFR (averaged over the last 10 Myr) correspond to the values within the MSA shutter of the NIRSpec observation.
\label{tab:properties}}
\begin{tabular*}{\columnwidth}{c@{\extracolsep{\fill}}c}
\hline
\hline
Quantity                             & Value \\
\hline
$Z_{\rm ISM} = Z_*$ [$Z_\odot$]      & $ 0.66$ \\
$12 + \log(\text{O/H})$              & $ 8.50$ \\
$\log U$                             & $-2.03$ \\
$\tau_V$                             & $ 0.14$ \\
$\xi_d$                              & $ 0.34$ \\
SFR$_{\rm 10\ Myr}$ [$M_\odot~{\rm yr}^{-1}$]      & $  11$          \\
SFR$_{\rm 10-20\ Myr}$ [$M_\odot~{\rm yr}^{-1}$]   & $  91$          \\
SFR$_{\rm 10\ Myr}$/SFR$_{\rm 10-20\ Myr}$         & $ 0.12$ \\
sSFR [Gyr$^{-1}$, 10 Myr]                          & $   7$          \\
\hline
\end{tabular*}
\end{table}

\begin{figure}
    \centering
    \includegraphics[width=\linewidth]{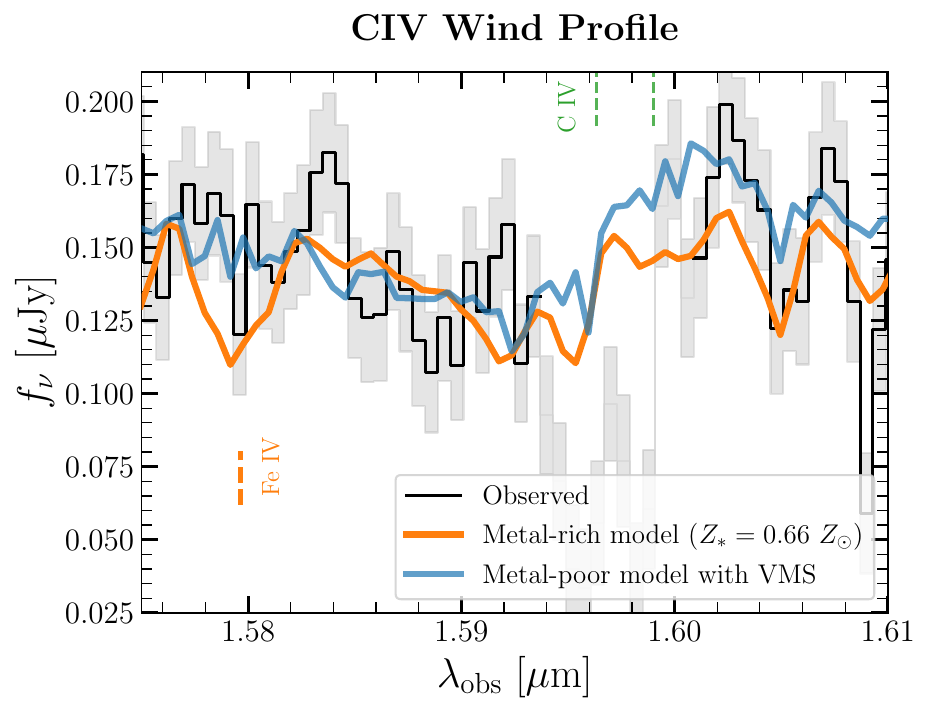}
    \caption{
    Zoom-in on the broad \civ{} wind absorption profile of the full spectrum fit.
    We show the observed profile in black (with uncertainty in gray), and mask the narrow interstellar component ([$-500$, 300]~\kms{} relative to the line center; see \S~\ref{subsec:ism_abs}) in gray. 
    We mark the systemic wavelength of \civ{} given the redshift in green.
    Our model with a moderately large metallicity ($Z_* = 0.66\,Z_\odot$, orange line) reproduces the observed wind absorption strength.
    The model additionally shows the \feiv{} photospheric lines at 1.58 um as it does not account for the effect of $\alpha$-enhancement.
    Alternatively, as we discuss in more detail in the appendix (\ref{appendix:low_stellar_metallicity}), the strong absorption can also be recovered at a low stellar metallicity ($Z_* = 0.03\,Z_\odot$) if we add more VMS (contributing 15\% of the continuum; blue line).
    }
    \label{fig:civ_fit}
\end{figure}

Equipped with knowledge of the ISM properties inferred from emission lines above, we now characterize the star formation history (SFH) and stellar populations derived from fitting the full grating spectrum.
As detailed in \S~\ref{subsec:modeling_method}, we apply Gaussian priors to both the ISM properties (metallicity, ionization parameter, dust-to-metal ratio, and V-band dust optical depth) and the SFRs within each SFH bin. 
The mean and standard deviation of these priors are set to the median and 68\% credible intervals, respectively, of the posteriors derived from the fit to the emission line fluxes and the pseudo-photometry. 
We present the resulting fit in Figure~\ref{fig:spectral_fit_fiducial} and report the median model parameters from the final full spectrum modeling in Table~\ref{tab:properties}.

We find the fit reproduces the overall continuum within the errors across most wavelengths, yielding a median $\chi^2_\nu$ of 1.04.
The modest amount of dust attenuation (V-band optical depth (median $\tau_{\rm V} = 0.14$) that fits the emission line fluxes also reproduces the continuum.
The data are well-fit by models in which  the stars are fairly enriched (median $Z_* = Z_{\rm ISM} = 0.66$).
{In Figure~\ref{fig:civ_fit}, we also show that this metallicity reproduces the \civ{} profile well, consistent with expectations from the dependence of \civ{} wind absorption on stellar metallicity as discussed in \S~\ref{subsec:wind}. 
We additionally notice a slight excess in the spectrum beyond the \beagle{} predictions at 1.92--1.95~\um{} (rest-frame 1860--1890~\AA{}; top panel of Figure~\ref{fig:spectral_fit_fiducial}), but we note that this is consistent with being blended \feii{} emission due to fluorescent pumping by \lya{}, which is expected if there is dense gas surrounding the ionizing sources \citep[e.g.][]{Sigut2003}.  So the picture that emerges is very similar to that from the emission line fits: \srcname{} is best fit by models that are fairly metal rich with  modest dust attenuation and  large ionization parameter.

To reproduce the weak emission line spectrum with these $Z_* = Z_{\rm ISM}$ models, we find that the SFH must be in a recent downturn.
As shown in the bottom right panel of Figure~\ref{fig:spectral_fit_fiducial}, the model SFH peaks 10--20~Myr prior to observation, with the median SFR subsequently declining by a factor of 8 over the last 10~Myr.
The downturn drives the weak emission lines in the model, with a low \hb{} EW (median model value $23$~\AA{}) comparable to that measured from the spectrum.
The model SFH also suggests some ongoing star formation, as would be expected from the presence of the \nv{} wind P-Cygni profile in our spectrum.
However, the median specific star formation rate is only $7$~Gyr$^{-1}$ (averaged over the last 10 Myr), a factor of 2 lower than what is typically seen at this redshift (e.g., \citealt{Tang2025}).
As the burst occurred very recently in the implied SFH, the model Balmer break remains weak (median $f_{\nu, 4050}/f_{\nu, 3500}=1.25$), consistent with the observations.
The model suggests that this recent burst can still contribute to high UV luminosity in this galaxy, in spite of the observation of weak lines.
We note that this post-burst SFH is also not unique to our \beagle{} modeling, and as we show in the appendix (\S~\ref{appendix:prospector_bursty} and \ref{appendix:prospector_continuity}), similar SFHs are also found if we use a different code (i.e., \prospector{}; \citealt{Johnson2021}).

\begin{figure*}
    \centering
    \includegraphics[width=1\linewidth]{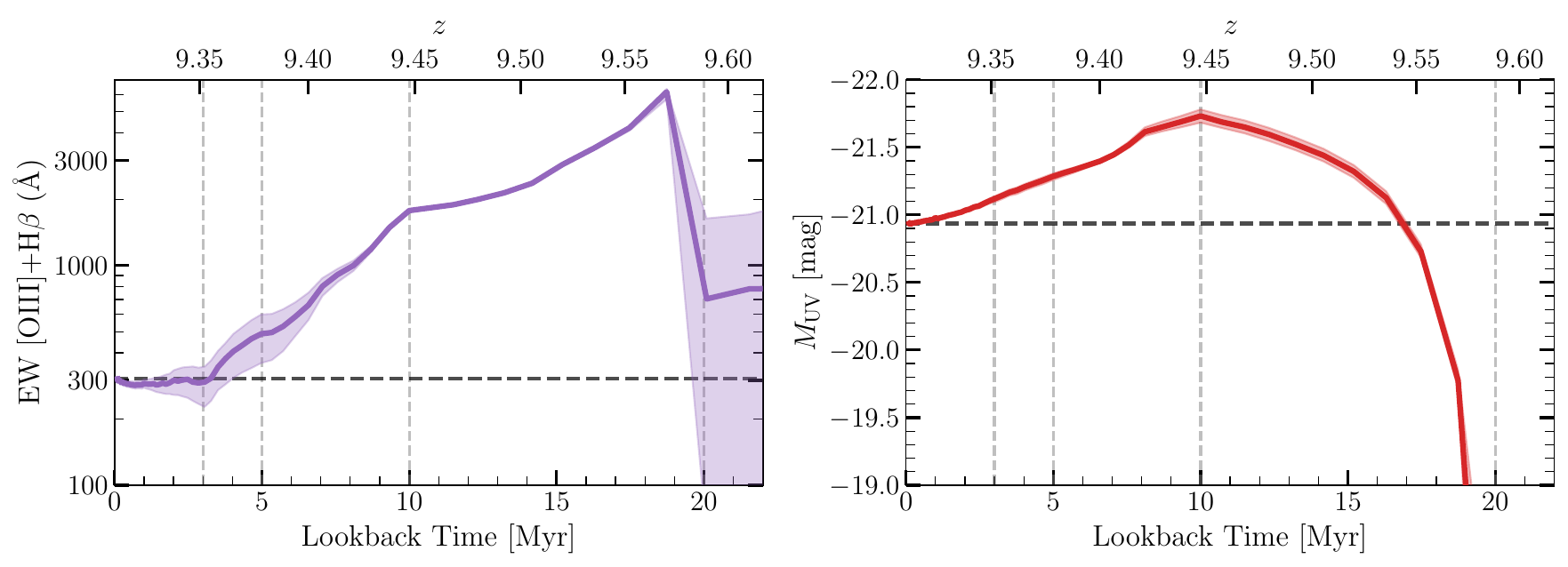}
    \caption{Recent history of EW [\oiii{}]+\hb{} (left panel) and absolute UV magnitude ($M_{\rm UV}$, right panel) reconstructed from the inferred SFH.
    Here, we consider the full posterior distribution of the SFH from our spectrum fit.
    In each panel, the solid line corresponds to the median EW [\oiii{}]+\hb{} (left) or $M_{\rm UV}$ (right) history in the last 20 Myr.
    We indicate the present-day measured value of EW [\oiii{}]+\hb{} and $M_{\rm UV}$ as black horizontal dashed lines for reference.
    The gray vertical dashed lines correspond to the edge of the age bins used in the non-parametric SFH modeling.
    }
    \label{fig:history}
\end{figure*}

At lower redshifts, galaxies with H$\beta$ EWs similar to \srcname{} are  well-fit with CSFH models and relatively extended star formation. While we have shown that a post-burst SFH provides a good fit to our NIRSpec observations, we must now consider whether a CSFH can also reproduce the data.
To do so, we fit the spectrum with the \beagle{} models ($Z_* = Z_{\rm ISM}$), assuming the corresponding priors on ISM properties derived from emission line and pseudo SED fitting (see \S~\ref{appendix:csfh} for more details).
In trying to simultaneously reproduce the weak emission lines and the continuum shape in the spectrum, the CSFH model favors an extremely old stellar population age (CSFH duration $198$~Myr).
However, even with such an old age, this model still overpredicts the fluxes of \hb{} and [\oiii{}]~$\lambda\lambda4960,5008$ by $\approx20\%$, while exhibiting a significantly stronger Balmer break (median $f_{\nu, 4050}/f_{\nu, 3500}=1.48$) than what we measure.
The fit is overall worse, with a $\chi^2_\nu$ (median $=1.15$) significantly larger than the above models characterized by a SFR downturn.
Therefore, we conclude that the weak lines in \srcname{} are  unlikely to have their origin in  continuous star formation that has lasted over a very extended period ($\gtrsim100$~Myr).

Ionizing photon escape provides another viable explanation for the weak emission lines. To investigate what escape fractions would be required to reproduce the spectrum, we fit the grating spectrum with \beagle{} models that allow the ionizing photon escape fraction to vary as a free parameter. 
Here, we focus on the CSFH fit with \beagle{} ionization-bounded models, where ionizing photons escape through sightlines with low-\hi{} column density, but we will also discuss the case in the density-bounded regime in \S~\ref{appendix:csfh_fesc}.
we find the spectrum can be best fit with an extremely large escape fraction value (median $f_{\rm esc}=0.84$) along with a young CSFH age of $16$~Myr.
This yields an improved fit ($\chi_\nu^2=0.96$) than the CSFH model without ionizing photon escape, confirming that a high escape fraction can fit the weak emission lines and the overall continuum shape.
However, as discussed in \S~\ref{subsec:ism_abs}, very high ionizing photon escape fractions ($f_{\rm esc}>0.4$) are unexpected given the deep rest-UV LIS absorption lines, which empirically trace high \hi{} covering fractions \citep[e.g.,][]{Reddy2016,Reddy2022,Gazagnes2018,Saldana-Lopez2022}.
As we will show in the appendix (\S~\ref{appendix:csfh_fesc}), assuming a moderate ionizing photon escape fraction ($f_{\rm esc}=0.2$) consistent with expectations from the LIS lines yields results very similar to the CSFH fit with zero escape fraction described above, and fails to match the spectrum.
The comparison between these different models suggests that a downturn in the SFH provides the most straightforward explanation for both the weak emission lines and the observed continuum.

\subsection{Implications for Star Formation History}\label{subsec:sfh}

In the previous subsection, we have motivated that the weak emission lines of \srcname{} are likely to reflect a post burst SFH. 
To illustrate the past  history of \srcname{} in the context of our model, we reconstruct the evolution of the [\oiii{}]+\hb{} EW and the absolute UV magnitude as a function of lookback time based on the full posterior SFH distributions (Figure~\ref{fig:history}).
Here, we consider the non-parametric SFH derived from the \beagle{} model fixing $Z_* = Z_{\rm ISM}$. 
The [\oiii{}]+\hb{} EW peaks strongly shortly after the starburst begins (approximately 20~Myr prior to observation in the models described above). If \srcname{} was observed at this period, it would have been an intense line emitting galaxy, plausibly similar to the extreme spectra that have been seen at $z>10$ with {\it JWST} \citep{Bunker2023,Castellano2024,Naidu2026_MoMz14}. The [\oiii{}]+\hb{} EW declines rapidly after the initial upturn in star formation. In contrast, the UV continuum builds up rapidly during the burst, reaching a peak luminosity ($M_{\rm UV}\approx-21.7$) roughly 10 Myr prior to observation. Since the UV continuum luminosity density does not decline as rapidly as emission lines,  \srcname{} is able to remain UV luminous while also having weak emission lines. If bursty star formation is at least partially responsible for the super-luminous galaxies that have been seen at very high redshifts, we should expect the population to comprise objects that are both currently bursting and those with weaker emission lines that have undergone very recent ($\lesssim 5$ Myr) declines in SFR.
We suggest that \srcname{} is likely in the latter category, similar to several other recently discovered sources \citep{Helton2025,Harikane2026}.  We note that if \srcname{} remains in a lull of star formation for much longer, its UV luminosity will decline significantly. So objects that have remained in SFR downturns for longer timescales will preferentially be found among less luminous galaxies (as is seen at $z\simeq 6$, e.g. \citealt{Endsley2025_bursty}).

While our \beagle{} models indicate that recent star formation strongly peaks in a burst 10--20~Myr prior to observation, it is still possible that there is an older stellar population that is outshined by the recent burst.
To investigate this possibility quantitatively, we take the median \beagle{} model (with $Z_* = Z_{\rm ISM}$), add synthetic SEDs from older stellar populations (fixed to the same metallicity), and quantify the additional mass tolerated before significantly altering the Balmer break strength ($f_{\nu, 4050}/f_{\nu, 3500} $) and emission line EWs.
We find that a moderate contribution from stars formed within the 50--100~Myr bin is allowed, but this population's mass must be less than half of the main burst to keep the Balmer break strength and emission line EWs (\hb{}, [\oiii{}]) consistent with our observations within $2\sigma$.
This suggests that star formation over the past 100~Myr is relatively concentrated in the recent burst episode.
On the other hand, a much older population ($>100$~Myr) can contribute up to the same mass as the main burst before driving these spectral features into $>2\sigma$ tension with the data.
However, while we cannot completely rule out the additional presence of a genuinely old ($>100$~Myr) stellar population, we emphasize that the recent burst alone is sufficient to explain the observed spectrum. We of course note that older star formation activity may have occurred in spatially distinct regions from the main UV-emitting clump we have probed with spectroscopy.

\section{Discussion}\label{sec:discussion}

\subsection{Implications for Galaxy Assembly at $z>9$}

\begin{figure*}
    \centering
    \includegraphics[width=1\linewidth]{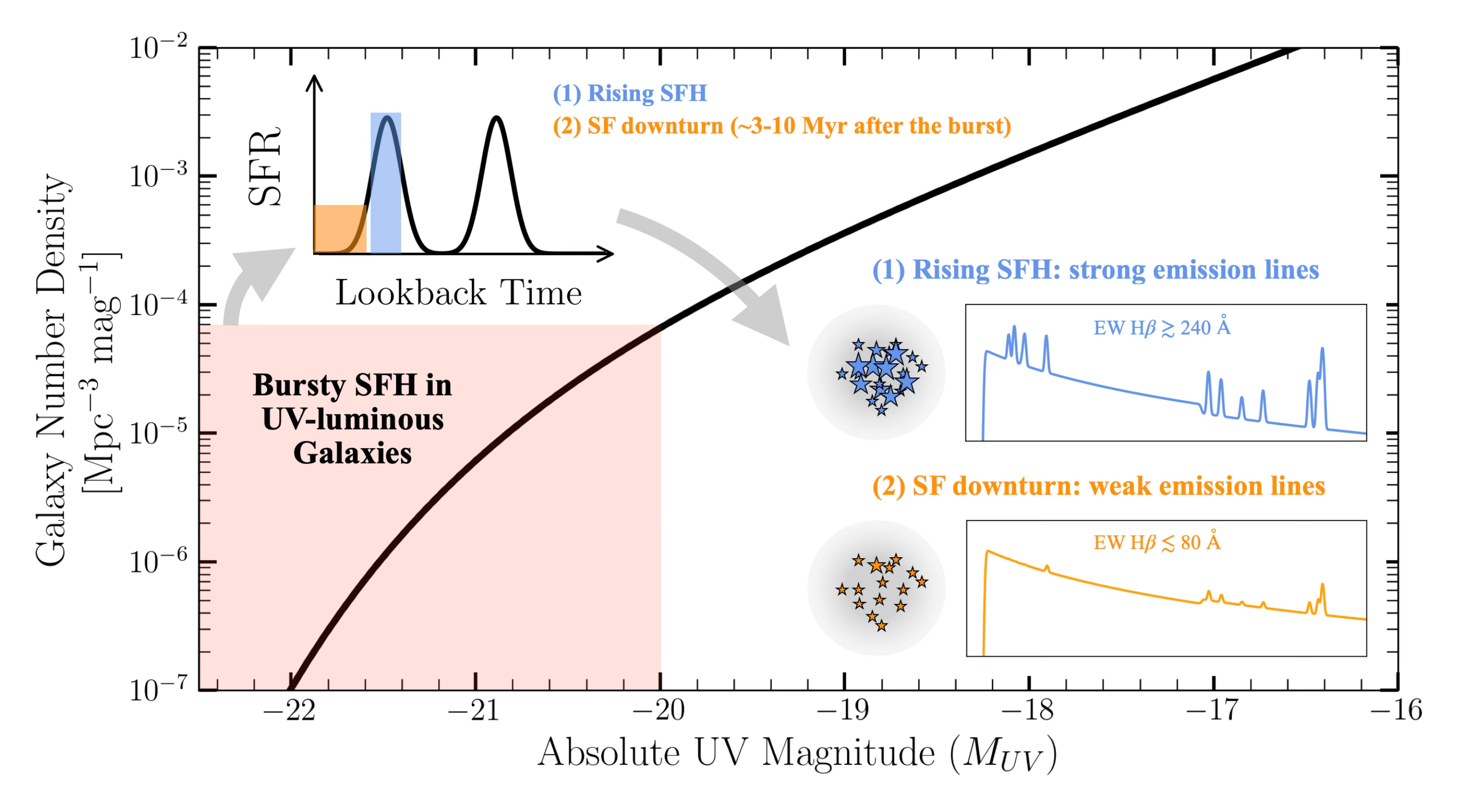}
    \caption{Schematic demonstrating the two different spectra expected in galaxies undergoing bursty star formation and observed at the bright end of the UV luminosity function at $z>9$.
    Galaxies observed near the peaks of the star formation have spectra characterized by extreme rest-UV and optical emission lines (EW \hb{} $\geq240$~\AA{}), similar to what we have seen in GN-z11 \citep{Bunker2023} and GHZ2 \citep{Castellano2024}.
    On the other hand, those with a recent star formation downturn will show significantly weaker emission lines, similar to what is seen in \srcname{} and several other sources \citep[e.g.,][]{Helton2025,Harikane2026}.
    }
    \label{fig:uvlf_illustration}
\end{figure*}

The discovery of a surprisingly abundant population of luminous $z>10$ galaxies has motivated a deeper investigation of the physics that governs the growth of the earliest galaxies. In this paper, we have advocated that emission line spectroscopy provides a valuable route toward isolating the relative importance of stochasticity (here parameterized as $\sigma_{\rm UV}$, the scatter of galaxy UV magnitude at fixed dark matter halo mass) and star formation efficiency (SFE) in regulating the observed early galaxy census. 
The emission line EWs are very sensitive to the specific star formation rates. The increase in baryon accretion rates with redshift is expected to boost the specific star formation rates of $z>10$ galaxies  (e.g., \citealt{McBride2009,Weinmann2011}), allowing lower mass halos to be luminous enough to come into view above observational detection limits. This should cause $z>10$ galaxies to exhibit more prominent (larger EW) recombination lines in the rest-optical (i.e., H$\beta$, H$\gamma$) at higher redshifts, as well as larger EW rest-UV emission lines. 
However, if star formation is bursty enough at $z>10$ to explain the observed luminosity function (i.e., $\sigma_{\rm{UV}}\gtrsim 1$; \citealt{Mirocha2023,Shen2023,Sun2023,Kravtsov2024,Gelli2024}), then there will also be a significant subset of the $z\gtrsim 10$ population that is caught in recent SFR downturns, with modest specific star formation rates and weak emission lines.

The first statistical studies at $z\gtrsim10$ based on NIRSpec low-resolution prism observations have revealed that very strong emission lines are indeed more common than at $z\simeq 6$ (e.g., \citealt{Tang2025,Roberts-Borsani2025}), with many galaxies showing extreme EW nebular emission lines (\citealt{Bunker2023,Castellano2024,Naidu2026_MoMz14}). 
Indeed, \citet{Tang2025} showed that the median H$\beta$ EW at $z>9$ is 150~\AA\ ($\simeq 1.5\times$ greater than that at $z\simeq 6$).
These are likely galaxies experiencing strong upturns in star formation, perhaps caught at the beginning of a burst episode. 
However, as more spectra have been obtained, a subset of early galaxies with weaker emission lines in their rest-UV and optical spectra has also been uncovered (e.g., \citealt{Carniani2024_z14,Wu2025,Donnan2026,Harikane2026,MarquesChaves2026}). 
These objects are not a negligible fraction of the early galaxy population: 
based on the sample in \citet{Tang2025}, 37\% of the $z>9$ population are found with relatively low H$\beta$ EWs ($\leq 80$~\AA), relative to 38\% at $z\sim6$ \citep{Endsley2024_jades}. 
If these are post-burst galaxies, the lack of significant evolution in the low EW sources, despite an increase in the median EW, may indicate that the H$\beta$ EW distribution is broadening between $z\simeq 6$ and $z\gtrsim 9$. This is exactly what would be expected if the stochasticity in the halos we are probing at $z\gtrsim 9$ is larger relative to $z\simeq 6$ \citep{Gelli2024,Munoz2026}. 
However, it is not clear whether the weak line sources actually reflect post-burst star formation histories. As noted previously, it is possible that the weak line EWs reflect old stellar populations \citep[e.g.,][]{Helton2025_MIRIphot} or large ionizing photon escape fractions \citep[e.g.,][]{Yanagisawa2024_EBG1,MarquesChaves2026}. 
The low spectral resolution of the prism data has made it challenging to break these degeneracies in the interpretation of weak emission line sources.

Alongside these studies, very detailed characterizations of individual sources at grating resolution enable identification of the origin of the weak emission lines, and offer a direct path toward constraining the typical SFHs of early galaxies.
In this paper, we have presented an investigation of the stars and gas in one of the weakest line emitters known among luminous galaxies at $z>9$, \srcname{}.
Our work shows that an old, mature stellar population is inconsistent with the modest Balmer break we observe, while a high escape fraction appears unlikely given the large covering fraction of low-ionization absorption lines in the rest-UV.
Instead, the combined weak emission lines and the rest-UV to optical continuum are best reproduced by a bursty SFH, where the star formation peaks 10--20~Myr prior to the observation and then declines rapidly.
These results demonstrate the importance of spectroscopic data in helping to constrain the SFHs at $z>9$, providing a complementary yet key step forward from previous modeling efforts that relied on photometry alone.

Our modeling of \srcname{} provides evidence that weak emission lines may represent a common phase expected from a bursty SFH (see also \citealt{Roberts-Borsani2025}), which we illustrate in Figure~\ref{fig:uvlf_illustration}.
For galaxies observed during a recent upturn in star formation, the large population of massive O stars produces spectra with intense nebular emission, with \hb{} EWs reaching extremely large values ($\gtrsim240$~\AA{}).
At the peak of the most intense bursts, conditions may also favor the formation of hard ionizing sources and dense stellar clusters, potentially giving rise to the strong \civ{} and \niv{}] emission observed in the most extreme $z>9$ spectra \citep[e.g.,][]{Bunker2023,Castellano2024,Naidu2026_MoMz14}.
Within a short window ($\sim10$~Myr) after the burst, the nebular emission weakens rapidly as O stars disappear, while the UV continuum remains bright due to the still large number of B stars.
This produces the population of metal-enriched weak emission line spectra with modest Balmer breaks, which have started to emerge in recent observations  \citep[e.g.,][]{Helton2025,Donnan2026,Harikane2026}.
In this picture, the broadening of the emission line EW distribution compared to that at lower redshifts (e.g., $z\sim6$) arises from the increased burstiness of the halos that are being sampled in $z>9$ observations of super-luminous galaxies.

Characterizing the fraction of galaxies in SFR downturns provides a route toward constraining the level of stochasticity in star formation and thus its role in the formation of UV-luminous galaxies at $z>9$.
To demonstrate this, we consider the expected fraction of $M_{\rm UV}<-20$ $z>10$ galaxies experiencing recent SFR downturns using models of stochastic star formation from Gelli et al.\ (in prep.).
These models are calibrated to match the observed UV luminosity function at $z\sim14$, which requires either a large stochasticity in star formation (large $\sigma_{\rm{UV}}$, defined at halo mass of $10^9~M_\odot$), enhanced star formation efficiency, or a combination of both.
In models with highly stochastic star formation ($\sigma_{\rm{UV}} \geq 1.6$~mag), a significant fraction ($>28\%$) of UV-luminous galaxies are predicted to be in a recent downturn in SFR ($\rm SFR_{3\ Myr} / SFR_{3-50\ Myr}<0.2$).
Conversely, in models with steadier star formation ($\sigma_{\rm{UV}} = 1.2$~mag) but higher SFE, the fraction of sources in downturns is much smaller (5\%).
We find a sample size of $N\gtrsim20$ galaxies (at $z>9$) will start to distinguish between these two scenarios at $\gtrsim 2\,\sigma$, which is within reach in the next few years with deep NIRSpec and MIRI observations.

The continuous expansion of spectroscopic samples of $z>9$ galaxies is transforming our ability to statistically study their physical properties and star formation histories.
While the majority of these programs focus on the bright end of the UVLF, initial efforts are beginning to constrain star formation at fainter magnitudes \citep[e.g.,][]{AlvarezMarquez2026}.
If the UV-luminous galaxies stay in a lull phase for a long time, we should also detect them fading into the faint end of the UV luminosity function.
Furthermore, as star formation is theoretically expected to be increasingly burstier in lower-mass halos \citep[e.g.,][]{Sparre2017,FaucherGiguere2018,Furlanetto2022,Gelli2024}, the fraction of galaxies caught in a recent SFR downturn should rise at lower UV luminosities.
Consequently, extending the characterization of SFHs down to faint UV magnitudes will also be crucial to constrain the timescale of the off mode and thus the duty cycles of bursty star formation at $z>9$.
Ultimately, a comprehensive census of star formation histories across a broad range of UV magnitudes will be essential to unravel the physical processes driving galaxy assembly within the first few hundred million years of cosmic time.

\subsection{The Potential of Ultra-Deep Grating Spectra for Characterizing the IGM at $z>9$}

Deep rest-UV continuum spectroscopy at $z>9$ also presents a key step forward in constraining the onset of reionization, by enabling direct measurements of the Ly$\alpha$ IGM damping wing \citep{Miralda-Escude1998}.
Characterizing the earliest stages of reionization is important not only for understanding star formation in the bulk of the early galaxy population \citep[e.g.,][]{Gelli2024,Whitler2025}, but also for providing independent constraints on the CMB electron scattering optical depth, which some recent analyses suggest may be higher than the value reported by \citet{PlanckCollaboration2020} and could thereby alleviate current cosmological tensions \citep[e.g.,][]{Giare2024,Allali2025,Jhaveri2025,Sailer2026}.

While \jwst{} observations over the past three cycles have revealed many sources show damped \lya{} breaks in prism spectra \citep[e.g.,][]{CurtisLake2023,Heintz2025_PRIMAL}, interpretation of this feature has been limited by the prism's $R\sim40$ resolution around the break, where the IGM damping wing is blended with both \lya{} emission and local DLA absorption over only a few pixels \citep[e.g.,][]{Keating2023, Huberty2025,Umeda2024,Umeda2026,Mason2026}.
Our spectrum of \srcname{} demonstrates that these different components can be resolved over $>200$ pixels with deep medium-resolution spectroscopy, breaking the degeneracies seen in prism spectra. 
Specifically, the detection of metal LIS absorption lines indicates a high covering fraction of neutral gas in the ISM/CGM of \srcname{}, with peak velocities $\approx - 161$ km/s from systemic, enabling us to use the LIS lines as priors for the DLA component to better isolate the IGM contribution. Meanwhile, the lack of a strong Ly$\alpha$ absorption trough indicates the local DLA HI column density does not exceed $N_\mathrm{HI}\gtrsim10^{21}$\,cm$^{-2}$, and we resolve a tentative Ly$\alpha$ emission line which would have been blended with the break in the prism spectrum.

To reproduce the observed \lya{} absorption profile requires that \srcname{} sits in a small ionized bubble ($D_b = 0.29_{-0.09}^{+0.11}$~pMpc), in addition to local DLA absorption.
The strong Ly$\alpha$ damping and small inferred ionized bubble size is in contrast to what is found for similarly UV-luminous galaxies at $z\sim7$, which tend to sit in overdensities and show no strong evolution in Ly$\alpha$ visibility or damping wing strength relative to $z\sim5-6$ \citep[implying they reside in the first large bubbles, $>1$~pMpc; e.g.,][]{Stark2017,Tilvi2020,Endsley2022_overdensity,Tang2023,Umeda2026}.
That one of the brightest and most massive galaxies known at $z>9$ appears not to trace a large ionized region indicates we may be observing the earliest stages of reionization, as suggested by the large IGM neutral fraction we infer ($\bar{x}_{\rm HI}=0.81_{-0.21}^{+0.14}$).
We note that there is an overdensity of spectroscopically confirmed galaxies at $z\sim10$ in this field, some of which do not show strongly damped Ly$\alpha$ breaks in their prism spectra \citep{Napolitano2024_7wonders}, hinting they may trace longer ionized sightlines. 
It is not yet clear if the $z\sim10$ overdensity extends to \srcname{}, nor if any of these sources show Ly$\alpha$ emission.
Deep grating spectroscopy of other confirmed and photometrically selected $z>9$ sources nearby \srcname{} \citep[e.g.,][]{Atek2023,Castellano2023} will help better characterize the morphology of the ionized IGM in this field.

While our analysis above is based on a single source, the sample of bright high redshift sources with deep UV grating spectra will continue to grow over Cycle 4 and 5 from SPURS and other programs.
This expanding database will provide critical cross-validation for \lya{} emission and prism studies, and start to enable statistically meaningful constraints on the IGM neutrality at the highest redshifts. 
The IGM neutral fraction at $z\gtrsim9$ is still very unconstrained, due to systematic uncertainties and small sample sizes, with current estimates from \lya{} emission and prism damping wing studies spanning $\bar{x}_{\rm HI} \approx 0.6-1.0$ \citep[e.g.,][]{Tang2024_nirspec,Kageura2025,Mason2026,Umeda2026}.
However, \citet{Mason2026} forecasted that already with damping wing fits to $\gtrsim 6$ galaxies per redshift bin with similar quality G140M spectra to \srcname{} (i.e. SNR$\gtrsim$5 per pixel) we will be able to distinguish between a neutral fraction $\bar{x}_{\rm HI}=0.7$ and $\bar{x}_{\rm HI}=0.9$ at $>1\sigma$ significance.
Continued wide-area searches for bright $z>9$ galaxies, and subsequent spectroscopic follow-up, will be essential to capture the inhomogeneity of the reionizing IGM \citep[e.g.,][]{Bruton2023,Mason2026} and provide the sample necessary to reveal properties of the IGM during the onset of reionization.

\section{Conclusions}\label{sec:conclusion}

In this work, we present ultra-deep \jwst{}/NIRSpec $R\sim1000$ spectroscopy of \srcname{}, one of the most UV-luminous galaxies confirmed at $z>9$, obtained as part of the SPURS Cycle 4 Large Program. 
The combined spectrum, comprising 29.18~hr in G140M, 7.90~hr in G235M, and 2.92~hr in G395M, provides continuous wavelength coverage from the rest-frame UV through the rest-optical and serves as a benchmark template for understanding early galaxy formation.
With this spectrum, we characterize the stellar population properties and ISM conditions in detail.
Our findings are summarized as follows.

(i) The deep observations reveal a rest-UV spectrum characterized by extremely weak emission lines.
We detect emission from the \ciii{}] $\lambda\lambda1907,1909$ doublet, with a small total EW of $1.6_{-0.3}^{+0.3}$~\AA{}.
The doublet flux ratio implies a relatively low electron density ($<10^3$~cm$^{-3}$), significantly lower than that of other $z>9$ systems with extreme UV lines.
We find non-detections of other high-ionization emission lines, with EW \civ{}$<0.29$~\AA{}, EW \heii{}$<0.77$~\AA{}, and EW \niv{}] $<0.80$~\AA{}, further supporting the weak line nature of this source.

(ii) The rest-optical spectrum is also dominated by weak emission lines, with EW~[\oiii{}] $=257_{-22}^{+23}$~\AA{} and EW~\hb{} $=27_{-4}^{+4}$~\AA{}, as expected from earlier low-resolution prism observations.
Despite the very low EW values, the emission line ratios indicate ionization conditions (O32=$8.4_{-0.9}^{+1.2}$, Ne3O2=$0.47_{-0.12}^{+0.13}$) comparable to what has been found in typical $z>9$ galaxies.
In addition, we detect the rest-frame optical continuum with a modest Balmer break ($f_{\nu, 4050}/f_{\nu, 3500} = 1.14_{-0.09}^{+0.07}$), which suggests that this source is still dominated by relatively young stellar populations ($\lesssim40$~Myr).

(iii) We confidently detect the rest-UV continuum at grating resolution (SNR$\approx10$ per pixel), enabling characterization of ISM properties via absorption lines at $z>9$.
We find strong absorption lines in both low- and high-ionization species that are blueshifted relative to systemic redshift, suggesting outflowing gas at $\sim-161$~\kms{}.
The low-ionization absorption lines appear saturated, from which we infer a covering fraction of $\approx$0.67.
Such a large covering fraction provides stringent constraints on the escape fractions of the Lyman continuum photons ($\lesssim0.20$), suggesting that the weak emission lines are unlikely to be due to very high escape fractions.

(iv) We detect broad stellar wind absorption in both \civ{} and \nv{}, which may be expected if there is a moderately metal-enriched (0.4--0.7$~Z_\odot$) massive star population or with overabundant very massive stars.
A relatively high gas-phase metallicity is also suggested by the rest-optical emission line ratios ($12 + \log(\mathrm{O/H}) \sim8.3\mbox{--}8.5$).
These findings indicate that significant chemical enrichment in the ISM, and possibly also in stars, is already in place at $z>9$.

(v) The detection of the \lya{} break at grating resolution makes it possible to disentangle the contribution from the IGM damping wing from potential damped \lya{} absorption.
We find significant damping wing absorption from a largely neutral IGM ($\bar{x}_{\rm HI}=0.81_{-0.21}^{+0.14}$) along with a small ionized bubble ($D_b = 0.29_{-0.09}^{+0.11}$~pMpc) is required to explain the observed absorption profile.
We additionally find a tentative weak emission feature at the \lya{} wavelength (EW $=0.46_{-0.14}^{+0.13}$~\AA{} measured from the spectrum), as expected given the increased IGM attenuation of \lya{} photons at such high neutral fractions.

(vi) We find that the grating spectrum can be explained by a post-burst star formation history that peaked 10--20~Myr prior to observation, with a downturn in the most recent 10 Myr.
This model simultaneously reproduces the Balmer break, weak emission lines, and stellar wind features, without requiring very large ionizing photon escape fractions, which are disfavored by the interstellar absorption lines.
Such an SFH is consistent with stochastic star formation expected at $z>9$, wherein galaxies caught during a star formation burst emit extreme nebular emission lines, while those observed shortly after the burst will remain UV luminous but with significantly weaker lines.
Ongoing campaigns obtaining deep MIRI and NIRSpec spectroscopy will be essential to statistically establish the distribution of the star formation histories in $z>9$ galaxies.

\begin{acknowledgments}

ZC acknowledges support by VILLUM FONDEN under grant 37459.
DPS acknowledges support by the National Science Foundation under Grant No. AST-2109066.
CAM acknowledges support by the European Union ERC grant RISES (101163035), Carlsberg Foundation (CF22-1322), and VILLUM FONDEN (37459).
VG acknowledges support by the Carlsberg Foundation (CF22-1322).
The Cosmic Dawn Center (DAWN) is funded by the Danish National Research Foundation under grant DNRF140.
MT acknowledges support by a Shanghai Jiao Tong University start-up grant. 
LW acknowledges support from the Gavin Boyle Fellowship at the Kavli Institute for Cosmology, Cambridge and from the Kavli Foundation. 

This work is based in part on observations made with the NASA/ESA/CSA \jwst{}. The data were obtained from the Mikulski Archive for Space Telescopes at the Space Telescope Science Institute, which is operated by the Association of Universities for Research in Astronomy, Inc., under NASA contract NAS 5-03127 for \jwst{}. 
These observations are associated with program GO 9214.
We thank our program coordinator, Christian Soto, and our NIRSpec viewer, Diane Karakla.
We thank Stéphane Charlot and Jacopo Chevallard for providing access to the \beagle{} tool used for photoionization modeling analysis.
The Tycho supercomputer hosted at the SCIENCE HPC center at the University of Copenhagen was used for supporting this work.


\software{{\sc Astropy}:\footnote{\url{http://www.astropy.org}} a community-developed core Python package and an ecosystem of tools and resources for astronomy \citep{astropy:2013, astropy:2018, astropy:2022}; \beagle{} \citep{Chevallard2016}; {\sc Emcee} \citep{Foreman-Mackey2013}; {\sc Jupyter} \citep{Kluyver2016}; {\sc Matplotlib} \citep{Hunter:2007}; {\sc Numpy} \citep{harris2020array}; {\sc Photutils}, an {\sc Astropy} package for detection and photometry of astronomical sources \citep{Bradley2022}; \prospector{} \citep{Johnson2021}; {\sc Scipy} \citep{2020SciPy-NMeth}; and {\sc Sedpy} \citep{Johnson2021_sedpy}
}

\end{acknowledgments}

%






\appendix

\section{Additional Continuum Fitting Models}\label{appendix:other_models}

To derive constraints on the star formation history from the spectrum, we take several different modeling approaches using both the \beagle{} (\citealt{Chevallard2016}) and \prospector{} \citep{Johnson2021} codes, in addition to the models described in \S~\ref{sec:modeling}.
We first investigate whether the observations can be described by a simple constant star formation history in \S~\ref{appendix:csfh} and then discuss the effect of ionizing photon leakage in \S~\ref{appendix:csfh_fesc}.
Next, we consider the non-parametric star formation history with alternative priors by modeling with \prospector{} (\S~\ref{appendix:prospector_bursty} and \ref{appendix:prospector_continuity}).
We comment on the impact of a lower stellar metallicity than that of the ISM in \S~\ref{appendix:low_stellar_metallicity}.

Unless otherwise stated, throughout the modeling process, we keep the assumptions other than the SFH as similar to those described in \S~\ref{subsec:modeling_method} as possible across different models and codes in this section.
These assumptions include the initial mass function of \cite{Chabrier2003} and the SMC extinction curve of \cite{Pei1992}.
We also adopt the same priors on metallicity, ionization parameter, dust-to-metal ratio, and optical depth in the V band, as described in \S~\ref{subsec:modeling_method}.
For the modeling with \beagle{}, by default, we also follow the approach described in \S~\ref{subsec:modeling_method}, first fitting with the rest-optical emission line fluxes and the pseudo photometry (synthesized from the spectrum in NIRCam broadband filters), and adopt the results as priors to fit the full spectrum.
We will primarily discuss the final results from this full spectrum fit below.
In what follows, we describe each of these model sets in turn.

\subsection{BEAGLE Constant SFH Models}\label{appendix:csfh}

\begin{figure*}
    \centering
    \includegraphics[width=1\linewidth]{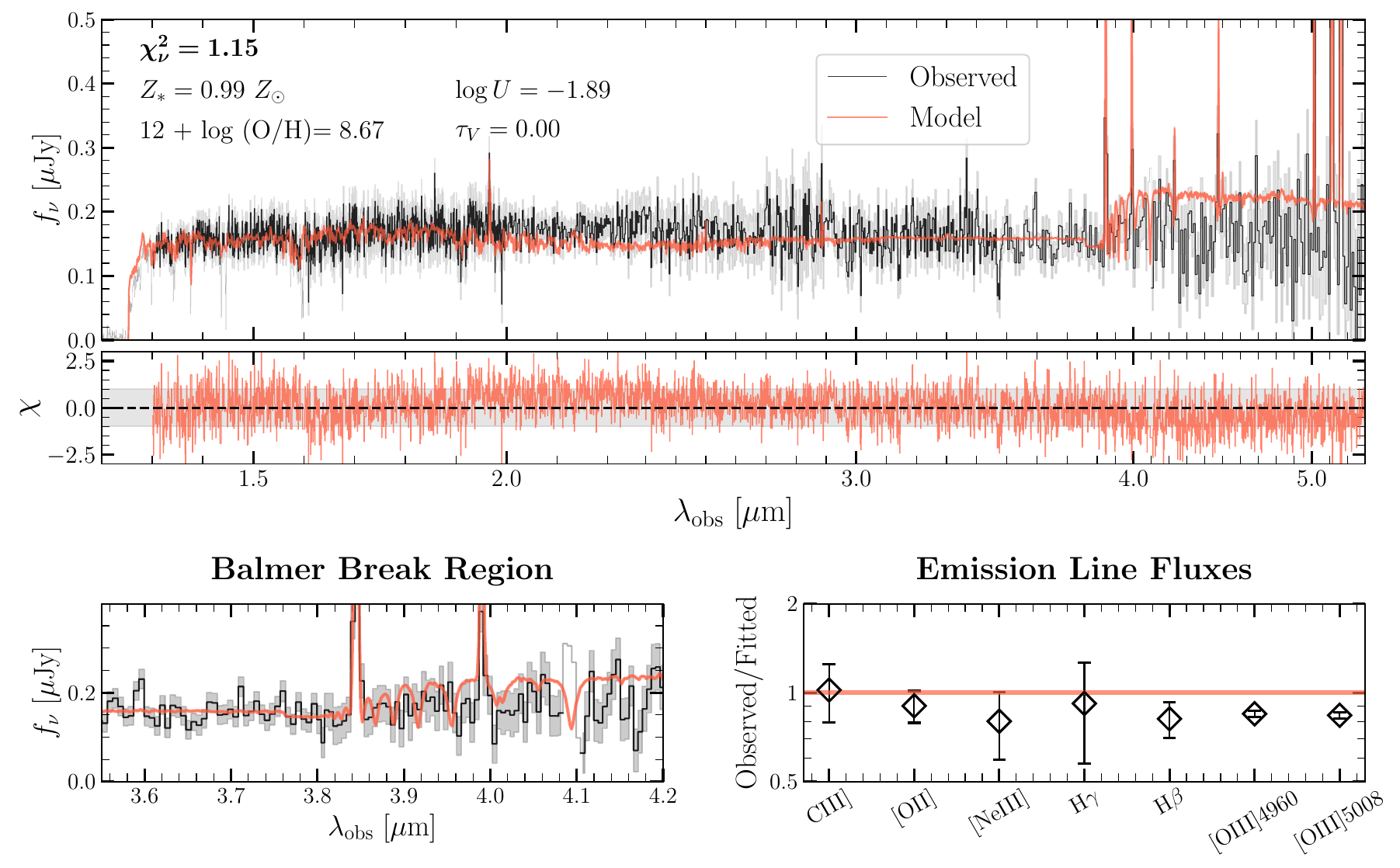}
    \caption{\beagle{} models fit to the observed grating spectrum, assuming a simple constant SFH (CSFH).
    These models do not allow the escape of ionizing photons.
    As in Figure~\ref{fig:spectral_fit_fiducial}, in the top panel, we plot the observed spectrum ($3\mbox{--}6\times$ binning at wavelengths longer than 2~\um{} for better visualization) with a solid black line.
    The model spectrum is shown in red (without binning).
    As shown in the bottom panel, these simple CSFH models predict a Balmer break slightly stronger than observed, with most of the emission line fluxes deviating significantly from our measurements.}
    \label{fig:beagle_fit_csfh}
\end{figure*}

We consider whether the high luminosity and weak emission lines in \srcname{} are due to extended star formation over a long period ($\gtrsim 100$~Myr).
To do so, we fit the spectrum with the \beagle{} models ($Z_*=Z_{\rm ISM}$) assuming a simple constant SFH (CSFH), as commonly done in fitting early galaxies \citep[e.g.,][]{Tang2025,Helton2025_MIRIphot}.
The CSFH is described by two parameters: the star formation duration and the total stellar mass. 
We assume log-uniform priors on both parameters, allowing the star formation duration to vary log-uniformly between 1~Myr and the age of the Universe at the source redshift, and the stellar mass to vary from $10^5$ to $10^{12}$~$M_\odot$.
In these simple CSFH models, we do not allow the escape of ionizing photons, which we will discuss separately in the next subsection.

We first consider the results derived from assuming the priors on ISM parameters and SFH derived from fitting emission lines together with the pseudo photometry.
As shown in Figure~\ref{fig:beagle_fit_csfh}, the resulting fit is much worse than the models with SFH downturns as described in \S~\ref{subsubsec:fit_continuum}, with $\chi^2_\nu = 1.15$.
As already discussed in the main text, this is due to the discrepancy in the rest-optical, where the model predicts a stronger rest-optical continuum resulting in a much more prominent Balmer break ($f_{\nu, 4050}/f_{\nu, 3500}=1.48$) than what we observe.
This is because the model favors an extremely old CSFH age ($198$~Myr) in order to match the rest-frame UV continuum as well as the weak emission lines.
It additionally adopts a near solar metallicity ($Z_* = 0.99~Z_\odot$) to further suppress the emission lines.
However, the model still overpredicts the rest-optical emission line fluxes for both \hb{} and [\oiii{}] (by $\approx20\%$) and fails to reproduce the continuum at 2--2.5~\um{} (corresponding to 1940--2400~\AA{} in the rest-frame).

To explore the full potential for the CSFH models to reproduce the spectrum, we additionally repeat the fitting, but allowing the ISM parameters (metallicity, ionization parameter, dust-to-metal ratio, and dust attenuation) to vary freely.
We adopt the wide log-uniform priors on metallicity and ionization parameter, with uniform priors on dust-to-metal ratio and dust attenuation, as those used in fitting the emission lines (see \S~\ref{subsubsec:fit_line})
As expected, to match the weak [\oiii{}]~$\lambda5008$ emission line, the CSFH models resort to a very old stellar population age (CSFH duration $511$~Myr) that is comparable to the age of the Universe at $z=9.311$.
In the meantime, as an attempt to reproduce the weak Balmer break in the spectrum, the models require a very low stellar metallicity ($\log (Z/Z_\odot) = -1.77$, or 1.7\%~$Z_\odot$) as the strength of Balmer breaks for old stellar populations becomes weaker as the metallicity decreases.
While a low metallicity also helps weaken the emission lines, an old stellar population age is required to bring the emission line EW low (EW~[\oiii{}]+\hb{} = 259~\AA{}) while also matching the shape of the continuum.
Still, the combination of a very old age with a low metallicity results in an overall $\chi^2_\nu = 1.34$ that is noticeably larger than that of our models with a recent downturn described in \S~\ref{subsubsec:fit_continuum}.

Indeed, we find discrepancies in the other emission lines both in the rest optical and UV.
While the low metallicity helps to weaken the [\oiii{}] emission line, the Balmer emission lines are expected to be strong due to the low stellar opacity \citep[e.g.,][]{Endsley2024_jades}.
The resulting \hb{} and \hg{} emission line fluxes are $4\times$ stronger than the observed values.
The low-metallicity solutions also produce significantly stronger nebular \ciii{}] (a factor of $2\times$ larger) and \civ{} (a factor of $8\times$ than the measured $3\sigma$ upper limit) emission line fluxes than observed, as expected from the hardening of the ionizing spectrum at smaller metallicities \citep[e.g.,][]{Senchyna2017}.
In addition, the resulting stellar mass for the main clump targeted by our observations is very large, $\log (M/M_*)=9.69$.
If scaled to that of the entire galaxy, assuming the same mass-to-light ratio in (at rest-frame 1500~\AA{}), we expect an integrated stellar mass of $\approx 7.7\times10^9~M_\odot$, which is unlikely at such early epochs \citep[e.g.,][]{BoylanKolchin2023,Boyett2024_z9p3}.
Our experiments suggest that simple CSFH models struggle to simultaneously match the full continuum and emission line features in the SPURS spectrum, and the weak emission lines in \srcname{} are unlikely to be explained by the presence of a mature stellar population.

\subsection{BEAGLE Constant SFH Models with Ionizing Photon Escape}\label{appendix:csfh_fesc}

\begin{figure*}
    \centering
    \includegraphics[width=1\linewidth]{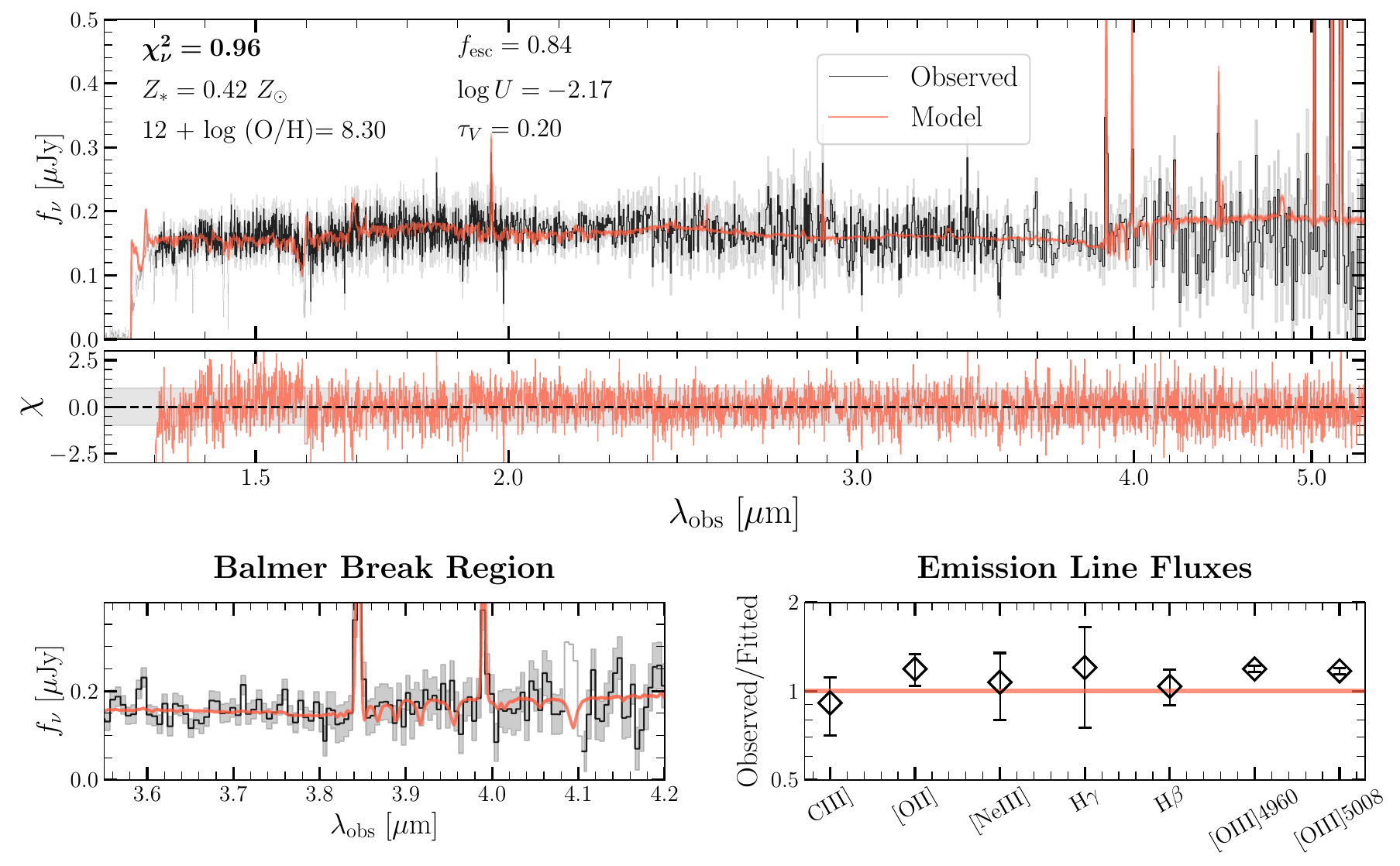}
    \caption{BEAGLE models fit to the observed grating spectrum, assuming a constant SFH (CSFH) and allowing the ionizing photon escape fraction to vary freely.
    As in Figure~\ref{fig:spectral_fit_fiducial}, we plot the observed spectrum ($3\mbox{--}6\times$ binning at wavelengths longer than 2~\um{} for better visualization) with a solid black line and models in red in the top panel.
    The comparison of Balmer break and emission line fluxes between the model and observed values are given in the bottom panels.
    The model reproduces the Balmer break and the weakness of \hb{} emission line with a high ionizing photon escape fraction, though it overpredicts the [\oiii{}] fluxes.
    However, given the presence of strong LIS absorption lines in the rest-UV spectrum, the neutral gas covering fraction is expected to be large, and a high escape fraction of ionizing photons is unlikely.
    }
    \label{fig:beagle_fit_csfh_fesc}
\end{figure*}

We investigate whether ionizing photon leakage can help explain the weak emission lines in \srcname{} without requiring a very old stellar population age.
To do so, we repeat the \beagle{} CSFH modeling with the same setup as in \S~\ref{appendix:csfh}, except that we now allow a non-zero fraction of Lyman continuum photons to escape.
We consider two scenarios for ionizing-photon escape: (i) the ``picket-fence'' (ionization-bounded) geometry, in which ionizing photons escape through low-\hi{}-column-density sightlines \citep[e.g.,][]{Heckman2001,Heckman2011,RiveraThorsen2017,Gazagnes2018}, and (ii) density-bounded \hii{} regions \citep{Plat2019}.
We first consider each of these cases, allowing the escape fraction to vary uniformly over $f_{\rm esc}\in[0,1]$.
We will also discuss the results if we conservatively allow a moderate escape fraction ($f_{\rm esc}=0.20$), which is still consistent with expectations from the interstellar absorption lines.
We describe each set of models in turn below.

In the models allowing a varying fraction of ionizing photons to escape through low-density channels, the weak emission lines can be easily reproduced with a very large escape fraction ($f_{\rm esc}=0.84$; Figure~\ref{fig:beagle_fit_csfh_fesc}).
In these models, the weak Balmer break can be matched by an extremely young stellar population (median CSFH age $16$~Myr), and the presence of a very old stellar population is no longer required.
At very large escape fractions, we expect the nebular continuum contribution in the near UV to reduce even at young stellar population ages, which usually will lead to very blue UV slopes dominated by the continuum from massive stars \citep[e.g.,][]{Topping2022_ceers}.
Accordingly, the model also uses some a small amount of dust attenuation (V-band dust optical depth $\tau_{\rm V}=0.20$) to redden the spectrum back to the observed flat UV slope ($\beta = -1.94$).
Overall, these CSFH models with ionizing photon escape yield a best-fit  $\chi_\nu^2$ ($0.96$) that is significantly better than that without it.

We also fit the spectrum with the models of \citep{Plat2019} that allow the escape of ionizing photons in the density-bound regime.
Compared to the ionization-bounded models, the density-bounded models not only reduce the absolute emission line fluxes due to escape of ionizing photons, but also affect the emission line flux ratios, due to the weaker low-ionization emission lines \citep[e.g.,][]{Jaskot2013,Nakajima2014,Izotov2018_higho32}.
We adopt a uniform prior on the escape fraction over $f_{\rm esc}\in[0,1]$.
We find these density-bound models yield qualitatively similar fits ($\chi_\nu^2 = 0.97$), requiring a large escape fraction ($f_{\rm esc}=0.77$) to weaken the emission lines.
Again, the large escape fraction allows the fit to converge to a relatively young stellar population age ($37$~Myr) to match the observed rest-optical continuum without producing a strong Balmer break.

While the models presented here demonstrate that substantial leakage of ionizing photons is capable of reproducing the overall continuum shape and weak emission lines, we consider whether such solutions are physically plausible given the observations of rest-UV absorption lines.
The deep rest-UV grating spectroscopy provides stringent constraints on the covering fraction of low-ionization absorption lines, which directly traces \hi{} covering fraction, and, in turn, is related to the ionizing photon escape fraction.
As we have shown in \S~\ref{subsec:ism_abs}, the SPURS spectrum reveals the presence of the deep low-ionization absorption lines from multiple species, including \siii{}, \cii{}, \oi{}, and \alii{}, likely indicating a large covering fraction ($\approx0.67$) of these ions and thus a very high covering fraction of neutral gas.
If the empirical relations between these parameters found at lower redshifts ($z\sim0$ and $z\sim2$; \citealt{Reddy2016,Reddy2022,Gazagnes2018,Saldana-Lopez2022}) hold at $z\sim9$, we expect the absorption line covering fraction in \srcname{} translates to only modest escape fractions ($0.1\mbox{--}0.2$). 
Indeed, the presence of both a high escape fraction and a large neutral gas covering fraction is only expected in rare conditions where the ionizing O stars are offset from the slightly older B stars, leading to different effective covering fractions for the ionizing and non-ionizing continuum.
This indicates that a high escape fraction ($\sim 0.80$) is disfavored as the primary cause of the weak emission lines in \srcname{}, emphasizing the necessity of deep absorption line spectroscopy to confirm or rule out extreme Lyman continuum leakage.

\begin{figure*}
    \centering
    \includegraphics[width=1\linewidth]{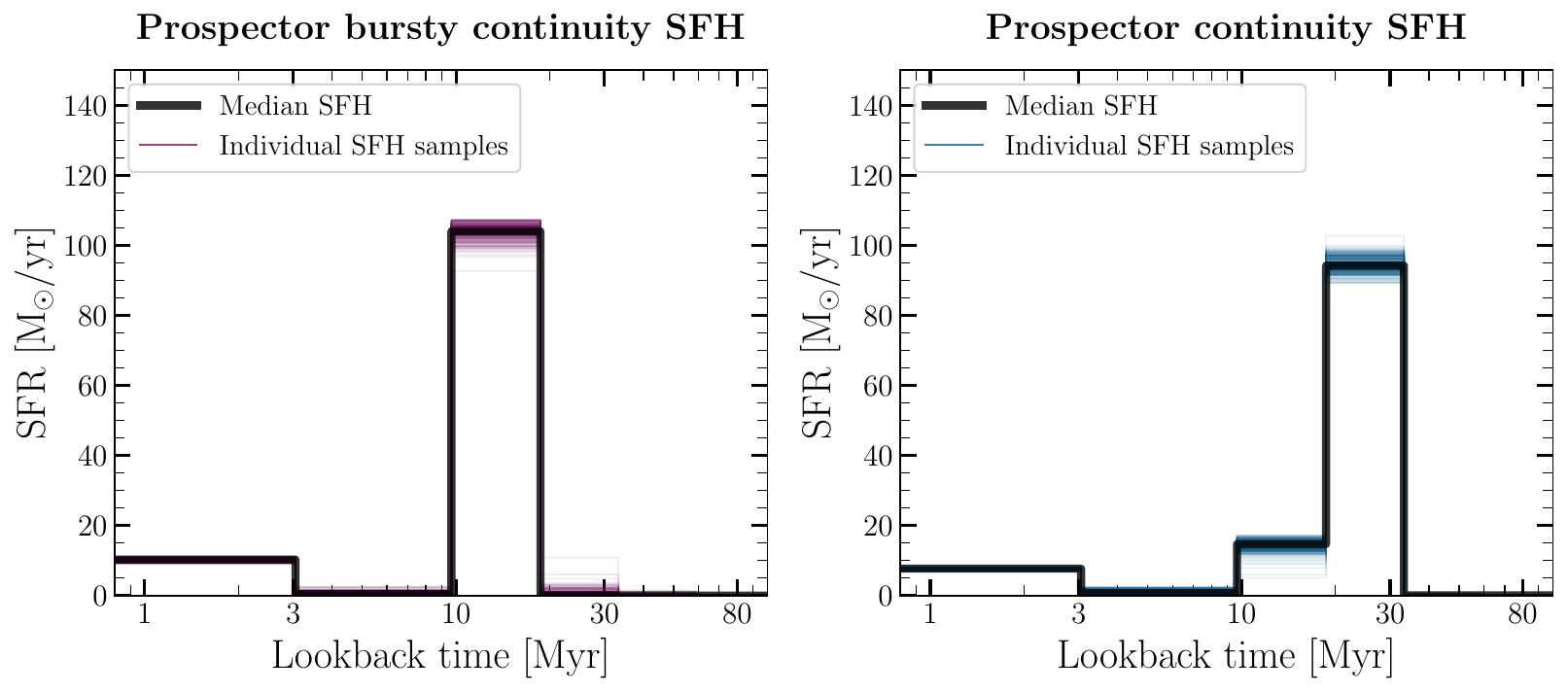}
    \caption{The binned star formation history from the \prospector{} non-parametric fit to the spectrum assuming the `bursty' continuity prior (left) and the continuity prior (right).
    The thin colored lines correspond to individual SFH samples, while the thick black line is the median SFH.
    }
    \label{fig:sfh_fiducial_prospector}
\end{figure*}

Motivated by the expectations based on the absorption line covering fraction, we further evaluate the CSFH models in which we fix the ionizing photon escape fraction to $f_{\rm esc}=0.20$, consistent with the upper limit allowed by the absorption lines.
Here, we focus on discussing the ionizing photon escape in the context of ionization-bounded models, but we note that adopting the density-bounded models yields similar results.
In an attempt to match the observations, these models still require a very old stellar population age (CSFH duration $153$~Myr) along with a very low metallicity ($\log(Z/Z_\odot) = -1.68$, or $0.02~Z_\odot$).
These values are only slightly less extreme compared to what we obtained in the models without ionizing photon escape (\S~\ref{appendix:csfh}), which is also expected, as the assumed escape fraction is small and thus will not significantly affect the emission line strength (only by 20\%).
We find similar inconsistency in the emission line fluxes in the Balmer emission lines, both of which are 3--4 times larger than observed, as well as a \civ{} profile pure in emission as seen in the models without ionizing photon escape.
These new models also require an extreme stellar mass, which is $6.5\times10^9~M_\odot$ when scaled to the integrated stellar mass of the galaxy (assuming the mass-to-light ratio in F150W).
Our results again suggest a simple constant star formation history is insufficient to describe the observed spectrum, even when we allow an escape fraction of ionizing photons that is consistent with the absorption lines.

\subsection{\prospector{} `Bursty' Prior SFH Models}\label{appendix:prospector_bursty}

To investigate whether our fit is sensitive to the choice of stellar population models or priors, we also compare our \beagle{} non-parametric SFH models with those from the \prospector{} fits.
\prospector{} \citep{Johnson2021} is based on a different set of stellar population models computed with the Flexible Stellar Population Synthesis code \citep{Conroy2009,Conroy2010}, and we adopt the default stellar population models assuming the MIST isochrones \citep{Choi2016} and the MILES spectral library \citep{SanchezBlazquez2006}.
The stellar emission is then combined with the nebular emission from \cloudy{} as described in \cite{Byler2017}. 

Similar to our approach described in \S~\ref{subsec:modeling_method}, we assume a non-parametric SFH, with the age bins chosen to follow the previous works in \prospector{} fitting \citep[e.g.,][]{Whitler2023_ceers,Tacchella2023}.
Specifically, we divide the age bins into eight intervals, where the first two are fixed to 0--3 Myr and 3--10 Myr.
The remaining age bins are spaced evenly in logarithmic time until a formation redshift of $z_{\rm f}=20$.
We assume a `bursty’ prior that allows for rapid changes in SFR over short timescales, which is achieved by fitting the logarithm of the SFR ratios between adjacent age bins with a Student's $t$-distribution prior with dispersion $\sigma$=1.
We will also explore different priors in the next subsection.

For simplicity, we directly fit with \prospector{} the spectrum that is binned every 50~\AA{} from 1350~\AA{} to 5100~\AA{} in the rest frame, instead of first fitting the line fluxes and pseudo broadband photometry and then the full spectrum.
This simpler approach allows us to sample the continuum shape from rest-frame UV to optical wavelengths, while the EWs of the emission lines are also recorded by the flux excess in the corresponding bin.
In doing so, we focus on the constraints on the star formation history that are required to be consistent with the observations.

The \prospector{} models employing the `bursty' continuity priors on SFHs successfully match both the continuum shape from the UV to optical and the flux excess due to emission lines in longer wavelength bins, albeit with a best-fit $\chi^2_\nu$ ($1.36$) that is larger than that of the \beagle{} non-parametric models.
The resulting constraints on the SFH are remarkably consistent with the \beagle{} findings.
As expected, the \prospector{} models require a strong burst of star formation in the age bin of 10--20 Myr at an average rate SFR $= 104$~$M_\odot$~yr$^{-1}$, in order to reproduce the observed strength of the Balmer break (Figure~\ref{fig:sfh_fiducial_prospector}).
The fits also suggest that, following this main epoch, the star formation continues at a much lower rate ($10$~$M_\odot$~yr$^{-1}$), with the nebular emission lines weakening in EWs due to the significant contribution to the continuum from the B and A stars formed earlier.
Our results suggest that a post-burst solution in \srcname{} is preferred by both SED fitting codes to simultaneously explain both the weak emission lines and the Balmer break continuum.

\subsection{\prospector{} Continuity Prior SFH Models}\label{appendix:prospector_continuity}

We specifically examine a different prior setup with \prospector{}: the continuity prior that distributes the stellar mass formed equally over time.
In these models, we fit the logarithm of the SFR ratios between adjacent age bins with a Student's $t$-distribution prior, but now with dispersion $\sigma=$0.2 \citep[e.g.,][]{Whitler2023_ceers,Tacchella2023}.
If the observed spectrum were consistent with a steadier star formation history, we might expect models that weigh against rapid SFR variations to be able to recover such a solution.

The \prospector{} continuity SFH models fit the spectrum slightly worse compared to the `bursty' continuity SFH models, with a somewhat larger reduced $\chi^2$ value ($\chi^2_\nu = 1.67$).
While both models suggest SFHs that peak very recently, the peak SFR in the continuity SFH model occurred in a slightly older age bin (20--30~Myr prior to observation).
However, even when adopting priors that explicitly discourage an abrupt decline in SFR, the continuity models still converge to an SFH solution that peaks very recently (20--30~Myr prior to observation), which then declines by a similar factor ($\sim12\times$) over the past 10~Myr (see Figure~\ref{fig:sfh_fiducial_prospector}), as we have seen in the \prospector{} `bursty' SFH models and \beagle{} non-parametric SFH models.

\begin{figure*}
    \centering
    \includegraphics[width=1\linewidth]{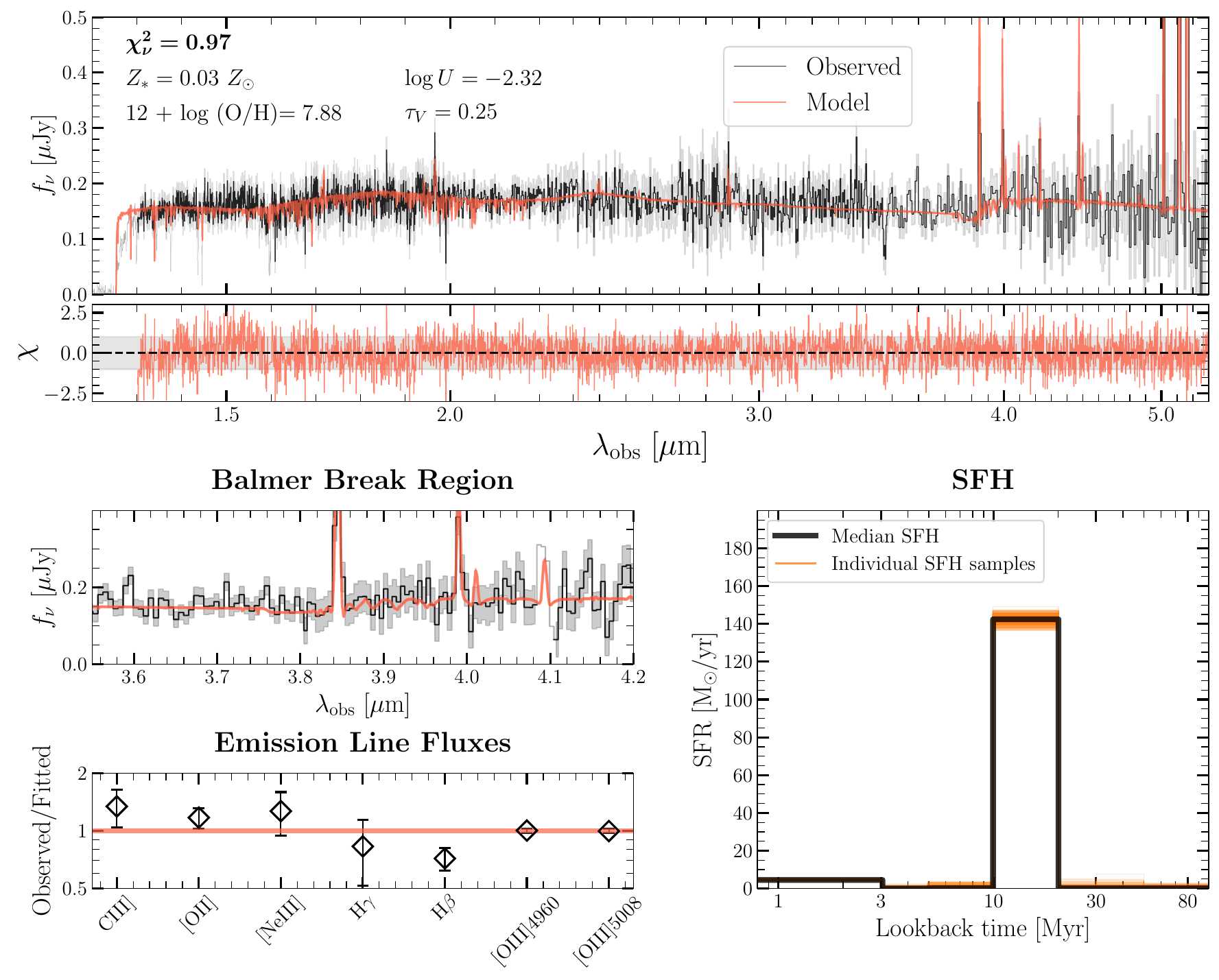}
    \caption{BEAGLE models fit to the observed grating spectrum, assuming a non-parametric SFH and a lower stellar metallicity than ISM ($Z_* = 0.2\,Z_{\rm ISM}$).
    As in Figure~\ref{fig:spectral_fit_fiducial}, we plot the observed spectrum ($3\mbox{--}6\times$ binning at wavelengths longer than 2~\um{} for better visualization) with a solid black line and models in red in the top panel. 
    We show the comparison of Balmer break and line fluxes in the bottom panels.
    Compared to the model with $Z_* = Z_{\rm ISM}$ discussed in \S~\ref{subsubsec:fit_continuum}, this model better fits the rest-optical continuum with a very low stellar metallicity ($Z_*=0.03\,Z_\odot$).
    However, this metallicity leads to discrepancies in emission line fluxes and requires VMS to reproduce the strong \civ{} wind absorption (see main text).
    }
    \label{fig:beagle_fit_alphaEnhanced}
\end{figure*}

\subsection{BEAGLE Non-parametric Models: Impact of Lower Stellar Metallicity than ISM}\label{appendix:low_stellar_metallicity}

Motivated by results at lower redshifts \citep[e.g.,][]{Steidel2016,Sanders2020,Topping2020,Runco2021}, we also fit with \beagle{} models in which the stellar metallicity is allowed to be lower than interstellar metallicity to investigate its impact on the inferred properties and star formation history.
Since the stellar metallicity is set by iron and iron-peak elements, and the gas-phase metallicity 
is largely set by oxygen and other $\alpha$ elements, this  
approximates the effects of $\alpha$ enhancement  \citep[e.g.,][]{Steidel2016,Sanders2020,Topping2020,Runco2021}, producing a harder ionizing spectrum at fixed interstellar metallicity.
Following \cite{Topping2022_ceers}, we specifically test models computed with the stellar metallicity fixed at $Z_* = 0.2\,Z_{\rm ISM}$, corresponding to the theoretical limit expected for Type~II supernova enrichment \citep{Nomoto2006}.
To be consistent with the $Z_* = Z_{\rm ISM}$ presented in \S~\ref{sec:modeling}, we adopt the same non-parametric form of SFH and assume a zero escape fraction of ionizing photons.

We first fit the rest-optical line fluxes and the pseudo-photometry using these $Z_* = 0.2 Z_{\rm ISM}$ \beagle{} models.
Since these models produce a harder ionizing spectrum at the metallicities that are relevant for this source, we may expect it to affect the observed emission line ratios sensitive to gas ionization (e.g., O32).
As expected, to reproduce the O32 and Ne3O2 line ratios, the model fixing $Z_* = 0.2 Z_{\rm ISM}$ requires a slightly lower ionization parameter (log~$U=-2.32_{-0.07}^{+0.10}$) than that of the $Z_* = Z_{\rm ISM}$ model presented in \S~\ref{subsubsec:fit_line}.
This $Z_* = 0.2 Z_{\rm ISM}$ model also indicates small but non-negligible dust attenuation, though with a slightly higher V-band dust optical depth ($\tau_{\rm V}=0.24_{-0.02}^{+0.02}$) than the $Z_* = Z_{\rm ISM}$ model.
It yields an ISM metallicity of $Z_{\rm ISM} = 0.30_{-0.06}^{+0.08}~Z_\odot$, which translates to a gas-phase abundance of $12 + \log(\mathrm{O/H}) = 8.17_{-0.10}^{+0.10}$.
Compared to that of the $Z_* = Z_{\rm ISM}$ model, this metallicity is slightly lower, which is also expected from the relatively smaller inferred ionization parameter and its associated degeneracy with metallicity.
Nevertheless, both models consistently suggest that moderately enriched ionized gas is required to reproduce the observed emission line fluxes.

We also investigate how the inferred SFH changes if the stellar metallicity is lower than that of the ISM. 
To do so, we fit the grating spectrum using Gaussian priors based on our emission line + pseudo-photometry fits above, with the result showing in Figure~\ref{fig:beagle_fit_alphaEnhanced}.
We find that this model produces a slightly bluer rest-optical continuum in better agreement with the observations by resorting to a very low stellar metallicity ($Z_* = 0.03\,Z_\odot$) than the $Z_* = Z_{\rm ISM}$ model.
Due to the coupling between the stellar metallicity and ISM metallicity, we also find a low $Z_{\rm ISM} = 0.15\,Z_\odot$, corresponding to a gas-phase oxygen abundance of $12 + \log(\mathrm{O/H}) = 7.88$.
However, these values lead to some discrepancies in the rest-optical emission line fluxes when compared to our observed values (with model \hb{} flux $\sim3\sigma$ lower, but other lines consistent within 1.2$\sigma$).
Regardless, the match in continuum also requires an SFH characterized by a prominent peak at 10--20~Myr prior to observations and then a strong recent decline (by a factor of 33) in SFR over the last 10 Myr, similar to what we have seen from the $Z_* = Z_{\rm ISM}$ models.

While the very low stellar metallicity in the $Z_* = 0.2\,Z_{\rm ISM}$ model matches the rest-optical continuum, it also leads to underprediction of the \civ{} wind absorption strength due to the weakening of stellar winds at lower metallicities. One possibility is that if we adopted true $\alpha$-enhanced models, we may expect stronger CIV absorption at low stellar (i.e. iron) metallicity.  Alternatively, one other possibility is an overabundance of very massive stars (VMS; $M>100~M_\odot$) with boosted mass loss rates \citep[e.g.,][]{Vink2018}.
Here, we seek to illustrate the effect of overabundant VMS by performing a simple experiment by combining the low-metallicity model assuming $Z_* = 0.2\,Z_{\rm ISM}$ with a synthetic VMS spectrum. 
We use the VMS model spectra drawn from the POLLUX database\footnote{\url{https://pollux.oreme.org}} \citep{Martins2025} without metallicity scaling (stars at 150--300~$M_\odot$, weighted by a \citealt{Salpeter1955} IMF).
model spectrum (stars at 150--300~$M_\odot$, weighted by a \citealt{Salpeter1955} IMF) from \cite{Martins2021}.
We find that a 15\% contribution from a VMS model at 2~Myr with metallicity of $0.1~Z_\odot$ to the rest-UV continuum at 1600~\AA{} (keeping the total continuum flux unchanged) is sufficient to recover the observed \civ{} wind absorption, which is also shown in Figure~\ref{fig:civ_fit}.
Ultimately, deep observations in the rest-frame optical will allow confirmation of the continuum slope, which will help further constrain whether very low stellar metallicity is required.



\bibliography{main}{}

@ARTICLE{Tang2023,
       author = {{Tang}, Mengtao and {Stark}, Daniel P. and {Chen}, Zuyi and {Mason}, Charlotte and {Topping}, Michael and {Endsley}, Ryan and {Senchyna}, Peter and {Plat}, Ad{\`e}le and {Lu}, Ting-Yi and {Whitler}, Lily and {Robertson}, Brant and {Charlot}, St{\'e}phane},
        title = "{JWST/NIRSpec spectroscopy of z = 7-9 star-forming galaxies with CEERS: new insight into bright Ly{\ensuremath{\alpha}} emitters in ionized bubbles}",
      journal = {\mnras},
         year = 2023,
        month = dec,
       volume = {526},
       number = {2},
        pages = {1657-1686},
          doi = {10.1093/mnras/stad2763},
archivePrefix = {arXiv},
       eprint = {2301.07072},
 primaryClass = {astro-ph.GA},
       adsurl = {https://ui.adsabs.harvard.edu/abs/2023MNRAS.526.1657T}
}

@software{Bradley2022,
author       = {Larry Bradley and
                Brigitta Sipőcz and
                Thomas Robitaille and
                Erik Tollerud and
                Zé Vinícius and
                Christoph Deil and
                Kyle Barbary and
                Tom J Wilson and
                Ivo Busko and
                Axel Donath and
                Hans Moritz Günther and
                Mihai Cara and
                P. L. Lim and
                Sebastian Meßlinger and
                Simon Conseil and
                Azalee Bostroem and
                Michael Droettboom and
                E. M. Bray and
                Lars Andersen Bratholm and
                Geert Barentsen and
                Matt Craig and
                Shivangee Rathi and
                Sergio Pascual and
                Gabriel Perren and
                Iskren Y. Georgiev and
                Miguel de Val-Borro and
                Wolfgang Kerzendorf and
                Yoonsoo P. Bach and
                Bruno Quint and
                Harrison Souchereau},
title        = {astropy/photutils: 1.5.0},
month        = jul,
year         = 2022,
publisher    = {Zenodo},
version      = {1.5.0},
doi          = {10.5281/zenodo.6825092},
url          = {https://doi.org/10.5281/zenodo.6825092}
}

@ARTICLE{Chevallard2016,
       author = {{Chevallard}, Jacopo and {Charlot}, St{\'e}phane},
        title = "{Modelling and interpreting spectral energy distributions of galaxies with BEAGLE}",
      journal = {\mnras},
         year = 2016,
        month = oct,
       volume = {462},
       number = {2},
        pages = {1415-1443},
          doi = {10.1093/mnras/stw1756},
archivePrefix = {arXiv},
       eprint = {1603.03037},
 primaryClass = {astro-ph.GA},
       adsurl = {https://ui.adsabs.harvard.edu/abs/2016MNRAS.462.1415C}
}

@ARTICLE{Gutkin2016,
       author = {{Gutkin}, Julia and {Charlot}, St{\'e}phane and {Bruzual}, Gustavo},
        title = "{Modelling the nebular emission from primeval to present-day star-forming galaxies}",
      journal = {\mnras},
         year = 2016,
        month = oct,
       volume = {462},
       number = {2},
        pages = {1757-1774},
          doi = {10.1093/mnras/stw1716},
archivePrefix = {arXiv},
       eprint = {1607.06086},
 primaryClass = {astro-ph.GA},
       adsurl = {https://ui.adsabs.harvard.edu/abs/2016MNRAS.462.1757G}
}

@ARTICLE{Bruzual2003,
       author = {{Bruzual}, G. and {Charlot}, S.},
        title = "{Stellar population synthesis at the resolution of 2003}",
      journal = {\mnras},
         year = 2003,
        month = oct,
       volume = {344},
       number = {4},
        pages = {1000-1028},
          doi = {10.1046/j.1365-8711.2003.06897.x},
archivePrefix = {arXiv},
       eprint = {astro-ph/0309134},
 primaryClass = {astro-ph},
       adsurl = {https://ui.adsabs.harvard.edu/abs/2003MNRAS.344.1000B}
}

@ARTICLE{Ferland2013,
       author = {{Ferland}, G.~J. and {Porter}, R.~L. and {van Hoof}, P.~A.~M. and {Williams}, R.~J.~R. and {Abel}, N.~P. and {Lykins}, M.~L. and {Shaw}, G. and {Henney}, W.~J. and {Stancil}, P.~C.},
        title = "{The 2013 Release of Cloudy}",
      journal = {\rmxaa},
         year = 2013,
        month = apr,
       volume = {49},
        pages = {137-163},
archivePrefix = {arXiv},
       eprint = {1302.4485},
 primaryClass = {astro-ph.GA},
       adsurl = {https://ui.adsabs.harvard.edu/abs/2013RMxAA..49..137F}
}

@ARTICLE{Chabrier2003,
       author = {{Chabrier}, Gilles},
        title = "{Galactic Stellar and Substellar Initial Mass Function}",
      journal = {\pasp},
         year = 2003,
        month = jul,
       volume = {115},
       number = {809},
        pages = {763-795},
          doi = {10.1086/376392},
archivePrefix = {arXiv},
       eprint = {astro-ph/0304382},
 primaryClass = {astro-ph},
       adsurl = {https://ui.adsabs.harvard.edu/abs/2003PASP..115..763C}
}

@ARTICLE{Chen2015,
       author = {{Chen}, Yang and {Bressan}, Alessandro and {Girardi}, L{\'e}o and {Marigo}, Paola and {Kong}, Xu and {Lanza}, Antonio},
        title = "{PARSEC evolutionary tracks of massive stars up to 350 M$_{{\ensuremath{\odot}}}$ at metallicities 0.0001 {\ensuremath{\leq}} Z {\ensuremath{\leq}} 0.04}",
      journal = {\mnras},
         year = 2015,
        month = sep,
       volume = {452},
       number = {1},
        pages = {1068-1080},
          doi = {10.1093/mnras/stv1281},
archivePrefix = {arXiv},
       eprint = {1506.01681},
 primaryClass = {astro-ph.SR},
       adsurl = {https://ui.adsabs.harvard.edu/abs/2015MNRAS.452.1068C}
}

@ARTICLE{Bressan2012,
       author = {{Bressan}, Alessandro and {Marigo}, Paola and {Girardi}, L{\'e}o. and {Salasnich}, Bernardo and {Dal Cero}, Claudia and {Rubele}, Stefano and {Nanni}, Ambra},
        title = "{PARSEC: stellar tracks and isochrones with the PAdova and TRieste Stellar Evolution Code}",
      journal = {\mnras},
         year = 2012,
        month = nov,
       volume = {427},
       number = {1},
        pages = {127-145},
          doi = {10.1111/j.1365-2966.2012.21948.x},
archivePrefix = {arXiv},
       eprint = {1208.4498},
 primaryClass = {astro-ph.SR},
       adsurl = {https://ui.adsabs.harvard.edu/abs/2012MNRAS.427..127B}
}

@ARTICLE{Pei1992,
       author = {{Pei}, Yichuan C.},
        title = "{Interstellar Dust from the Milky Way to the Magellanic Clouds}",
      journal = {\apj},
         year = 1992,
        month = aug,
       volume = {395},
        pages = {130},
          doi = {10.1086/171637},
       adsurl = {https://ui.adsabs.harvard.edu/abs/1992ApJ...395..130P}
}

@ARTICLE{deGraaff2024_kinematics,
       author = {{de Graaff}, Anna and {Rix}, Hans-Walter and {Carniani}, Stefano and {Suess}, Katherine A. and {Charlot}, St{\'e}phane and {Curtis-Lake}, Emma and {Arribas}, Santiago and {Baker}, William M. and {Boyett}, Kristan and {Bunker}, Andrew J. and {Cameron}, Alex J. and {Chevallard}, Jacopo and {Curti}, Mirko and {Eisenstein}, Daniel J. and {Franx}, Marijn and {Hainline}, Kevin and {Hausen}, Ryan and {Ji}, Zhiyuan and {Johnson}, Benjamin D. and {Jones}, Gareth C. and {Maiolino}, Roberto and {Maseda}, Michael V. and {Nelson}, Erica and {Parlanti}, Eleonora and {Rawle}, Tim and {Robertson}, Brant and {Tacchella}, Sandro and {{\"U}bler}, Hannah and {Williams}, Christina C. and {Willmer}, Christopher N.~A. and {Willott}, Chris},
        title = "{Ionised gas kinematics and dynamical masses of z {\ensuremath{\gtrsim}} 6 galaxies from JADES/NIRSpec high-resolution spectroscopy}",
      journal = {\aap},
         year = 2024,
        month = apr,
       volume = {684},
          eid = {A87},
        pages = {A87},
          doi = {10.1051/0004-6361/202347755},
archivePrefix = {arXiv},
       eprint = {2308.09742},
 primaryClass = {astro-ph.GA},
       adsurl = {https://ui.adsabs.harvard.edu/abs/2024A&A...684A..87D}
}

@ARTICLE{Treu2022,
       author = {{Treu}, T. and {Roberts-Borsani}, G. and {Bradac}, M. and {Brammer}, G. and {Fontana}, A. and {Henry}, A. and {Mason}, C. and {Morishita}, T. and {Pentericci}, L. and {Wang}, X. and {Acebron}, A. and {Bagley}, M. and {Bergamini}, P. and {Belfiori}, D. and {Bonchi}, A. and {Boyett}, K. and {Boutsia}, K. and {Calabr{\'o}}, A. and {Caminha}, G.~B. and {Castellano}, M. and {Dressler}, A. and {Glazebrook}, K. and {Grillo}, C. and {Jacobs}, C. and {Jones}, T. and {Kelly}, P.~L. and {Leethochawalit}, N. and {Malkan}, M.~A. and {Marchesini}, D. and {Mascia}, S. and {Mercurio}, A. and {Merlin}, E. and {Nanayakkara}, T. and {Nonino}, M. and {Paris}, D. and {Poggianti}, B. and {Rosati}, P. and {Santini}, P. and {Scarlata}, C. and {Shipley}, H.~V. and {Strait}, V. and {Trenti}, M. and {Tubthong}, C. and {Vanzella}, E. and {Vulcani}, B. and {Yang}, L.},
        title = "{The GLASS-JWST Early Release Science Program. I. Survey Design and Release Plans}",
      journal = {\apj},
         year = 2022,
        month = aug,
       volume = {935},
       number = {2},
          eid = {110},
        pages = {110},
          doi = {10.3847/1538-4357/ac8158},
archivePrefix = {arXiv},
       eprint = {2206.07978},
 primaryClass = {astro-ph.GA},
       adsurl = {https://ui.adsabs.harvard.edu/abs/2022ApJ...935..110T}
}

@ARTICLE{Finkelstein2024,
       author = {{Finkelstein}, Steven L. and {Leung}, Gene C.~K. and {Bagley}, Micaela B. and {Dickinson}, Mark and {Ferguson}, Henry C. and {Papovich}, Casey and {Akins}, Hollis B. and {Arrabal Haro}, Pablo and {Dav{\'e}}, Romeel and {Dekel}, Avishai and {Kartaltepe}, Jeyhan S. and {Kocevski}, Dale D. and {Koekemoer}, Anton M. and {Pirzkal}, Nor and {Somerville}, Rachel S. and {Yung}, L.~Y. Aaron and {Amor{\'\i}n}, Ricardo O. and {Backhaus}, Bren E. and {Behroozi}, Peter and {Bisigello}, Laura and {Bromm}, Volker and {Casey}, Caitlin M. and {Ch{\'a}vez Ortiz}, {\'O}scar A. and {Cheng}, Yingjie and {Chworowsky}, Katherine and {Cleri}, Nikko J. and {Cooper}, M.~C. and {Davis}, Kelcey and {de la Vega}, Alexander and {Elbaz}, David and {Franco}, Maximilien and {Fontana}, Adriano and {Fujimoto}, Seiji and {Giavalisco}, Mauro and {Grogin}, Norman A. and {Holwerda}, Benne W. and {Huertas-Company}, Marc and {Hirschmann}, Michaela and {Iyer}, Kartheik G. and {Jogee}, Shardha and {Jung}, Intae and {Larson}, Rebecca L. and {Lucas}, Ray A. and {Mobasher}, Bahram and {Morales}, Alexa M. and {Morley}, Caroline V. and {Mukherjee}, Sagnick and {P{\'e}rez-Gonz{\'a}lez}, Pablo G. and {Ravindranath}, Swara and {Rodighiero}, Giulia and {Rowland}, Melanie J. and {Tacchella}, Sandro and {Taylor}, Anthony J. and {Trump}, Jonathan R. and {Wilkins}, Stephen M.},
        title = "{The Complete CEERS Early Universe Galaxy Sample: A Surprisingly Slow Evolution of the Space Density of Bright Galaxies at z {\ensuremath{\sim}} 8.5{\textendash}14.5}",
      journal = {\apjl},
         year = 2024,
        month = jul,
       volume = {969},
       number = {1},
          eid = {L2},
        pages = {L2},
          doi = {10.3847/2041-8213/ad4495},
archivePrefix = {arXiv},
       eprint = {2311.04279},
 primaryClass = {astro-ph.GA},
       adsurl = {https://ui.adsabs.harvard.edu/abs/2024ApJ...969L...2F}
}

@ARTICLE{Bezanson2024,
       author = {{Bezanson}, Rachel and {Labbe}, Ivo and {Whitaker}, Katherine E. and {Leja}, Joel and {Price}, Sedona H. and {Franx}, Marijn and {Brammer}, Gabriel and {Marchesini}, Danilo and {Zitrin}, Adi and {Wang}, Bingjie and {Weaver}, John R. and {Furtak}, Lukas J. and {Atek}, Hakim and {Coe}, Dan and {Cutler}, Sam E. and {Dayal}, Pratika and {van Dokkum}, Pieter and {Feldmann}, Robert and {F{\"o}rster Schreiber}, Natascha M. and {Fujimoto}, Seiji and {Geha}, Marla and {Glazebrook}, Karl and {de Graaff}, Anna and {Greene}, Jenny E. and {Juneau}, St{\'e}phanie and {Kassin}, Susan and {Kriek}, Mariska and {Khullar}, Gourav and {Maseda}, Michael and {Mowla}, Lamiya A. and {Muzzin}, Adam and {Nanayakkara}, Themiya and {Nelson}, Erica J. and {Oesch}, Pascal A. and {Pacifici}, Camilla and {Pan}, Richard and {Papovich}, Casey and {Setton}, David J. and {Shapley}, Alice E. and {Smit}, Renske and {Stefanon}, Mauro and {Taylor}, Edward N. and {Williams}, Christina C.},
        title = "{The JWST UNCOVER Treasury Survey: Ultradeep NIRSpec and NIRCam Observations before the Epoch of Reionization}",
      journal = {\apj},
         year = 2024,
        month = oct,
       volume = {974},
       number = {1},
          eid = {92},
        pages = {92},
          doi = {10.3847/1538-4357/ad66cf},
archivePrefix = {arXiv},
       eprint = {2212.04026},
 primaryClass = {astro-ph.GA},
       adsurl = {https://ui.adsabs.harvard.edu/abs/2024ApJ...974...92B}
}

@ARTICLE{Furtak2023_lesning_model,
       author = {{Furtak}, Lukas J. and {Zitrin}, Adi and {Weaver}, John R. and {Atek}, Hakim and {Bezanson}, Rachel and {Labb{\'e}}, Ivo and {Whitaker}, Katherine E. and {Leja}, Joel and {Price}, Sedona H. and {Brammer}, Gabriel B. and {Wang}, Bingjie and {Marchesini}, Danilo and {Pan}, Richard and {Dayal}, Pratika and {van Dokkum}, Pieter and {Feldmann}, Robert and {Fujimoto}, Seiji and {Franx}, Marijn and {Khullar}, Gourav and {Nelson}, Erica J. and {Mowla}, Lamiya A.},
        title = "{UNCOVERing the extended strong lensing structures of Abell 2744 with the deepest JWST imaging}",
      journal = {\mnras},
         year = 2023,
        month = aug,
       volume = {523},
       number = {3},
        pages = {4568-4582},
          doi = {10.1093/mnras/stad1627},
archivePrefix = {arXiv},
       eprint = {2212.04381},
 primaryClass = {astro-ph.GA},
       adsurl = {https://ui.adsabs.harvard.edu/abs/2023MNRAS.523.4568F}
}

@ARTICLE{Valentino2025,
       author = {{Valentino}, F. and {Heintz}, K.~E. and {Brammer}, G. and {Ito}, K. and {Kokorev}, V. and {Whitaker}, K.~E. and {Gallazzi}, A. and {de Graaff}, A. and {Weibel}, A. and {Frye}, B.~L. and {Kamieneski}, P.~S. and {Jin}, S. and {Ceverino}, D. and {Faisst}, A. and {Farcy}, M. and {Fujimoto}, S. and {Gillman}, S. and {Gottumukkala}, R. and {Hamadouche}, M. and {Harrington}, K.~C. and {Hirschmann}, M. and {Jespersen}, C.~K. and {Kakimoto}, T. and {Kubo}, M. and {Lagos}, C. d. P. and {Lee}, M. and {Magdis}, G.~E. and {Man}, A.~W.~S. and {Onodera}, M. and {Rizzo}, F. and {Shimakawa}, R. and {Setton}, D.~J. and {Tanaka}, M. and {Toft}, S. and {Wu}, P.-F. and {Zhu}, P.},
        title = "{Gas outflows in two recently quenched galaxies at z = 4 and 7}",
      journal = {\aap},
         year = 2025,
        month = jul,
       volume = {699},
          eid = {A358},
        pages = {A358},
          doi = {10.1051/0004-6361/202553908},
archivePrefix = {arXiv},
       eprint = {2503.01990},
 primaryClass = {astro-ph.GA},
       adsurl = {https://ui.adsabs.harvard.edu/abs/2025A&A...699A.358V}
}

@ARTICLE{Carniani2024_z14,
       author = {{Carniani}, Stefano and {Hainline}, Kevin and {D'Eugenio}, Francesco and {Eisenstein}, Daniel J. and {Jakobsen}, Peter and {Witstok}, Joris and {Johnson}, Benjamin D. and {Chevallard}, Jacopo and {Maiolino}, Roberto and {Helton}, Jakob M. and {Willott}, Chris and {Robertson}, Brant and {Alberts}, Stacey and {Arribas}, Santiago and {Baker}, William M. and {Bhatawdekar}, Rachana and {Boyett}, Kristan and {Bunker}, Andrew J. and {Cameron}, Alex J. and {Cargile}, Phillip A. and {Charlot}, St{\'e}phane and {Curti}, Mirko and {Curtis-Lake}, Emma and {Egami}, Eiichi and {Giardino}, Giovanna and {Isaak}, Kate and {Ji}, Zhiyuan and {Jones}, Gareth C. and {Kumari}, Nimisha and {Maseda}, Michael V. and {Parlanti}, Eleonora and {P{\'e}rez-Gonz{\'a}lez}, Pablo G. and {Rawle}, Tim and {Rieke}, George and {Rieke}, Marcia and {Del Pino}, Bruno Rodr{\'\i}guez and {Saxena}, Aayush and {Scholtz}, Jan and {Smit}, Renske and {Sun}, Fengwu and {Tacchella}, Sandro and {{\"U}bler}, Hannah and {Venturi}, Giacomo and {Williams}, Christina C. and {Willmer}, Christopher N.~A.},
        title = "{Spectroscopic confirmation of two luminous galaxies at a redshift of 14}",
      journal = {\nat},
         year = 2024,
        month = sep,
       volume = {633},
       number = {8029},
        pages = {318-322},
          doi = {10.1038/s41586-024-07860-9},
archivePrefix = {arXiv},
       eprint = {2405.18485},
 primaryClass = {astro-ph.GA},
       adsurl = {https://ui.adsabs.harvard.edu/abs/2024Natur.633..318C}
}

@ARTICLE{Miralda-Escude1998,
       author = {{Miralda-Escud{\'e}}, Jordi},
        title = "{Reionization of the Intergalactic Medium and the Damping Wing of the Gunn-Peterson Trough}",
      journal = {\apj},
         year = 1998,
        month = jul,
       volume = {501},
       number = {1},
        pages = {15-22},
          doi = {10.1086/305799},
archivePrefix = {arXiv},
       eprint = {astro-ph/9708253},
 primaryClass = {astro-ph},
       adsurl = {https://ui.adsabs.harvard.edu/abs/1998ApJ...501...15M}
}

@ARTICLE{Mason2018_transmission,
       author = {{Mason}, Charlotte A. and {Treu}, Tommaso and {de Barros}, Stephane and {Dijkstra}, Mark and {Fontana}, Adriano and {Mesinger}, Andrei and {Pentericci}, Laura and {Trenti}, Michele and {Vanzella}, Eros},
        title = "{Beacons into the Cosmic Dark Ages: Boosted Transmission of Ly{\ensuremath{\alpha}} from UV Bright Galaxies at z {\ensuremath{\gtrsim}} 7}",
      journal = {\apjl},
         year = 2018,
        month = apr,
       volume = {857},
       number = {2},
          eid = {L11},
        pages = {L11},
          doi = {10.3847/2041-8213/aabbab},
archivePrefix = {arXiv},
       eprint = {1801.01891},
 primaryClass = {astro-ph.CO},
       adsurl = {https://ui.adsabs.harvard.edu/abs/2018ApJ...857L..11M}
}

@ARTICLE{Mason2018,
       author = {{Mason}, Charlotte A. and {Treu}, Tommaso and {Dijkstra}, Mark and {Mesinger}, Andrei and {Trenti}, Michele and {Pentericci}, Laura and {de Barros}, Stephane and {Vanzella}, Eros},
        title = "{The Universe Is Reionizing at z {\ensuremath{\sim}} 7: Bayesian Inference of the IGM Neutral Fraction Using Ly{\ensuremath{\alpha}} Emission from Galaxies}",
      journal = {\apj},
         year = 2018,
        month = mar,
       volume = {856},
       number = {1},
          eid = {2},
        pages = {2},
          doi = {10.3847/1538-4357/aab0a7},
archivePrefix = {arXiv},
       eprint = {1709.05356},
 primaryClass = {astro-ph.CO},
       adsurl = {https://ui.adsabs.harvard.edu/abs/2018ApJ...856....2M}
}

@ARTICLE{PlanckCollaboration2020,
       author = {{Planck Collaboration} and {Aghanim}, N. and {Akrami}, Y. and {Ashdown}, M. and {Aumont}, J. and {Baccigalupi}, C. and {Ballardini}, M. and {Banday}, A.~J. and {Barreiro}, R.~B. and {Bartolo}, N. and {Basak}, S. and {Battye}, R. and {Benabed}, K. and {Bernard}, J. -P. and {Bersanelli}, M. and {Bielewicz}, P. and {Bock}, J.~J. and {Bond}, J.~R. and {Borrill}, J. and {Bouchet}, F.~R. and {Boulanger}, F. and {Bucher}, M. and {Burigana}, C. and {Butler}, R.~C. and {Calabrese}, E. and {Cardoso}, J. -F. and {Carron}, J. and {Challinor}, A. and {Chiang}, H.~C. and {Chluba}, J. and {Colombo}, L.~P.~L. and {Combet}, C. and {Contreras}, D. and {Crill}, B.~P. and {Cuttaia}, F. and {de Bernardis}, P. and {de Zotti}, G. and {Delabrouille}, J. and {Delouis}, J. -M. and {Di Valentino}, E. and {Diego}, J.~M. and {Dor{\'e}}, O. and {Douspis}, M. and {Ducout}, A. and {Dupac}, X. and {Dusini}, S. and {Efstathiou}, G. and {Elsner}, F. and {En{\ss}lin}, T.~A. and {Eriksen}, H.~K. and {Fantaye}, Y. and {Farhang}, M. and {Fergusson}, J. and {Fernandez-Cobos}, R. and {Finelli}, F. and {Forastieri}, F. and {Frailis}, M. and {Fraisse}, A.~A. and {Franceschi}, E. and {Frolov}, A. and {Galeotta}, S. and {Galli}, S. and {Ganga}, K. and {G{\'e}nova-Santos}, R.~T. and {Gerbino}, M. and {Ghosh}, T. and {Gonz{\'a}lez-Nuevo}, J. and {G{\'o}rski}, K.~M. and {Gratton}, S. and {Gruppuso}, A. and {Gudmundsson}, J.~E. and {Hamann}, J. and {Handley}, W. and {Hansen}, F.~K. and {Herranz}, D. and {Hildebrandt}, S.~R. and {Hivon}, E. and {Huang}, Z. and {Jaffe}, A.~H. and {Jones}, W.~C. and {Karakci}, A. and {Keih{\"a}nen}, E. and {Keskitalo}, R. and {Kiiveri}, K. and {Kim}, J. and {Kisner}, T.~S. and {Knox}, L. and {Krachmalnicoff}, N. and {Kunz}, M. and {Kurki-Suonio}, H. and {Lagache}, G. and {Lamarre}, J. -M. and {Lasenby}, A. and {Lattanzi}, M. and {Lawrence}, C.~R. and {Le Jeune}, M. and {Lemos}, P. and {Lesgourgues}, J. and {Levrier}, F. and {Lewis}, A. and {Liguori}, M. and {Lilje}, P.~B. and {Lilley}, M. and {Lindholm}, V. and {L{\'o}pez-Caniego}, M. and {Lubin}, P.~M. and {Ma}, Y. -Z. and {Mac{\'\i}as-P{\'e}rez}, J.~F. and {Maggio}, G. and {Maino}, D. and {Mandolesi}, N. and {Mangilli}, A. and {Marcos-Caballero}, A. and {Maris}, M. and {Martin}, P.~G. and {Martinelli}, M. and {Mart{\'\i}nez-Gonz{\'a}lez}, E. and {Matarrese}, S. and {Mauri}, N. and {McEwen}, J.~D. and {Meinhold}, P.~R. and {Melchiorri}, A. and {Mennella}, A. and {Migliaccio}, M. and {Millea}, M. and {Mitra}, S. and {Miville-Desch{\^e}nes}, M. -A. and {Molinari}, D. and {Montier}, L. and {Morgante}, G. and {Moss}, A. and {Natoli}, P. and {N{\o}rgaard-Nielsen}, H.~U. and {Pagano}, L. and {Paoletti}, D. and {Partridge}, B. and {Patanchon}, G. and {Peiris}, H.~V. and {Perrotta}, F. and {Pettorino}, V. and {Piacentini}, F. and {Polastri}, L. and {Polenta}, G. and {Puget}, J. -L. and {Rachen}, J.~P. and {Reinecke}, M. and {Remazeilles}, M. and {Renzi}, A. and {Rocha}, G. and {Rosset}, C. and {Roudier}, G. and {Rubi{\~n}o-Mart{\'\i}n}, J.~A. and {Ruiz-Granados}, B. and {Salvati}, L. and {Sandri}, M. and {Savelainen}, M. and {Scott}, D. and {Shellard}, E.~P.~S. and {Sirignano}, C. and {Sirri}, G. and {Spencer}, L.~D. and {Sunyaev}, R. and {Suur-Uski}, A. -S. and {Tauber}, J.~A. and {Tavagnacco}, D. and {Tenti}, M. and {Toffolatti}, L. and {Tomasi}, M. and {Trombetti}, T. and {Valenziano}, L. and {Valiviita}, J. and {Van Tent}, B. and {Vibert}, L. and {Vielva}, P. and {Villa}, F. and {Vittorio}, N. and {Wandelt}, B.~D. and {Wehus}, I.~K. and {White}, M. and {White}, S.~D.~M. and {Zacchei}, A. and {Zonca}, A.},
        title = "{Planck 2018 results. VI. Cosmological parameters}",
      journal = {\aap},
         year = 2020,
        month = sep,
       volume = {641},
          eid = {A6},
        pages = {A6},
          doi = {10.1051/0004-6361/201833910},
archivePrefix = {arXiv},
       eprint = {1807.06209},
 primaryClass = {astro-ph.CO},
       adsurl = {https://ui.adsabs.harvard.edu/abs/2020A&A...641A...6P}
}

@ARTICLE{Lu2024,
       author = {{Lu}, Ting-Yi and {Mason}, Charlotte A. and {Hutter}, Anne and {Mesinger}, Andrei and {Qin}, Yuxiang and {Stark}, Daniel P. and {Endsley}, Ryan},
        title = "{The reionizing bubble size distribution around galaxies}",
      journal = {\mnras},
         year = 2024,
        month = mar,
       volume = {528},
       number = {3},
        pages = {4872-4890},
          doi = {10.1093/mnras/stae266},
archivePrefix = {arXiv},
       eprint = {2304.11192},
 primaryClass = {astro-ph.GA},
       adsurl = {https://ui.adsabs.harvard.edu/abs/2024MNRAS.528.4872L}
}

@ARTICLE{Tang2024_nirspec,
       author = {{Tang}, Mengtao and {Stark}, Daniel P. and {Topping}, Michael W. and {Mason}, Charlotte and {Ellis}, Richard S.},
        title = "{JWST/NIRSpec Observations of Lyman {\ensuremath{\alpha}} Emission in Star-forming Galaxies at 6.5 {\ensuremath{\lesssim}} z {\ensuremath{\lesssim}} 13}",
      journal = {\apj},
         year = 2024,
        month = nov,
       volume = {975},
       number = {2},
          eid = {208},
        pages = {208},
          doi = {10.3847/1538-4357/ad7eb7},
archivePrefix = {arXiv},
       eprint = {2408.01507},
 primaryClass = {astro-ph.GA},
       adsurl = {https://ui.adsabs.harvard.edu/abs/2024ApJ...975..208T}
}

@ARTICLE{Tang2024_z56,
       author = {{Tang}, Mengtao and {Stark}, Daniel P. and {Ellis}, Richard S. and {Sun}, Fengwu and {Topping}, Michael and {Robertson}, Brant and {Tacchella}, Sandro and {Arribas}, Santiago and {Baker}, William M. and {Bhatawdekar}, Rachana and {Boyett}, Kristan and {Bunker}, Andrew J. and {Charlot}, St{\'e}phane and {Chen}, Zuyi and {Chevallard}, Jacopo and {Jones}, Gareth C. and {Kumari}, Nimisha and {Lyu}, Jianwei and {Maiolino}, Roberto and {Maseda}, Michael V. and {Saxena}, Aayush and {Whitler}, Lily and {Williams}, Christina C. and {Willott}, Chris and {Witstok}, Joris},
        title = "{Ly{\ensuremath{\alpha}} emission in galaxies at z ≃ 5-6: new insight from JWST into the statistical distributions of Ly{\ensuremath{\alpha}} properties at the end of reionization}",
      journal = {\mnras},
         year = 2024,
        month = jun,
       volume = {531},
       number = {2},
        pages = {2701-2730},
          doi = {10.1093/mnras/stae1338},
archivePrefix = {arXiv},
       eprint = {2402.06070},
 primaryClass = {astro-ph.GA},
       adsurl = {https://ui.adsabs.harvard.edu/abs/2024MNRAS.531.2701T}
}

@ARTICLE{Jones2025,
       author = {{Jones}, Gareth C. and {Bunker}, Andrew J. and {Saxena}, Aayush and {Arribas}, Santiago and {Bhatawdekar}, Rachana and {Boyett}, Kristan and {Cameron}, Alex J. and {Carniani}, Stefano and {Charlot}, Stephane and {Curtis-Lake}, Emma and {Hainline}, Kevin and {Johnson}, Benjamin D. and {Kumari}, Nimisha and {Maseda}, Michael V. and {Rix}, Hans-Walter and {Robertson}, Brant E. and {Tacchella}, Sandro and {{\"U}bler}, Hannah and {Williams}, Christina C. and {Willott}, Chris and {Witstok}, Joris and {Zhu}, Yongda},
        title = "{JADES: measuring reionization properties using Lyman-alpha emission}",
      journal = {\mnras},
         year = 2025,
        month = jan,
       volume = {536},
       number = {3},
        pages = {2355-2380},
          doi = {10.1093/mnras/stae2670},
archivePrefix = {arXiv},
       eprint = {2409.06405},
 primaryClass = {astro-ph.GA},
       adsurl = {https://ui.adsabs.harvard.edu/abs/2025MNRAS.536.2355J}
}

@ARTICLE{Bunker2023,
       author = {{Bunker}, Andrew J. and {Saxena}, Aayush and {Cameron}, Alex J. and {Willott}, Chris J. and {Curtis-Lake}, Emma and {Jakobsen}, Peter and {Carniani}, Stefano and {Smit}, Renske and {Maiolino}, Roberto and {Witstok}, Joris and {Curti}, Mirko and {D'Eugenio}, Francesco and {Jones}, Gareth C. and {Ferruit}, Pierre and {Arribas}, Santiago and {Charlot}, Stephane and {Chevallard}, Jacopo and {Giardino}, Giovanna and {de Graaff}, Anna and {Looser}, Tobias J. and {L{\"u}tzgendorf}, Nora and {Maseda}, Michael V. and {Rawle}, Tim and {Rix}, Hans-Walter and {Del Pino}, Bruno Rodr{\'\i}guez and {Alberts}, Stacey and {Egami}, Eiichi and {Eisenstein}, Daniel J. and {Endsley}, Ryan and {Hainline}, Kevin and {Hausen}, Ryan and {Johnson}, Benjamin D. and {Rieke}, George and {Rieke}, Marcia and {Robertson}, Brant E. and {Shivaei}, Irene and {Stark}, Daniel P. and {Sun}, Fengwu and {Tacchella}, Sandro and {Tang}, Mengtao and {Williams}, Christina C. and {Willmer}, Christopher N.~A. and {Baker}, William M. and {Baum}, Stefi and {Bhatawdekar}, Rachana and {Bowler}, Rebecca and {Boyett}, Kristan and {Chen}, Zuyi and {Circosta}, Chiara and {Helton}, Jakob M. and {Ji}, Zhiyuan and {Kumari}, Nimisha and {Lyu}, Jianwei and {Nelson}, Erica and {Parlanti}, Eleonora and {Perna}, Michele and {Sandles}, Lester and {Scholtz}, Jan and {Suess}, Katherine A. and {Topping}, Michael W. and {{\"U}bler}, Hannah and {Wallace}, Imaan E.~B. and {Whitler}, Lily},
        title = "{JADES NIRSpec Spectroscopy of GN-z11: Lyman-{\ensuremath{\alpha}} emission and possible enhanced nitrogen abundance in a z = 10.60 luminous galaxy}",
      journal = {\aap},
         year = 2023,
        month = sep,
       volume = {677},
          eid = {A88},
        pages = {A88},
          doi = {10.1051/0004-6361/202346159},
archivePrefix = {arXiv},
       eprint = {2302.07256},
 primaryClass = {astro-ph.GA},
       adsurl = {https://ui.adsabs.harvard.edu/abs/2023A&A...677A..88B}
}

@ARTICLE{Topping2024,
       author = {{Topping}, Michael W. and {Stark}, Daniel P. and {Endsley}, Ryan and {Whitler}, Lily and {Hainline}, Kevin and {Johnson}, Benjamin D. and {Robertson}, Brant and {Tacchella}, Sandro and {Chen}, Zuyi and {Alberts}, Stacey and {Baker}, William M. and {Bunker}, Andrew J. and {Carniani}, Stefano and {Charlot}, Stephane and {Chevallard}, Jacopo and {Curtis-Lake}, Emma and {DeCoursey}, Christa and {Egami}, Eiichi and {Eisenstein}, Daniel J. and {Ji}, Zhiyuan and {Maiolino}, Roberto and {Williams}, Christina C. and {Willmer}, Christopher N.~A. and {Willott}, Chris and {Witstok}, Joris},
        title = "{The UV continuum slopes of early star-forming galaxies in JADES}",
      journal = {\mnras},
         year = 2024,
        month = apr,
       volume = {529},
       number = {4},
        pages = {4087-4103},
          doi = {10.1093/mnras/stae800},
archivePrefix = {arXiv},
       eprint = {2307.08835},
 primaryClass = {astro-ph.GA},
       adsurl = {https://ui.adsabs.harvard.edu/abs/2024MNRAS.529.4087T}
}

@ARTICLE{Jones2013,
       author = {{Jones}, Tucker A. and {Ellis}, Richard S. and {Schenker}, Matthew A. and {Stark}, Daniel P.},
        title = "{Keck Spectroscopy of Gravitationally Lensed z \raisebox{-0.5ex}\textasciitilde= 4 Galaxies: Improved Constraints on the Escape Fraction of Ionizing Photons}",
      journal = {\apj},
         year = 2013,
        month = dec,
       volume = {779},
       number = {1},
          eid = {52},
        pages = {52},
          doi = {10.1088/0004-637X/779/1/52},
archivePrefix = {arXiv},
       eprint = {1304.7015},
 primaryClass = {astro-ph.CO},
       adsurl = {https://ui.adsabs.harvard.edu/abs/2013ApJ...779...52J}
}

@ARTICLE{Heintz2025_PRIMAL,
       author = {{Heintz}, K.~E. and {Brammer}, G.~B. and {Watson}, D. and {Oesch}, P.~A. and {Keating}, L.~C. and {Hayes}, M.~J. and {Abdurro'uf} and {Arellano-C{\'o}rdova}, K.~Z. and {Carnall}, A.~C. and {Christiansen}, C.~R. and {Cullen}, F. and {Dav{\'e}}, R. and {Dayal}, P. and {Ferrara}, A. and {Finlator}, K. and {Fynbo}, J.~P.~U. and {Flury}, S.~R. and {Gelli}, V. and {Gillman}, S. and {Gottumukkala}, R. and {Gould}, K. and {Greve}, T.~R. and {Hardin}, S.~E. and {Hsiao}, T.~Y. -Y. and {Hutter}, A. and {Jakobsson}, P. and {Killi}, M. and {Khosravaninezhad}, N. and {Laursen}, P. and {Lee}, M.~M. and {Magdis}, G.~E. and {Matthee}, J. and {Naidu}, R.~P. and {Narayanan}, D. and {Pollock}, C. and {Prescott}, M.~K.~M. and {Rusakov}, V. and {Shuntov}, M. and {Sneppen}, A. and {Smit}, R. and {Tanvir}, N.~R. and {Terp}, C. and {Toft}, S. and {Valentino}, F. and {Vijayan}, A.~P. and {Weaver}, J.~R. and {Wise}, J.~H. and {Witstok}, J.},
        title = "{The JWST-PRIMAL archival survey: A JWST/NIRSpec reference sample for the physical properties and Lyman-{\ensuremath{\alpha}} absorption and emission of {\ensuremath{\sim}}600 galaxies at z = 5.0 ‑ 13.4}",
      journal = {\aap},
         year = 2025,
        month = jan,
       volume = {693},
          eid = {A60},
        pages = {A60},
          doi = {10.1051/0004-6361/202450243},
archivePrefix = {arXiv},
       eprint = {2404.02211},
 primaryClass = {astro-ph.GA},
       adsurl = {https://ui.adsabs.harvard.edu/abs/2025A&A...693A..60H}
}

@ARTICLE{Endsley2024_jades,
       author = {{Endsley}, Ryan and {Stark}, Daniel P. and {Whitler}, Lily and {Topping}, Michael W. and {Johnson}, Benjamin D. and {Robertson}, Brant and {Tacchella}, Sandro and {Alberts}, Stacey and {Baker}, William M. and {Bhatawdekar}, Rachana and {Boyett}, Kristan and {Bunker}, Andrew J. and {Cameron}, Alex J. and {Carniani}, Stefano and {Charlot}, Stephane and {Chen}, Zuyi and {Chevallard}, Jacopo and {Curtis-Lake}, Emma and {Danhaive}, A. Lola and {Egami}, Eiichi and {Eisenstein}, Daniel J. and {Hainline}, Kevin and {Helton}, Jakob M. and {Ji}, Zhiyuan and {Looser}, Tobias J. and {Maiolino}, Roberto and {Nelson}, Erica and {Pusk{\'a}s}, D{\'a}vid and {Rieke}, George and {Rieke}, Marcia and {Rix}, Hans-Walter and {Sandles}, Lester and {Saxena}, Aayush and {Simmonds}, Charlotte and {Smit}, Renske and {Sun}, Fengwu and {Williams}, Christina C. and {Willmer}, Christopher N.~A. and {Willott}, Chris and {Witstok}, Joris},
        title = "{The star-forming and ionizing properties of dwarf z 6-9 galaxies in JADES: insights on bursty star formation and ionized bubble growth}",
      journal = {\mnras},
         year = 2024,
        month = sep,
       volume = {533},
       number = {1},
        pages = {1111-1142},
          doi = {10.1093/mnras/stae1857},
archivePrefix = {arXiv},
       eprint = {2306.05295},
 primaryClass = {astro-ph.GA},
       adsurl = {https://ui.adsabs.harvard.edu/abs/2024MNRAS.533.1111E}
}

@ARTICLE{Cameron2023,
       author = {{Cameron}, Alex J. and {Saxena}, Aayush and {Bunker}, Andrew J. and {D'Eugenio}, Francesco and {Carniani}, Stefano and {Maiolino}, Roberto and {Curtis-Lake}, Emma and {Ferruit}, Pierre and {Jakobsen}, Peter and {Arribas}, Santiago and {Bonaventura}, Nina and {Charlot}, Stephane and {Chevallard}, Jacopo and {Curti}, Mirko and {Looser}, Tobias J. and {Maseda}, Michael V. and {Rawle}, Tim and {Rodr{\'\i}guez Del Pino}, Bruno and {Smit}, Renske and {{\"U}bler}, Hannah and {Willott}, Chris and {Witstok}, Joris and {Egami}, Eiichi and {Eisenstein}, Daniel J. and {Johnson}, Benjamin D. and {Hainline}, Kevin and {Rieke}, Marcia and {Robertson}, Brant E. and {Stark}, Daniel P. and {Tacchella}, Sandro and {Williams}, Christina C. and {Willmer}, Christopher N.~A. and {Bhatawdekar}, Rachana and {Bowler}, Rebecca and {Boyett}, Kristan and {Circosta}, Chiara and {Helton}, Jakob M. and {Jones}, Gareth C. and {Kumari}, Nimisha and {Ji}, Zhiyuan and {Nelson}, Erica and {Parlanti}, Eleonora and {Sandles}, Lester and {Scholtz}, Jan and {Sun}, Fengwu},
        title = "{JADES: Probing interstellar medium conditions at z {\ensuremath{\sim}} 5.5-9.5 with ultra-deep JWST/NIRSpec spectroscopy}",
      journal = {\aap},
         year = 2023,
        month = sep,
       volume = {677},
          eid = {A115},
        pages = {A115},
          doi = {10.1051/0004-6361/202346107},
archivePrefix = {arXiv},
       eprint = {2302.04298},
 primaryClass = {astro-ph.GA},
       adsurl = {https://ui.adsabs.harvard.edu/abs/2023A&A...677A.115C}
}

@ARTICLE{Castellano2024,
       author = {{Castellano}, Marco and {Napolitano}, Lorenzo and {Fontana}, Adriano and {Roberts-Borsani}, Guido and {Treu}, Tommaso and {Vanzella}, Eros and {Zavala}, Jorge A. and {Arrabal Haro}, Pablo and {Calabr{\`o}}, Antonello and {Llerena}, Mario and {Mascia}, Sara and {Merlin}, Emiliano and {Paris}, Diego and {Pentericci}, Laura and {Santini}, Paola and {Bakx}, Tom J.~L.~C. and {Bergamini}, Pietro and {Cupani}, Guido and {Dickinson}, Mark and {Filippenko}, Alexei V. and {Glazebrook}, Karl and {Grillo}, Claudio and {Kelly}, Patrick L. and {Malkan}, Matthew A. and {Mason}, Charlotte A. and {Morishita}, Takahiro and {Nanayakkara}, Themiya and {Rosati}, Piero and {Sani}, Eleonora and {Wang}, Xin and {Yoon}, Ilsang},
        title = "{JWST NIRSpec Spectroscopy of the Remarkable Bright Galaxy GHZ2/GLASS-z12 at Redshift 12.34}",
      journal = {\apj},
         year = 2024,
        month = sep,
       volume = {972},
       number = {2},
          eid = {143},
        pages = {143},
          doi = {10.3847/1538-4357/ad5f88},
archivePrefix = {arXiv},
       eprint = {2403.10238},
 primaryClass = {astro-ph.GA},
       adsurl = {https://ui.adsabs.harvard.edu/abs/2024ApJ...972..143C}
}

@ARTICLE{Saxena2024_slope,
       author = {{Saxena}, Aayush and {Cameron}, Alex J. and {Katz}, Harley and {Bunker}, Andrew J. and {Chevallard}, Jacopo and {D'Eugenio}, Francesco and {Arribas}, Santiago and {Bhatawdekar}, Rachana and {Boyett}, Kristan and {Cargile}, Phillip A. and {Carniani}, Stefano and {Charlot}, Stephane and {Curti}, Mirko and {Curtis-Lake}, Emma and {Hainline}, Kevin and {Ji}, Zhiyuan and {Johnson}, Benjamin D. and {Jones}, Gareth C. and {Kumari}, Nimisha and {Laseter}, Isaac and {Maseda}, Michael V. and {Robertson}, Brant and {Simmonds}, Charlotte and {Tacchella}, Sandro and {Ubler}, Hannah and {Williams}, Christina C. and {Willott}, Chris and {Witstok}, Joris and {Zhu}, Yongda},
        title = "{Hitting the slopes: A spectroscopic view of UV continuum slopes of galaxies reveals a reddening at z > 9.5}",
      journal = {arXiv e-prints},
         year = 2024,
        month = nov,
          eid = {arXiv:2411.14532},
        pages = {arXiv:2411.14532},
          doi = {10.48550/arXiv.2411.14532},
archivePrefix = {arXiv},
       eprint = {2411.14532},
 primaryClass = {astro-ph.GA},
       adsurl = {https://ui.adsabs.harvard.edu/abs/2024arXiv241114532S}
}

@ARTICLE{Tang2025,
       author = {{Tang}, Mengtao and {Stark}, Daniel P. and {Mason}, Charlotte A. and {Gelli}, Viola and {Chen}, Zuyi and {Topping}, Michael W.},
        title = "{The JWST Spectroscopic Properties of Galaxies at $z=9-14$}",
      journal = {arXiv e-prints},
         year = 2025,
        month = jul,
          eid = {arXiv:2507.08245},
        pages = {arXiv:2507.08245},
          doi = {10.48550/arXiv.2507.08245},
archivePrefix = {arXiv},
       eprint = {2507.08245},
 primaryClass = {astro-ph.GA},
       adsurl = {https://ui.adsabs.harvard.edu/abs/2025arXiv250708245T}
}

@ARTICLE{Roberts-Borsani2025,
       author = {{Roberts-Borsani}, Guido and {Oesch}, Pascal and {Ellis}, Richard and {Weibel}, Andrea and {Giovinazzo}, Emma and {Bouwens}, Rychard and {Dayal}, Pratika and {Fontana}, Adriano and {Heintz}, Kasper and {Matthee}, Jorryt and {Meyer}, Romain and {Pentericci}, Laura and {Shapley}, Alice and {Tacchella}, Sandro and {Treu}, Tommaso and {Walter}, Fabian and {Atek}, Hakim and {Bose}, Sownak and {Castellano}, Marco and {Fudamoto}, Yoshinobu and {Morishita}, Takahiro and {Naidu}, Rohan and {Sanders}, Ryan and {van der Wel}, Arjen},
        title = "{JWST Spectroscopic Insights Into the Diversity of Galaxies in the First 500 Myr: Short-Lived Snapshots Along a Common Evolutionary Pathway}",
      journal = {arXiv e-prints},
         year = 2025,
        month = aug,
          eid = {arXiv:2508.21708},
        pages = {arXiv:2508.21708},
          doi = {10.48550/arXiv.2508.21708},
archivePrefix = {arXiv},
       eprint = {2508.21708},
 primaryClass = {astro-ph.GA},
       adsurl = {https://ui.adsabs.harvard.edu/abs/2025arXiv250821708R}
}

@ARTICLE{Shapley2025_z7,
       author = {{Shapley}, Alice E. and {Sanders}, Ryan L. and {Topping}, Michael W. and {Reddy}, Naveen A. and {Pahl}, Anthony J. and {Oesch}, Pascal A. and {Berg}, Danielle A. and {Bouwens}, Rychard J. and {Brammer}, Gabriel and {Carnall}, Adam C. and {Cullen}, Fergus and {Dav{\'e}}, Romeel and {Dunlop}, James S. and {Ellis}, Richard S. and {F{\"o}rster Schreiber}, N.~M. and {Furlanetto}, Steven R. and {Glazebrook}, Karl and {Illingworth}, Garth D. and {Jones}, Tucker and {Kriek}, Mariska and {McLeod}, Derek J. and {McLure}, Ross J. and {Narayanan}, Desika and {Pettini}, Max and {Schaerer}, Daniel and {Stark}, Daniel P. and {Steidel}, Charles C. and {Tang}, Mengtao and {Clarke}, Leonardo and {Donnan}, Callum T. and {Kehoe}, Emily},
        title = "{The AURORA Survey: An Extraordinarily Mature, Star-forming Galaxy at z {\ensuremath{\sim}} 7}",
      journal = {\apj},
         year = 2025,
        month = mar,
       volume = {981},
       number = {2},
          eid = {167},
        pages = {167},
          doi = {10.3847/1538-4357/adaf98},
archivePrefix = {arXiv},
       eprint = {2410.00110},
 primaryClass = {astro-ph.GA},
       adsurl = {https://ui.adsabs.harvard.edu/abs/2025ApJ...981..167S}
}

@ARTICLE{Tang2019,
       author = {{Tang}, Mengtao and {Stark}, Daniel P. and {Chevallard}, Jacopo and {Charlot}, St{\'e}phane},
        title = "{MMT/MMIRS spectroscopy of z = 1.3 - 2.4 extreme [O III] emitters: implications for galaxies in the reionization era}",
      journal = {\mnras},
         year = 2019,
        month = oct,
       volume = {489},
       number = {2},
        pages = {2572-2594},
          doi = {10.1093/mnras/stz2236},
archivePrefix = {arXiv},
       eprint = {1809.09637},
 primaryClass = {astro-ph.GA},
       adsurl = {https://ui.adsabs.harvard.edu/abs/2019MNRAS.489.2572T}
}

@ARTICLE{Reddy2016,
       author = {{Reddy}, Naveen A. and {Steidel}, Charles C. and {Pettini}, Max and {Bogosavljevi{\'c}}, Milan and {Shapley}, Alice E.},
        title = "{The Connection Between Reddening, Gas Covering Fraction, and the Escape of Ionizing Radiation at High Redshift}",
      journal = {\apj},
         year = 2016,
        month = sep,
       volume = {828},
       number = {2},
          eid = {108},
        pages = {108},
          doi = {10.3847/0004-637X/828/2/108},
archivePrefix = {arXiv},
       eprint = {1606.03452},
 primaryClass = {astro-ph.GA},
       adsurl = {https://ui.adsabs.harvard.edu/abs/2016ApJ...828..108R}
}

@ARTICLE{Du2018,
       author = {{Du}, Xinnan and {Shapley}, Alice E. and {Reddy}, Naveen A. and {Jones}, Tucker and {Stark}, Daniel P. and {Steidel}, Charles C. and {Strom}, Allison L. and {Rudie}, Gwen C. and {Erb}, Dawn K. and {Ellis}, Richard S. and {Pettini}, Max},
        title = "{The Redshift Evolution of Rest-UV Spectroscopic Properties in Lyman-break Galaxies at z {\ensuremath{\sim}} 2-4}",
      journal = {\apj},
         year = 2018,
        month = jun,
       volume = {860},
       number = {1},
          eid = {75},
        pages = {75},
          doi = {10.3847/1538-4357/aabfcf},
archivePrefix = {arXiv},
       eprint = {1803.05912},
 primaryClass = {astro-ph.GA},
       adsurl = {https://ui.adsabs.harvard.edu/abs/2018ApJ...860...75D}
}

@ARTICLE{Jones2012,
       author = {{Jones}, Tucker and {Stark}, Daniel P. and {Ellis}, Richard S.},
        title = "{Keck Spectroscopy of Faint 3 < z < 7 Lyman Break Galaxies. III. The Mean Ultraviolet Spectrum at z \raisebox{-0.5ex}\textasciitilde= 4}",
      journal = {\apj},
         year = 2012,
        month = may,
       volume = {751},
       number = {1},
          eid = {51},
        pages = {51},
          doi = {10.1088/0004-637X/751/1/51},
archivePrefix = {arXiv},
       eprint = {1111.5102},
 primaryClass = {astro-ph.CO},
       adsurl = {https://ui.adsabs.harvard.edu/abs/2012ApJ...751...51J}
}

@ARTICLE{Xu2022,
       author = {{Xu}, Xinfeng and {Heckman}, Timothy and {Henry}, Alaina and {Berg}, Danielle A. and {Chisholm}, John and {James}, Bethan L. and {Martin}, Crystal L. and {Stark}, Daniel P. and {Aloisi}, Alessandra and {Amor{\'\i}n}, Ricardo O. and {Arellano-C{\'o}rdova}, Karla Z. and {Bordoloi}, Rongmon and {Charlot}, St{\'e}phane and {Chen}, Zuyi and {Hayes}, Matthew and {Mingozzi}, Matilde and {Sugahara}, Yuma and {Kewley}, Lisa J. and {Ouchi}, Masami and {Scarlata}, Claudia and {Steidel}, Charles C.},
        title = "{CLASSY III. The Properties of Starburst-driven Warm Ionized Outflows}",
      journal = {\apj},
         year = 2022,
        month = jul,
       volume = {933},
       number = {2},
          eid = {222},
        pages = {222},
          doi = {10.3847/1538-4357/ac6d56},
archivePrefix = {arXiv},
       eprint = {2204.09181},
 primaryClass = {astro-ph.GA},
       adsurl = {https://ui.adsabs.harvard.edu/abs/2022ApJ...933..222X}
}

@ARTICLE{Hu2023,
       author = {{Hu}, Weida and {Martin}, Crystal L. and {Gronke}, Max and {Gazagnes}, Simon and {Hayes}, Matthew and {Chisholm}, John and {Heckman}, Timothy and {Mingozzi}, Matilde and {Roy}, Namrata and {Senchyna}, Peter and {Xu}, Xinfeng and {Berg}, Danielle A. and {James}, Bethan L. and {Stark}, Daniel P. and {Arellano-C{\'o}rdova}, Karla Z. and {Henry}, Alaina and {Jaskot}, Anne E. and {Kumari}, Nimisha and {Parker}, Kaelee S. and {Scarlata}, Claudia and {Wofford}, Aida and {Amor{\'\i}n}, Ricardo O. and {Leonhardes-Barboza}, Naunet and {Brinchmann}, Jarle and {Carr}, Cody and {Aloisi}, Alessandra},
        title = "{CLASSY VII Ly{\ensuremath{\alpha}} Profiles: The Structure and Kinematics of Neutral Gas and Implications for LyC Escape in Reionization-era Analogs}",
      journal = {\apj},
         year = 2023,
        month = oct,
       volume = {956},
       number = {1},
          eid = {39},
        pages = {39},
          doi = {10.3847/1538-4357/aceefd},
archivePrefix = {arXiv},
       eprint = {2307.04911},
 primaryClass = {astro-ph.GA},
       adsurl = {https://ui.adsabs.harvard.edu/abs/2023ApJ...956...39H}
}

@ARTICLE{Vidal-Garcia2017,
       author = {{Vidal-Garc{\'\i}a}, A. and {Charlot}, S. and {Bruzual}, G. and {Hubeny}, I.},
        title = "{Modelling ultraviolet-line diagnostics of stars, the ionized and the neutral interstellar medium in star-forming galaxies}",
      journal = {\mnras},
         year = 2017,
        month = sep,
       volume = {470},
       number = {3},
        pages = {3532-3556},
          doi = {10.1093/mnras/stx1324},
archivePrefix = {arXiv},
       eprint = {1705.10320},
 primaryClass = {astro-ph.GA},
       adsurl = {https://ui.adsabs.harvard.edu/abs/2017MNRAS.470.3532V}
}

@ARTICLE{Rix2004,
       author = {{Rix}, Samantha A. and {Pettini}, Max and {Leitherer}, Claus and {Bresolin}, Fabio and {Kudritzki}, Rolf-Peter and {Steidel}, Charles C.},
        title = "{Spectral Modeling of Star-forming Regions in the Ultraviolet: Stellar Metallicity Diagnostics for High-Redshift Galaxies}",
      journal = {\apj},
         year = 2004,
        month = nov,
       volume = {615},
       number = {1},
        pages = {98-117},
          doi = {10.1086/424031},
archivePrefix = {arXiv},
       eprint = {astro-ph/0407296},
 primaryClass = {astro-ph},
       adsurl = {https://ui.adsabs.harvard.edu/abs/2004ApJ...615...98R}
}

@BOOK{Osterbrock2006,
       author = {{Osterbrock}, Donald E. and {Ferland}, Gary J.},
        title = "{Astrophysics of gaseous nebulae and active galactic nuclei}",
         year = 2006,
       adsurl = {https://ui.adsabs.harvard.edu/abs/2006agna.book.....O}
}

@ARTICLE{Johnson2021,
       author = {{Johnson}, Benjamin D. and {Leja}, Joel and {Conroy}, Charlie and {Speagle}, Joshua S.},
        title = "{Stellar Population Inference with Prospector}",
      journal = {\apjs},
         year = 2021,
        month = jun,
       volume = {254},
       number = {2},
          eid = {22},
        pages = {22},
          doi = {10.3847/1538-4365/abef67},
archivePrefix = {arXiv},
       eprint = {2012.01426},
 primaryClass = {astro-ph.GA},
       adsurl = {https://ui.adsabs.harvard.edu/abs/2021ApJS..254...22J}
}

@ARTICLE{Whitler2023_ceers,
       author = {{Whitler}, Lily and {Endsley}, Ryan and {Stark}, Daniel P. and {Topping}, Michael and {Chen}, Zuyi and {Charlot}, St{\'e}phane},
        title = "{On the ages of bright galaxies 500 Myr after the big bang: insights into star formation activity at z {\ensuremath{\gtrsim}} 15 with JWST}",
      journal = {\mnras},
         year = 2023,
        month = feb,
       volume = {519},
       number = {1},
        pages = {157-171},
          doi = {10.1093/mnras/stac3535},
archivePrefix = {arXiv},
       eprint = {2208.01599},
 primaryClass = {astro-ph.GA},
       adsurl = {https://ui.adsabs.harvard.edu/abs/2023MNRAS.519..157W}
}

@ARTICLE{Sanders2020,
       author = {{Sanders}, Ryan L. and {Shapley}, Alice E. and {Reddy}, Naveen A. and {Kriek}, Mariska and {Siana}, Brian and {Coil}, Alison L. and {Mobasher}, Bahram and {Shivaei}, Irene and {Freeman}, William R. and {Azadi}, Mojegan and {Price}, Sedona H. and {Leung}, Gene and {Fetherolf}, Tara and {de Groot}, Laura and {Zick}, Tom and {Fornasini}, Francesca M. and {Barro}, Guillermo},
        title = "{The MOSDEF survey: direct-method metallicities and ISM conditions at z {\ensuremath{\sim}} 1.5-3.5}",
      journal = {\mnras},
         year = 2020,
        month = jan,
       volume = {491},
       number = {1},
        pages = {1427-1455},
          doi = {10.1093/mnras/stz3032},
archivePrefix = {arXiv},
       eprint = {1907.00013},
 primaryClass = {astro-ph.GA},
       adsurl = {https://ui.adsabs.harvard.edu/abs/2020MNRAS.491.1427S}
}

@ARTICLE{Senchyna2017,
       author = {{Senchyna}, Peter and {Stark}, Daniel P. and {Vidal-Garc{\'\i}a}, Alba and {Chevallard}, Jacopo and {Charlot}, St{\'e}phane and {Mainali}, Ramesh and {Jones}, Tucker and {Wofford}, Aida and {Feltre}, Anna and {Gutkin}, Julia},
        title = "{Ultraviolet spectra of extreme nearby star-forming regions - approaching a local reference sample for JWST}",
      journal = {\mnras},
         year = 2017,
        month = dec,
       volume = {472},
       number = {3},
        pages = {2608-2632},
          doi = {10.1093/mnras/stx2059},
archivePrefix = {arXiv},
       eprint = {1706.00881},
 primaryClass = {astro-ph.GA},
       adsurl = {https://ui.adsabs.harvard.edu/abs/2017MNRAS.472.2608S}
}

@ARTICLE{Stark2017,
       author = {{Stark}, Daniel P. and {Ellis}, Richard S. and {Charlot}, St{\'e}phane and {Chevallard}, Jacopo and {Tang}, Mengtao and {Belli}, Sirio and {Zitrin}, Adi and {Mainali}, Ramesh and {Gutkin}, Julia and {Vidal-Garc{\'\i}a}, Alba and {Bouwens}, Rychard and {Oesch}, Pascal},
        title = "{Ly{\ensuremath{\alpha}} and C III] emission in z = 7-9 Galaxies: accelerated reionization around luminous star-forming systems?}",
      journal = {\mnras},
         year = 2017,
        month = jan,
       volume = {464},
       number = {1},
        pages = {469-479},
          doi = {10.1093/mnras/stw2233},
archivePrefix = {arXiv},
       eprint = {1606.01304},
 primaryClass = {astro-ph.GA},
       adsurl = {https://ui.adsabs.harvard.edu/abs/2017MNRAS.464..469S}
}

@ARTICLE{Llerena2022,
       author = {{Llerena}, M. and {Amor{\'\i}n}, R. and {Cullen}, F. and {Pentericci}, L. and {Calabr{\`o}}, A. and {McLure}, R. and {Carnall}, A. and {P{\'e}rez-Montero}, E. and {Marchi}, F. and {Bongiorno}, A. and {Castellano}, M. and {Fontana}, A. and {McLeod}, D.~J. and {Talia}, M. and {Hathi}, N.~P. and {Hibon}, P. and {Mannucci}, F. and {Saxena}, A. and {Schaerer}, D. and {Zamorani}, G.},
        title = "{The VANDELS survey: Global properties of CIII]{\ensuremath{\lambda}}1908 {\r{A}} emitting star-forming galaxies at z {\ensuremath{\sim}} 3}",
      journal = {\aap},
         year = 2022,
        month = mar,
       volume = {659},
          eid = {A16},
        pages = {A16},
          doi = {10.1051/0004-6361/202141651},
archivePrefix = {arXiv},
       eprint = {2107.00660},
 primaryClass = {astro-ph.GA},
       adsurl = {https://ui.adsabs.harvard.edu/abs/2022A&A...659A..16L}
}

@ARTICLE{Du2017,
       author = {{Du}, Xinnan and {Shapley}, Alice E. and {Martin}, Crystal L. and {Coil}, Alison L.},
        title = "{C III] Emission in Star-forming Galaxies at z {\ensuremath{\sim}} 1}",
      journal = {\apj},
         year = 2017,
        month = mar,
       volume = {838},
       number = {1},
          eid = {63},
        pages = {63},
          doi = {10.3847/1538-4357/aa64cf},
archivePrefix = {arXiv},
       eprint = {1612.06866},
 primaryClass = {astro-ph.GA},
       adsurl = {https://ui.adsabs.harvard.edu/abs/2017ApJ...838...63D}
}

@ARTICLE{Steidel2016,
       author = {{Steidel}, Charles C. and {Strom}, Allison L. and {Pettini}, Max and {Rudie}, Gwen C. and {Reddy}, Naveen A. and {Trainor}, Ryan F.},
        title = "{Reconciling the Stellar and Nebular Spectra of High-redshift Galaxies}",
      journal = {\apj},
         year = 2016,
        month = aug,
       volume = {826},
       number = {2},
          eid = {159},
        pages = {159},
          doi = {10.3847/0004-637X/826/2/159},
archivePrefix = {arXiv},
       eprint = {1605.07186},
 primaryClass = {astro-ph.GA},
       adsurl = {https://ui.adsabs.harvard.edu/abs/2016ApJ...826..159S}
}

@ARTICLE{Shapley2003,
       author = {{Shapley}, Alice E. and {Steidel}, Charles C. and {Pettini}, Max and {Adelberger}, Kurt L.},
        title = "{Rest-Frame Ultraviolet Spectra of z\raisebox{-0.5ex}\textasciitilde3 Lyman Break Galaxies}",
      journal = {\apj},
         year = 2003,
        month = may,
       volume = {588},
       number = {1},
        pages = {65-89},
          doi = {10.1086/373922},
archivePrefix = {arXiv},
       eprint = {astro-ph/0301230},
 primaryClass = {astro-ph},
       adsurl = {https://ui.adsabs.harvard.edu/abs/2003ApJ...588...65S}
}

@ARTICLE{Dijkstra2011,
       author = {{Dijkstra}, Mark and {Mesinger}, Andrei and {Wyithe}, J. Stuart B.},
        title = "{The detectability of Ly{\ensuremath{\alpha}} emission from galaxies during the epoch of reionization}",
      journal = {\mnras},
         year = 2011,
        month = jul,
       volume = {414},
       number = {3},
        pages = {2139-2147},
          doi = {10.1111/j.1365-2966.2011.18530.x},
archivePrefix = {arXiv},
       eprint = {1101.5160},
 primaryClass = {astro-ph.CO},
       adsurl = {https://ui.adsabs.harvard.edu/abs/2011MNRAS.414.2139D}
}

@ARTICLE{Endsley2022_overdensity,
       author = {{Endsley}, Ryan and {Stark}, Daniel P.},
        title = "{Strong Lyman-{\ensuremath{\alpha}} emission in an overdense region at z = 6.8: a very large (R   3 physical Mpc) ionized bubble in COSMOS?}",
      journal = {\mnras},
         year = 2022,
        month = apr,
       volume = {511},
       number = {4},
        pages = {6042-6054},
          doi = {10.1093/mnras/stac524},
archivePrefix = {arXiv},
       eprint = {2112.14779},
 primaryClass = {astro-ph.GA},
       adsurl = {https://ui.adsabs.harvard.edu/abs/2022MNRAS.511.6042E}
}

@ARTICLE{Nakajima2023,
       author = {{Nakajima}, Kimihiko and {Ouchi}, Masami and {Isobe}, Yuki and {Harikane}, Yuichi and {Zhang}, Yechi and {Ono}, Yoshiaki and {Umeda}, Hiroya and {Oguri}, Masamune},
        title = "{JWST Census for the Mass-Metallicity Star Formation Relations at z = 4-10 with Self-consistent Flux Calibration and Proper Metallicity Calibrators}",
      journal = {\apjs},
         year = 2023,
        month = dec,
       volume = {269},
       number = {2},
          eid = {33},
        pages = {33},
          doi = {10.3847/1538-4365/acd556},
archivePrefix = {arXiv},
       eprint = {2301.12825},
 primaryClass = {astro-ph.GA},
       adsurl = {https://ui.adsabs.harvard.edu/abs/2023ApJS..269...33N}
}

@ARTICLE{Tilvi2020,
       author = {{Tilvi}, V. and {Malhotra}, S. and {Rhoads}, J.~E. and {Coughlin}, A. and {Zheng}, Z. and {Finkelstein}, S.~L. and {Veilleux}, S. and {Mobasher}, B. and {Wang}, J. and {Probst}, R. and {Swaters}, R. and {Hibon}, P. and {Joshi}, B. and {Zabl}, J. and {Jiang}, T. and {Pharo}, J. and {Yang}, H.},
        title = "{Onset of Cosmic Reionization: Evidence of an Ionized Bubble Merely 680 Myr after the Big Bang}",
      journal = {\apjl},
         year = 2020,
        month = mar,
       volume = {891},
       number = {1},
          eid = {L10},
        pages = {L10},
          doi = {10.3847/2041-8213/ab75ec},
archivePrefix = {arXiv},
       eprint = {2001.00873},
 primaryClass = {astro-ph.GA},
       adsurl = {https://ui.adsabs.harvard.edu/abs/2020ApJ...891L..10T}
}

@ARTICLE{Topping2022_ceers,
       author = {{Topping}, Michael W. and {Stark}, Daniel P. and {Endsley}, Ryan and {Plat}, Adele and {Whitler}, Lily and {Chen}, Zuyi and {Charlot}, St{\'e}phane},
        title = "{Searching for Extremely Blue UV Continuum Slopes at z = 7-11 in JWST/NIRCam Imaging: Implications for Stellar Metallicity and Ionizing Photon Escape in Early Galaxies}",
      journal = {\apj},
         year = 2022,
        month = dec,
       volume = {941},
       number = {2},
          eid = {153},
        pages = {153},
          doi = {10.3847/1538-4357/aca522},
archivePrefix = {arXiv},
       eprint = {2208.01610},
 primaryClass = {astro-ph.GA},
       adsurl = {https://ui.adsabs.harvard.edu/abs/2022ApJ...941..153T}
}

@ARTICLE{Luridiana2015,
       author = {{Luridiana}, V. and {Morisset}, C. and {Shaw}, R.~A.},
        title = "{PyNeb: a new tool for analyzing emission lines. I. Code description and validation of results}",
      journal = {\aap},
         year = 2015,
        month = jan,
       volume = {573},
          eid = {A42},
        pages = {A42},
          doi = {10.1051/0004-6361/201323152},
archivePrefix = {arXiv},
       eprint = {1410.6662},
 primaryClass = {astro-ph.IM},
       adsurl = {https://ui.adsabs.harvard.edu/abs/2015A&A...573A..42L}
}

@ARTICLE{Mason2020_bubble,
       author = {{Mason}, Charlotte A. and {Gronke}, Max},
        title = "{Measuring the properties of reionized bubbles with resolved Ly{\ensuremath{\alpha}} spectra}",
      journal = {\mnras},
         year = 2020,
        month = nov,
       volume = {499},
       number = {1},
        pages = {1395-1405},
          doi = {10.1093/mnras/staa2910},
archivePrefix = {arXiv},
       eprint = {2004.13065},
 primaryClass = {astro-ph.GA},
       adsurl = {https://ui.adsabs.harvard.edu/abs/2020MNRAS.499.1395M}
}

@ARTICLE{Oesch2015,
       author = {{Oesch}, P.~A. and {van Dokkum}, P.~G. and {Illingworth}, G.~D. and {Bouwens}, R.~J. and {Momcheva}, I. and {Holden}, B. and {Roberts-Borsani}, G.~W. and {Smit}, R. and {Franx}, M. and {Labb{\'e}}, I. and {Gonz{\'a}lez}, V. and {Magee}, D.},
        title = "{A Spectroscopic Redshift Measurement for a Luminous Lyman Break Galaxy at z = 7.730 Using Keck/MOSFIRE}",
      journal = {\apjl},
         year = 2015,
        month = may,
       volume = {804},
       number = {2},
          eid = {L30},
        pages = {L30},
          doi = {10.1088/2041-8205/804/2/L30},
archivePrefix = {arXiv},
       eprint = {1502.05399},
 primaryClass = {astro-ph.GA},
       adsurl = {https://ui.adsabs.harvard.edu/abs/2015ApJ...804L..30O}
}

@ARTICLE{Oke1983,
       author = {{Oke}, J.~B. and {Gunn}, J.~E.},
        title = "{Secondary standard stars for absolute spectrophotometry.}",
      journal = {\apj},
         year = 1983,
        month = mar,
       volume = {266},
        pages = {713-717},
          doi = {10.1086/160817},
       adsurl = {https://ui.adsabs.harvard.edu/abs/1983ApJ...266..713O}
}

@ARTICLE{Tacchella2023,
       author = {{Tacchella}, Sandro and {Johnson}, Benjamin D. and {Robertson}, Brant E. and {Carniani}, Stefano and {D'Eugenio}, Francesco and {Kumari}, Nimisha and {Maiolino}, Roberto and {Nelson}, Erica J. and {Suess}, Katherine A. and {{\"U}bler}, Hannah and {Williams}, Christina C. and {Adebusola}, Alabi and {Alberts}, Stacey and {Arribas}, Santiago and {Bhatawdekar}, Rachana and {Bonaventura}, Nina and {Bowler}, Rebecca A.~A. and {Bunker}, Andrew J. and {Cameron}, Alex J. and {Curti}, Mirko and {Egami}, Eiichi and {Eisenstein}, Daniel J. and {Frye}, Brenda and {Hainline}, Kevin and {Helton}, Jakob M. and {Ji}, Zhiyuan and {Looser}, Tobias J. and {Lyu}, Jianwei and {Perna}, Michele and {Rawle}, Timothy and {Rieke}, George and {Rieke}, Marcia and {Saxena}, Aayush and {Sandles}, Lester and {Shivaei}, Irene and {Simmonds}, Charlotte and {Sun}, Fengwu and {Willmer}, Christopher N.~A. and {Willott}, Chris J. and {Witstok}, Joris},
        title = "{JWST NIRCam + NIRSpec: interstellar medium and stellar populations of young galaxies with rising star formation and evolving gas reservoirs}",
      journal = {\mnras},
         year = 2023,
        month = jul,
       volume = {522},
       number = {4},
        pages = {6236-6249},
          doi = {10.1093/mnras/stad1408},
archivePrefix = {arXiv},
       eprint = {2208.03281},
 primaryClass = {astro-ph.GA},
       adsurl = {https://ui.adsabs.harvard.edu/abs/2023MNRAS.522.6236T}
}

@ARTICLE{Heckman2001,
       author = {{Heckman}, T.~M. and {Sembach}, K.~R. and {Meurer}, G.~R. and {Leitherer}, C. and {Calzetti}, D. and {Martin}, C.~L.},
        title = "{On the Escape of Ionizing Radiation from Starbursts}",
      journal = {\apj},
         year = 2001,
        month = sep,
       volume = {558},
       number = {1},
        pages = {56-62},
          doi = {10.1086/322475},
archivePrefix = {arXiv},
       eprint = {astro-ph/0105012},
 primaryClass = {astro-ph},
       adsurl = {https://ui.adsabs.harvard.edu/abs/2001ApJ...558...56H}
}

@ARTICLE{Heckman2011,
       author = {{Heckman}, Timothy M. and {Borthakur}, Sanchayeeta and {Overzier}, Roderik and {Kauffmann}, Guinevere and {Basu-Zych}, Antara and {Leitherer}, Claus and {Sembach}, Ken and {Martin}, D. Chris and {Rich}, R. Michael and {Schiminovich}, David and {Seibert}, Mark},
        title = "{Extreme Feedback and the Epoch of Reionization: Clues in the Local Universe}",
      journal = {\apj},
         year = 2011,
        month = mar,
       volume = {730},
       number = {1},
          eid = {5},
        pages = {5},
          doi = {10.1088/0004-637X/730/1/5},
archivePrefix = {arXiv},
       eprint = {1101.4219},
 primaryClass = {astro-ph.CO},
       adsurl = {https://ui.adsabs.harvard.edu/abs/2011ApJ...730....5H}
}

@article{astropy:2013,
Adsurl = {http://adsabs.harvard.edu/abs/2013A%26A...558A..33A},
Archiveprefix = {arXiv},
Author = {{Astropy Collaboration} and {Robitaille}, T.~P. and {Tollerud}, E.~J. and {Greenfield}, P. and {Droettboom}, M. and {Bray}, E. and {Aldcroft}, T. and {Davis}, M. and {Ginsburg}, A. and {Price-Whelan}, A.~M. and {Kerzendorf}, W.~E. and {Conley}, A. and {Crighton}, N. and {Barbary}, K. and {Muna}, D. and {Ferguson}, H. and {Grollier}, F. and {Parikh}, M.~M. and {Nair}, P.~H. and {Unther}, H.~M. and {Deil}, C. and {Woillez}, J. and {Conseil}, S. and {Kramer}, R. and {Turner}, J.~E.~H. and {Singer}, L. and {Fox}, R. and {Weaver}, B.~A. and {Zabalza}, V. and {Edwards}, Z.~I. and {Azalee Bostroem}, K. and {Burke}, D.~J. and {Casey}, A.~R. and {Crawford}, S.~M. and {Dencheva}, N. and {Ely}, J. and {Jenness}, T. and {Labrie}, K. and {Lim}, P.~L. and {Pierfederici}, F. and {Pontzen}, A. and {Ptak}, A. and {Refsdal}, B. and {Servillat}, M. and {Streicher}, O.},
Doi = {10.1051/0004-6361/201322068},
Eid = {A33},
Eprint = {1307.6212},
Journal = {\aap},
Month = oct,
Pages = {A33},
Primaryclass = {astro-ph.IM},
Title = {{Astropy: A community Python package for astronomy}},
Volume = 558,
Year = 2013}

@ARTICLE{astropy:2018,
       author = {{Astropy Collaboration} and {Price-Whelan}, A.~M. and
         {Sip{\H{o}}cz}, B.~M. and {G{\"u}nther}, H.~M. and {Lim}, P.~L. and
         {Crawford}, S.~M. and {Conseil}, S. and {Shupe}, D.~L. and
         {Craig}, M.~W. and {Dencheva}, N. and {Ginsburg}, A. and {Vand
        erPlas}, J.~T. and {Bradley}, L.~D. and {P{\'e}rez-Su{\'a}rez}, D. and
         {de Val-Borro}, M. and {Aldcroft}, T.~L. and {Cruz}, K.~L. and
         {Robitaille}, T.~P. and {Tollerud}, E.~J. and {Ardelean}, C. and
         {Babej}, T. and {Bach}, Y.~P. and {Bachetti}, M. and {Bakanov}, A.~V. and
         {Bamford}, S.~P. and {Barentsen}, G. and {Barmby}, P. and
         {Baumbach}, A. and {Berry}, K.~L. and {Biscani}, F. and {Boquien}, M. and
         {Bostroem}, K.~A. and {Bouma}, L.~G. and {Brammer}, G.~B. and
         {Bray}, E.~M. and {Breytenbach}, H. and {Buddelmeijer}, H. and
         {Burke}, D.~J. and {Calderone}, G. and {Cano Rodr{\'\i}guez}, J.~L. and
         {Cara}, M. and {Cardoso}, J.~V.~M. and {Cheedella}, S. and {Copin}, Y. and
         {Corrales}, L. and {Crichton}, D. and {D'Avella}, D. and {Deil}, C. and
         {Depagne}, {\'E}. and {Dietrich}, J.~P. and {Donath}, A. and
         {Droettboom}, M. and {Earl}, N. and {Erben}, T. and {Fabbro}, S. and
         {Ferreira}, L.~A. and {Finethy}, T. and {Fox}, R.~T. and
         {Garrison}, L.~H. and {Gibbons}, S.~L.~J. and {Goldstein}, D.~A. and
         {Gommers}, R. and {Greco}, J.~P. and {Greenfield}, P. and
         {Groener}, A.~M. and {Grollier}, F. and {Hagen}, A. and {Hirst}, P. and
         {Homeier}, D. and {Horton}, A.~J. and {Hosseinzadeh}, G. and {Hu}, L. and
         {Hunkeler}, J.~S. and {Ivezi{\'c}}, {\v{Z}}. and {Jain}, A. and
         {Jenness}, T. and {Kanarek}, G. and {Kendrew}, S. and {Kern}, N.~S. and
         {Kerzendorf}, W.~E. and {Khvalko}, A. and {King}, J. and {Kirkby}, D. and
         {Kulkarni}, A.~M. and {Kumar}, A. and {Lee}, A. and {Lenz}, D. and
         {Littlefair}, S.~P. and {Ma}, Z. and {Macleod}, D.~M. and
         {Mastropietro}, M. and {McCully}, C. and {Montagnac}, S. and
         {Morris}, B.~M. and {Mueller}, M. and {Mumford}, S.~J. and {Muna}, D. and
         {Murphy}, N.~A. and {Nelson}, S. and {Nguyen}, G.~H. and
         {Ninan}, J.~P. and {N{\"o}the}, M. and {Ogaz}, S. and {Oh}, S. and
         {Parejko}, J.~K. and {Parley}, N. and {Pascual}, S. and {Patil}, R. and
         {Patil}, A.~A. and {Plunkett}, A.~L. and {Prochaska}, J.~X. and
         {Rastogi}, T. and {Reddy Janga}, V. and {Sabater}, J. and
         {Sakurikar}, P. and {Seifert}, M. and {Sherbert}, L.~E. and
         {Sherwood-Taylor}, H. and {Shih}, A.~Y. and {Sick}, J. and
         {Silbiger}, M.~T. and {Singanamalla}, S. and {Singer}, L.~P. and
         {Sladen}, P.~H. and {Sooley}, K.~A. and {Sornarajah}, S. and
         {Streicher}, O. and {Teuben}, P. and {Thomas}, S.~W. and
         {Tremblay}, G.~R. and {Turner}, J.~E.~H. and {Terr{\'o}n}, V. and
         {van Kerkwijk}, M.~H. and {de la Vega}, A. and {Watkins}, L.~L. and
         {Weaver}, B.~A. and {Whitmore}, J.~B. and {Woillez}, J. and
         {Zabalza}, V. and {Astropy Contributors}},
        title = "{The Astropy Project: Building an Open-science Project and Status of the v2.0 Core Package}",
      journal = {\aj},
         year = 2018,
        month = sep,
       volume = {156},
       number = {3},
          eid = {123},
        pages = {123},
          doi = {10.3847/1538-3881/aabc4f},
archivePrefix = {arXiv},
       eprint = {1801.02634},
 primaryClass = {astro-ph.IM},
       adsurl = {https://ui.adsabs.harvard.edu/abs/2018AJ....156..123A}
}

@ARTICLE{astropy:2022,
       author = {{Astropy Collaboration} and {Price-Whelan}, Adrian M. and {Lim}, Pey Lian and {Earl}, Nicholas and {Starkman}, Nathaniel and {Bradley}, Larry and {Shupe}, David L. and {Patil}, Aarya A. and {Corrales}, Lia and {Brasseur}, C.~E. and {N{"o}the}, Maximilian and {Donath}, Axel and {Tollerud}, Erik and {Morris}, Brett M. and {Ginsburg}, Adam and {Vaher}, Eero and {Weaver}, Benjamin A. and {Tocknell}, James and {Jamieson}, William and {van Kerkwijk}, Marten H. and {Robitaille}, Thomas P. and {Merry}, Bruce and {Bachetti}, Matteo and {G{"u}nther}, H. Moritz and {Aldcroft}, Thomas L. and {Alvarado-Montes}, Jaime A. and {Archibald}, Anne M. and {B{'o}di}, Attila and {Bapat}, Shreyas and {Barentsen}, Geert and {Baz{'a}n}, Juanjo and {Biswas}, Manish and {Boquien}, M{'e}d{'e}ric and {Burke}, D.~J. and {Cara}, Daria and {Cara}, Mihai and {Conroy}, Kyle E. and {Conseil}, Simon and {Craig}, Matthew W. and {Cross}, Robert M. and {Cruz}, Kelle L. and {D'Eugenio}, Francesco and {Dencheva}, Nadia and {Devillepoix}, Hadrien A.~R. and {Dietrich}, J{"o}rg P. and {Eigenbrot}, Arthur Davis and {Erben}, Thomas and {Ferreira}, Leonardo and {Foreman-Mackey}, Daniel and {Fox}, Ryan and {Freij}, Nabil and {Garg}, Suyog and {Geda}, Robel and {Glattly}, Lauren and {Gondhalekar}, Yash and {Gordon}, Karl D. and {Grant}, David and {Greenfield}, Perry and {Groener}, Austen M. and {Guest}, Steve and {Gurovich}, Sebastian and {Handberg}, Rasmus and {Hart}, Akeem and {Hatfield-Dodds}, Zac and {Homeier}, Derek and {Hosseinzadeh}, Griffin and {Jenness}, Tim and {Jones}, Craig K. and {Joseph}, Prajwel and {Kalmbach}, J. Bryce and {Karamehmetoglu}, Emir and {Ka{l}uszy{'n}ski}, Miko{l}aj and {Kelley}, Michael S.~P. and {Kern}, Nicholas and {Kerzendorf}, Wolfgang E. and {Koch}, Eric W. and {Kulumani}, Shankar and {Lee}, Antony and {Ly}, Chun and {Ma}, Zhiyuan and {MacBride}, Conor and {Maljaars}, Jakob M. and {Muna}, Demitri and {Murphy}, N.~A. and {Norman}, Henrik and {O'Steen}, Richard and {Oman}, Kyle A. and {Pacifici}, Camilla and {Pascual}, Sergio and {Pascual-Granado}, J. and {Patil}, Rohit R. and {Perren}, Gabriel I. and {Pickering}, Timothy E. and {Rastogi}, Tanuj and {Roulston}, Benjamin R. and {Ryan}, Daniel F. and {Rykoff}, Eli S. and {Sabater}, Jose and {Sakurikar}, Parikshit and {Salgado}, Jes{'u}s and {Sanghi}, Aniket and {Saunders}, Nicholas and {Savchenko}, Volodymyr and {Schwardt}, Ludwig and {Seifert-Eckert}, Michael and {Shih}, Albert Y. and {Jain}, Anany Shrey and {Shukla}, Gyanendra and {Sick}, Jonathan and {Simpson}, Chris and {Singanamalla}, Sudheesh and {Singer}, Leo P. and {Singhal}, Jaladh and {Sinha}, Manodeep and {Sip{H{o}}cz}, Brigitta M. and {Spitler}, Lee R. and {Stansby}, David and {Streicher}, Ole and {{{S}}umak}, Jani and {Swinbank}, John D. and {Taranu}, Dan S. and {Tewary}, Nikita and {Tremblay}, Grant R. and {Val-Borro}, Miguel de and {Van Kooten}, Samuel J. and {Vasovi{'c}}, Zlatan and {Verma}, Shresth and {de Miranda Cardoso}, Jos{'e} Vin{'i}cius and {Williams}, Peter K.~G. and {Wilson}, Tom J. and {Winkel}, Benjamin and {Wood-Vasey}, W.~M. and {Xue}, Rui and {Yoachim}, Peter and {Zhang}, Chen and {Zonca}, Andrea and {Astropy Project Contributors}},
        title = "{The Astropy Project: Sustaining and Growing a Community-oriented Open-source Project and the Latest Major Release (v5.0) of the Core Package}",
      journal = {apj},
         year = 2022,
        month = aug,
       volume = {935},
       number = {2},
          eid = {167},
        pages = {167},
          doi = {10.3847/1538-4357/ac7c74},
archivePrefix = {arXiv},
       eprint = {2206.14220},
 primaryClass = {astro-ph.IM},
       adsurl = {https://ui.adsabs.harvard.edu/abs/2022ApJ...935..167A}
}

@Article{Hunter:2007,
  Author    = {Hunter, J. D.},
  Title     = {Matplotlib: A 2D graphics environment},
  Journal   = {Computing in Science \& Engineering},
  Volume    = {9},
  Number    = {3},
  Pages     = {90--95},
  publisher = {IEEE COMPUTER SOC},
  doi       = {10.1109/MCSE.2007.55},
  year      = 2007
}

@Article{         harris2020array,
 title         = {Array programming with {NumPy}},
 author        = {Charles R. Harris and K. Jarrod Millman and St{\'{e}}fan J.
                 van der Walt and Ralf Gommers and Pauli Virtanen and David
                 Cournapeau and Eric Wieser and Julian Taylor and Sebastian
                 Berg and Nathaniel J. Smith and Robert Kern and Matti Picus
                 and Stephan Hoyer and Marten H. van Kerkwijk and Matthew
                 Brett and Allan Haldane and Jaime Fern{\'{a}}ndez del
                 R{\'{i}}o and Mark Wiebe and Pearu Peterson and Pierre
                 G{\'{e}}rard-Marchant and Kevin Sheppard and Tyler Reddy and
                 Warren Weckesser and Hameer Abbasi and Christoph Gohlke and
                 Travis E. Oliphant},
 year          = {2020},
 month         = sep,
 journal       = {Nature},
 volume        = {585},
 number        = {7825},
 pages         = {357--362},
 doi           = {10.1038/s41586-020-2649-2},
 publisher     = {Springer Science and Business Media {LLC}},
 url           = {https://doi.org/10.1038/s41586-020-2649-2}
}

@ARTICLE{2020SciPy-NMeth,
  author  = {Virtanen, Pauli and Gommers, Ralf and Oliphant, Travis E. and
            Haberland, Matt and Reddy, Tyler and Cournapeau, David and
            Burovski, Evgeni and Peterson, Pearu and Weckesser, Warren and
            Bright, Jonathan and {van der Walt}, St{\'e}fan J. and
            Brett, Matthew and Wilson, Joshua and Millman, K. Jarrod and
            Mayorov, Nikolay and Nelson, Andrew R. J. and Jones, Eric and
            Kern, Robert and Larson, Eric and Carey, C J and
            Polat, {\.I}lhan and Feng, Yu and Moore, Eric W. and
            {VanderPlas}, Jake and Laxalde, Denis and Perktold, Josef and
            Cimrman, Robert and Henriksen, Ian and Quintero, E. A. and
            Harris, Charles R. and Archibald, Anne M. and
            Ribeiro, Ant{\^o}nio H. and Pedregosa, Fabian and
            {van Mulbregt}, Paul and {SciPy 1.0 Contributors}},
  title   = {{{SciPy} 1.0: Fundamental Algorithms for Scientific
            Computing in Python}},
  journal = {Nature Methods},
  year    = {2020},
  volume  = {17},
  pages   = {261--272},
  adsurl  = {https://rdcu.be/b08Wh},
  doi     = {10.1038/s41592-019-0686-2},
}

@MISC{Johnson2021_sedpy,
       author = {{Johnson}, Benjamin D.},
        title = "{bd-j/sedpy: sedpy v0.2.0}",
 howpublished = {Zenodo},
         year = 2021,
        month = mar,
          eid = {10.5281/zenodo.4582723},
          doi = {10.5281/zenodo.4582723},
      version = {v0.2.0},
    publisher = {Zenodo},
       adsurl = {https://ui.adsabs.harvard.edu/abs/2021zndo...4582723J}
}

@ARTICLE{Steidel2010,
       author = {{Steidel}, Charles C. and {Erb}, Dawn K. and {Shapley}, Alice E. and {Pettini}, Max and {Reddy}, Naveen and {Bogosavljevi{\'c}}, Milan and {Rudie}, Gwen C. and {Rakic}, Olivera},
        title = "{The Structure and Kinematics of the Circumgalactic Medium from Far-ultraviolet Spectra of z \raisebox{-0.5ex}\textasciitilde= 2-3 Galaxies}",
      journal = {\apj},
         year = 2010,
        month = jul,
       volume = {717},
       number = {1},
        pages = {289-322},
          doi = {10.1088/0004-637X/717/1/289},
archivePrefix = {arXiv},
       eprint = {1003.0679},
 primaryClass = {astro-ph.CO},
       adsurl = {https://ui.adsabs.harvard.edu/abs/2010ApJ...717..289S}
}

@ARTICLE{Saldana-Lopez2022,
       author = {{Saldana-Lopez}, Alberto and {Schaerer}, Daniel and {Chisholm}, John and {Flury}, Sophia R. and {Jaskot}, Anne E. and {Worseck}, G{\'a}bor and {Makan}, Kirill and {Gazagnes}, Simon and {Mauerhofer}, Valentin and {Verhamme}, Anne and {Amor{\'\i}n}, Ricardo O. and {Ferguson}, Harry C. and {Giavalisco}, Mauro and {Grazian}, Andrea and {Hayes}, Matthew J. and {Heckman}, Timothy M. and {Henry}, Alaina and {Ji}, Zhiyuan and {Marques-Chaves}, Rui and {McCandliss}, Stephan R. and {Oey}, M. Sally and {{\"O}stlin}, G{\"o}ran and {Pentericci}, Laura and {Thuan}, Trinh X. and {Trebitsch}, Maxime and {Vanzella}, Eros and {Xu}, Xinfeng},
        title = "{The Low-Redshift Lyman Continuum Survey. Unveiling the ISM properties of low-z Lyman-continuum emitters}",
      journal = {\aap},
         year = 2022,
        month = jul,
       volume = {663},
          eid = {A59},
        pages = {A59},
          doi = {10.1051/0004-6361/202141864},
archivePrefix = {arXiv},
       eprint = {2201.11800},
 primaryClass = {astro-ph.GA},
       adsurl = {https://ui.adsabs.harvard.edu/abs/2022A&A...663A..59S}
}

@ARTICLE{Foreman-Mackey2013,
       author = {{Foreman-Mackey}, Daniel and {Hogg}, David W. and {Lang}, Dustin and {Goodman}, Jonathan},
        title = "{emcee: The MCMC Hammer}",
      journal = {\pasp},
         year = 2013,
        month = mar,
       volume = {125},
       number = {925},
        pages = {306},
          doi = {10.1086/670067},
archivePrefix = {arXiv},
       eprint = {1202.3665},
 primaryClass = {astro-ph.IM},
       adsurl = {https://ui.adsabs.harvard.edu/abs/2013PASP..125..306F}
}

@ARTICLE{Izotov2018_higho32,
       author = {{Izotov}, Y.~I. and {Worseck}, G. and {Schaerer}, D. and {Guseva}, N.~G. and {Thuan}, T.~X. and {Fricke}, Verhamme, A. and {Orlitov{\'a}}, I.},
        title = "{Low-redshift Lyman continuum leaking galaxies with high [O III]/[O II] ratios}",
      journal = {\mnras},
         year = 2018,
        month = aug,
       volume = {478},
       number = {4},
        pages = {4851-4865},
          doi = {10.1093/mnras/sty1378},
archivePrefix = {arXiv},
       eprint = {1805.09865},
 primaryClass = {astro-ph.GA},
       adsurl = {https://ui.adsabs.harvard.edu/abs/2018MNRAS.478.4851I}
}

@ARTICLE{Cullen2024,
       author = {{Cullen}, F. and {McLeod}, D.~J. and {McLure}, R.~J. and {Dunlop}, J.~S. and {Donnan}, C.~T. and {Carnall}, A.~C. and {Keating}, L.~C. and {Magee}, D. and {Arellano-Cordova}, K.~Z. and {Bowler}, R.~A.~A. and {Begley}, R. and {Flury}, S.~R. and {Hamadouche}, M.~L. and {Stanton}, T.~M.},
        title = "{The ultraviolet continuum slopes of high-redshift galaxies: evidence for the emergence of dust-free stellar populations at z > 10}",
      journal = {\mnras},
         year = 2024,
        month = jun,
       volume = {531},
       number = {1},
        pages = {997-1020},
          doi = {10.1093/mnras/stae1211},
archivePrefix = {arXiv},
       eprint = {2311.06209},
 primaryClass = {astro-ph.GA},
       adsurl = {https://ui.adsabs.harvard.edu/abs/2024MNRAS.531..997C}
}

@INCOLLECTION{Kluyver2016,
       author = {{Kluyver}, Thomas and {Ragan-Kelley}, Benjain and {P{\'e}rez}, Fernando and {Granger}, Brian and {Bussonnier}, Matthias and {Frederic}, Jonathan and {Kelley}, Kyle and {Hamrick}, Jessica and {Grout}, Jason and {Corlay}, Sylvain and {Ivanov}, Paul and {Avila}, Dami{\'a}n and {Abdalla}, Safia and {Willing}, Carol and {Jupyter Development Team}},
        title = "{Jupyter Notebooks{\textemdash}a publishing format for reproducible computational workflows}",
    booktitle = {IOS Press},
         year = 2016,
        pages = {87-90},
          doi = {10.3233/978-1-61499-649-1-87},
       adsurl = {https://ui.adsabs.harvard.edu/abs/2016ppap.book...87K}
}

@ARTICLE{Keating2023,
       author = {{Keating}, Laura C. and {Bolton}, James S. and {Cullen}, Fergus and {Haehnelt}, Martin G. and {Puchwein}, Ewald and {Kulkarni}, Girish},
        title = "{JWST observations of galaxy damping wings during reionization interpreted with cosmological simulations}",
      journal = {arXiv e-prints},
         year = 2023,
        month = aug,
          eid = {arXiv:2308.05800},
        pages = {arXiv:2308.05800},
          doi = {10.48550/arXiv.2308.05800},
archivePrefix = {arXiv},
       eprint = {2308.05800},
 primaryClass = {astro-ph.GA},
       adsurl = {https://ui.adsabs.harvard.edu/abs/2023arXiv230805800K}
}

@ARTICLE{Boyett2024_z9p3,
       author = {{Boyett}, Kristan and {Trenti}, Michele and {Leethochawalit}, Nicha and {Calabr{\'o}}, Antonello and {Metha}, Benjamin and {Roberts-Borsani}, Guido and {Dalmasso}, Nicol{\'o} and {Yang}, Lilan and {Santini}, Paola and {Treu}, Tommaso and {Jones}, Tucker and {Henry}, Alaina and {Mason}, Charlotte A. and {Morishita}, Takahiro and {Nanayakkara}, Themiya and {Roy}, Namrata and {Wang}, Xin and {Fontana}, Adriano and {Merlin}, Emiliano and {Castellano}, Marco and {Paris}, Diego and {Brada{\v{c}}}, Maru{\v{s}}a and {Malkan}, Matt and {Marchesini}, Danilo and {Mascia}, Sara and {Glazebrook}, Karl and {Pentericci}, Laura and {Vanzella}, Eros and {Vulcani}, Benedetta},
        title = "{A massive interacting galaxy 510 million years after the Big Bang}",
      journal = {Nature Astronomy},
         year = 2024,
        month = may,
       volume = {8},
        pages = {657-672},
          doi = {10.1038/s41550-024-02218-7},
archivePrefix = {arXiv},
       eprint = {2303.00306},
 primaryClass = {astro-ph.GA},
       adsurl = {https://ui.adsabs.harvard.edu/abs/2024NatAs...8..657B}
}

@ARTICLE{Napolitano2025_uds,
       author = {{Napolitano}, L. and {Pentericci}, L. and {Dickinson}, M. and {Arrabal Haro}, P. and {Taylor}, A.~J. and {Calabr{\`o}}, A. and {Bhagwat}, A. and {Santini}, P. and {Arevalo-Gonzalez}, F. and {Begley}, R. and {Castellano}, M. and {Ciardi}, B. and {Donnan}, C.~T. and {Dottorini}, D. and {Dunlop}, J.~S. and {Finkelstein}, S.~L. and {Fontana}, A. and {Giavalisco}, M. and {Hirschmann}, M. and {Jung}, I. and {Koekemoer}, A.~M. and {Kokorev}, V. and {Llerena}, M. and {Lucas}, R.~A. and {Mascia}, S. and {Merlin}, E. and {P{\'e}rez-Gonz{\'a}lez}, P.~G. and {Stanton}, T.~M. and {Tripodi}, R. and {Wang}, X. and {Weiner}, B.~J.},
        title = "{Ly$α$ visibility from z = 4.5 to 11 in the UDS field: evidence for a high neutral hydrogen fraction and small ionized bubbles at z $\sim$ 7}",
      journal = {arXiv e-prints},
         year = 2025,
        month = aug,
          eid = {arXiv:2508.14171},
        pages = {arXiv:2508.14171},
          doi = {10.48550/arXiv.2508.14171},
archivePrefix = {arXiv},
       eprint = {2508.14171},
 primaryClass = {astro-ph.GA},
       adsurl = {https://ui.adsabs.harvard.edu/abs/2025arXiv250814171N}
}

@Article{Donnan2024,
  author           = {Donnan, C.~T. and McLure, R.~J. and Dunlop, J.~S. and McLeod, D.~J. and Magee, D. and Arellano-C{\'o}rdova, K.~Z. and Barrufet, L. and Begley, R. and Bowler, R.~A.~A. and Carnall, A.~C. and Cullen, F. and Ellis, R.~S. and Fontana, A. and Illingworth, G.~D. and Grogin, N.~A. and Hamadouche, M.~L. and Koekemoer, A.~M. and Liu, F.-Y. and Mason, C. and Santini, P. and Stanton, T.~M.},
  journal          = {\mnras},
  title            = {JWST PRIMER: a new multifield determination of the evolving galaxy UV luminosity function at redshifts z ≃ 9 - 15},
  year             = {2024},
  number           = {3},
  pages            = {3222-3237},
  volume           = {533},
  date             = {2024-09},
  doi              = {10.1093/mnras/stae2037},
  eprint           = {2403.03171},
  eprintclass      = {astro-ph.GA},
  eprinttype       = {arXiv},
  journaltitle     = {\mnras},
  modificationdate = {2026-03-07T18:22:30},
  url              = {https://ui.adsabs.harvard.edu/abs/2024MNRAS.533.3222D},
}

@Article{Algera2025,
  author           = {Algera, Hiddo S. B. and Weaver, John R. and Bakx, Tom J. L. C. and Aravena, Manuel and Bouwens, Rychard J. and Cescon, Karin and Chen, Chian-Chou and da Cunha, Elisabete and Dayal, Pratika and Faisst, Andreas and Ferrara, Andrea and Fujimoto, Seiji and Hashimoto, Takuya and Heintz, Kasper and Herrera-Camus, Rodrigo and Hodge, Jacqueline and Inami, Hanae and Inoue, Akio K. and Matthee, Jorryt and Meyer, Romain and Mizukoshi, Shoichiro and Mondal, Chayan and Nanayakkara, Themiya and Oesch, Pascal A. and Pallottini, Andrea and Röttgering, Huub and Rowland, Lucie E. and Schouws, Sander and Smit, Renske and Sommovigo, Laura and Stark, Daniel P. and Sugahara, Yuma and Vallini, Livia and Vijarnwannaluk, Bovornpratch and van der Werf, Paul and Werner, Norbert and Witstok, Joris and Xiao, Mengyuan},
  title            = {A first systematic study of [OIII] 88$μ$m at $z>8$: two luminous oxygen lines and a powerful ionized outflow in the first 600 million years},
  year             = {2025},
  copyright        = {arXiv.org perpetual, non-exclusive license},
  date             = {2025-12-16},
  doi              = {10.48550/ARXIV.2512.14486},
  eprint           = {2512.14486},
  eprintclass      = {astro-ph.GA},
  eprinttype       = {arXiv},
  modificationdate = {2025-12-17T13:26:23},
  publisher        = {arXiv},
}

@Article{RobertsBorsani2024,
  author           = {Roberts-Borsani, Guido and Treu, Tommaso and Shapley, Alice and Fontana, Adriano and Pentericci, Laura and Castellano, Marco and Morishita, Takahiro and Bergamini, Pietro and Rosati, Piero},
  journal          = {ApJ},
  title            = {Between the Extremes: A JWST Spectroscopic Benchmark for High-redshift Galaxies Using ∼500 Confirmed Sources at z ≥ 5},
  year             = {2024},
  issn             = {1538-4357},
  month            = nov,
  number           = {2},
  pages            = {193},
  volume           = {976},
  doi              = {10.3847/1538-4357/ad85d3},
  modificationdate = {2025-12-28T00:37:53},
  publisher        = {American Astronomical Society},
}

@Article{Castellano2016a,
  author           = {Castellano, M. and Amor{\'\i}n, R. and Merlin, E. and Fontana, A. and McLure, R.~J. and M{\'a}rmol-Queralt{\'o}, E. and Mortlock, A. and Parsa, S. and Dunlop, J.~S. and Elbaz, D. and Balestra, I. and Boucaud, A. and Bourne, N. and Boutsia, K. and Brammer, G. and Bruce, V.~A. and Buitrago, F. and Capak, P. and Cappelluti, N. and Ciesla, L. and Comastri, A. and Cullen, F. and Derriere, S. and Faber, S.~M. and Giallongo, E. and Grazian, A. and Grillo, C. and Mercurio, A. and Micha{\l}owski, M.~J. and Nonino, M. and Paris, D. and Pentericci, L. and Pilo, S. and Rosati, P. and Santini, P. and Schreiber, C. and Shu, X. and Wang, T.},
  journal          = {\aap},
  title            = {The ASTRODEEP Frontier Fields catalogues. II. Photometric redshifts and rest frame properties in Abell-2744 and MACS-J0416},
  year             = {2016},
  month            = may,
  pages            = {A31},
  volume           = {590},
  archiveprefix    = {arXiv},
  doi              = {10.1051/0004-6361/201527514},
  eid              = {A31},
  eprint           = {1603.02461},
  modificationdate = {2025-12-27T16:34:43},
  primaryclass     = {astro-ph.GA},
  url              = {https://ui.adsabs.harvard.edu/abs/2016A&A...590A..31C},
}

@Article{Castellano2023,
  author           = {Castellano, Marco and Fontana, Adriano and Treu, Tommaso and Merlin, Emiliano and Santini, Paola and Bergamini, Pietro and Grillo, Claudio and Rosati, Piero and Acebron, Ana and Leethochawalit, Nicha and Paris, Diego and Bonchi, Andrea and Belfiori, Davide and Calabr{\`o}, Antonello and Correnti, Matteo and Nonino, Mario and Polenta, Gianluca and Trenti, Michele and Boyett, Kristan and Brammer, G. and Broadhurst, Tom and Caminha, Gabriel B. and Chen, Wenlei and Filippenko, Alexei V. and Fortuni, Flaminia and Glazebrook, Karl and Mascia, Sara and Mason, Charlotte A. and Menci, Nicola and Meneghetti, Massimo and Mercurio, Amata and Metha, Benjamin and Morishita, Takahiro and Nanayakkara, Themiya and Pentericci, Laura and Roberts-Borsani, Guido and Roy, Namrata and Vanzella, Eros and Vulcani, Benedetta and Yang, Lilan and Wang, Xin},
  journal          = {\apjl},
  title            = {Early Results from GLASS-JWST. XIX. A High Density of Bright Galaxies at z {\ensuremath{\approx}} 10 in the A2744 Region},
  year             = {2023},
  month            = may,
  number           = {2},
  pages            = {L14},
  volume           = {948},
  archiveprefix    = {arXiv},
  doi              = {10.3847/2041-8213/accea5},
  eid              = {L14},
  eprint           = {2212.06666},
  modificationdate = {2025-12-27T16:47:41},
  primaryclass     = {astro-ph.GA},
  url              = {https://ui.adsabs.harvard.edu/abs/2023ApJ...948L..14C},
}

@Article{Atek2023,
  author           = {Atek, Hakim and Chemerynska, Iryna and Wang, Bingjie and Furtak, Lukas J and Weibel, Andrea and Oesch, Pascal and Weaver, John R and Labbé, Ivo and Bezanson, Rachel and van Dokkum, Pieter and Zitrin, Adi and Dayal, Pratika and Williams, Christina C and Nannayakkara, Themiya and Price, Sedona H and Brammer, Gabriel and Goulding, Andy D and Leja, Joel and Marchesini, Danilo and Nelson, Erica J and Pan, Richard and Whitaker, Katherine E},
  journal          = {MNRAS},
  title            = {JWST UNCOVER: discovery of z \$>$ 9 galaxy candidates behind the lensing cluster Abell 2744},
  year             = {2023},
  issn             = {1365-2966},
  month            = jul,
  number           = {4},
  pages            = {5486--5496},
  volume           = {524},
  doi              = {10.1093/mnras/stad1998},
  modificationdate = {2025-12-28T00:34:47},
  publisher        = {Oxford University Press (OUP)},
}

@Article{Harikane2025,
  author           = {Harikane, Yuichi and Inoue, Akio K. and Ellis, Richard S. and Ouchi, Masami and Nakazato, Yurina and Yoshida, Naoki and Ono, Yoshiaki and Sun, Fengwu and Sato, Riku A. and Ferrami, Giovanni and Fujimoto, Seiji and Kashikawa, Nobunari and McLeod, Derek J. and Pérez-González, Pablo G. and Sawicki, Marcin and Sugahara, Yuma and Xu, Yi and Yamanaka, Satoshi and Carnall, Adam C. and Cullen, Fergus and Dunlop, James S. and Egami, Eiichi and Grogin, Norman and Isobe, Yuki and Koekemoer, Anton M. and Laporte, Nicolas and Lee, Chien-Hsiu and Magee, Dan and Matsuo, Hiroshi and Matsuoka, Yoshiki and Mawatari, Ken and Nakajima, Kimihiko and Nakane, Minami and Tamura, Yoichi and Umeda, Hiroya and Yanagisawa, Hiroto},
  journal          = {ApJ},
  title            = {JWST, ALMA, and Keck Spectroscopic Constraints on the UV Luminosity Functions at z ∼ 7–14: Clumpiness and Compactness of the Brightest Galaxies in the Early Universe},
  year             = {2025},
  issn             = {1538-4357},
  month            = feb,
  number           = {1},
  pages            = {138},
  volume           = {980},
  doi              = {10.3847/1538-4357/ad9b2c},
  modificationdate = {2025-12-28T00:35:23},
  publisher        = {American Astronomical Society},
}

@Article{Donnan2026,
  author           = {Donnan, Callum T. and McLeod, Derek J. and McLure, Ross J. and Dunlop, James S. and Cullen, Fergus and Dickinson, Mark and Haro, Pablo Arrabal and Taylor, Anthony J. and Bondestam, Cecilia and Liu, Feng-Yuan and Arellano-Córdova, Karla Z. and Barrufet, Laia and Begley, Ryan and Carnall, Adam C. and Golawska, Hanna and Leung, Ho-Hin and Scholte, Dirk and Stanton, Thomas M.},
  title            = {Spectroscopic confirmation of a large and luminous galaxy with weak emission lines at $\mathbf{z = 13.53}$},
  year             = {2026},
  month            = jan,
  archiveprefix    = {arXiv},
  copyright        = {Creative Commons Attribution 4.0 International},
  doi              = {10.48550/ARXIV.2601.11515},
  eprint           = {2601.11515},
  modificationdate = {2026-01-20T17:20:48},
  primaryclass     = {astro-ph.GA},
  publisher        = {arXiv},
}

@Article{Helton2025,
  author           = {Helton, Jakob M. and Morrison, Jane E. and Hainline, Kevin N. and D'Eugenio, Francesco and Rieke, George H. and Alberts, Stacey and Carniani, Stefano and Leja, Joel and Li, Yijia and Rinaldi, Pierluigi and Scholtz, Jan and Stone, Meredith and Willmer, Christopher N.~A. and Wu, Zihao and Baker, William M. and Bunker, Andrew J. and Charlot, Stephane and Chevallard, Jacopo and Cleri, Nikko J. and Curti, Mirko and Curtis-Lake, Emma and Egami, Eiichi and Eisenstein, Daniel J. and Jakobsen, Peter and Ji, Zhiyuan and Johnson, Benjamin D. and Kumari, Nimisha and Lin, Xiaojing and Lyu, Jianwei and Maiolino, Roberto and Maseda, Michael and P{\'e}rez-Gonz{\'a}lez, Pablo G. and Rieke, Marcia J. and Robertson, Brant and Saxena, Aayush and Sun, Fengwu and Tacchella, Sandro and {\"U}bler, Hannah and Venturi, Giacomo and Williams, Christina C. and Willott, Chris and Witstok, Joris and Zhu, Yongda},
  journal          = {arXiv e-prints},
  title            = {Ionizing Photon Production Efficiencies and Chemical Abundances at Cosmic Dawn Revealed by Ultra-Deep Rest-Frame Optical Spectroscopy of JADES-GS-z14-0},
  year             = {2025},
  month            = dec,
  pages            = {arXiv:2512.19695},
  archiveprefix    = {arXiv},
  doi              = {10.48550/arXiv.2512.19695},
  eid              = {arXiv:2512.19695},
  eprint           = {2512.19695},
  modificationdate = {2026-01-20T17:22:34},
  primaryclass     = {astro-ph.GA},
  url              = {https://ui.adsabs.harvard.edu/abs/2025arXiv251219695H},
}

@Article{Sanders2016,
  author           = {Sanders, Ryan L. and Shapley, Alice E. and Kriek, Mariska and Reddy, Naveen A. and Freeman, William R. and Coil, Alison L. and Siana, Brian and Mobasher, Bahram and Shivaei, Irene and Price, Sedona H. and de Groot, Laura},
  journal          = {\apj},
  title            = {The MOSDEF Survey: Electron Density and Ionization Parameter at z \raisebox{-0.5ex}\textasciitilde 2.3},
  year             = {2016},
  month            = jan,
  number           = {1},
  pages            = {23},
  volume           = {816},
  archiveprefix    = {arXiv},
  doi              = {10.3847/0004-637X/816/1/23},
  eid              = {23},
  eprint           = {1509.03636},
  modificationdate = {2026-01-21T15:57:17},
  primaryclass     = {astro-ph.GA},
  url              = {https://ui.adsabs.harvard.edu/abs/2016ApJ...816...23S},
}

@Article{Byler2017,
  author           = {Byler, Nell and Dalcanton, Julianne J. and Conroy, Charlie and Johnson, Benjamin D.},
  journal          = {\apj},
  title            = {Nebular Continuum and Line Emission in Stellar Population Synthesis Models},
  year             = {2017},
  month            = may,
  number           = {1},
  pages            = {44},
  volume           = {840},
  archiveprefix    = {arXiv},
  doi              = {10.3847/1538-4357/aa6c66},
  eid              = {44},
  eprint           = {1611.08305},
  modificationdate = {2026-01-21T16:23:19},
  primaryclass     = {astro-ph.GA},
  url              = {https://ui.adsabs.harvard.edu/abs/2017ApJ...840...44B},
}

@Article{Conroy2010,
  author           = {Conroy, Charlie and White, Martin and Gunn, James E.},
  journal          = {\apj},
  title            = {The Propagation of Uncertainties in Stellar Population Synthesis Modeling. II. The Challenge of Comparing Galaxy Evolution Models to Observations},
  year             = {2010},
  month            = jan,
  number           = {1},
  pages            = {58-70},
  volume           = {708},
  archiveprefix    = {arXiv},
  doi              = {10.1088/0004-637X/708/1/58},
  eprint           = {0904.0002},
  modificationdate = {2026-01-21T16:23:36},
  primaryclass     = {astro-ph.CO},
  url              = {https://ui.adsabs.harvard.edu/abs/2010ApJ...708...58C},
}

@Article{Conroy2009,
  author           = {Conroy, Charlie and Gunn, James E. and White, Martin},
  journal          = {\apj},
  title            = {The Propagation of Uncertainties in Stellar Population Synthesis Modeling. I. The Relevance of Uncertain Aspects of Stellar Evolution and the Initial Mass Function to the Derived Physical Properties of Galaxies},
  year             = {2009},
  month            = jul,
  number           = {1},
  pages            = {486-506},
  volume           = {699},
  archiveprefix    = {arXiv},
  doi              = {10.1088/0004-637X/699/1/486},
  eprint           = {0809.4261},
  modificationdate = {2026-01-21T16:23:50},
  primaryclass     = {astro-ph},
  url              = {https://ui.adsabs.harvard.edu/abs/2009ApJ...699..486C},
}

@Article{Choi2016,
  author           = {Choi, Jieun and Dotter, Aaron and Conroy, Charlie and Cantiello, Matteo and Paxton, Bill and Johnson, Benjamin D.},
  journal          = {\apj},
  title            = {Mesa Isochrones and Stellar Tracks (MIST). I. Solar-scaled Models},
  year             = {2016},
  month            = jun,
  number           = {2},
  pages            = {102},
  volume           = {823},
  archiveprefix    = {arXiv},
  doi              = {10.3847/0004-637X/823/2/102},
  eid              = {102},
  eprint           = {1604.08592},
  modificationdate = {2026-01-21T16:27:30},
  primaryclass     = {astro-ph.SR},
  url              = {https://ui.adsabs.harvard.edu/abs/2016ApJ...823..102C},
}

@Article{SanchezBlazquez2006,
  author           = {S{\'a}nchez-Bl{\'a}zquez, P. and Peletier, R.~F. and Jim{\'e}nez-Vicente, J. and Cardiel, N. and Cenarro, A.~J. and Falc{\'o}n-Barroso, J. and Gorgas, J. and Selam, S. and Vazdekis, A.},
  journal          = {\mnras},
  title            = {Medium-resolution Isaac Newton Telescope library of empirical spectra},
  year             = {2006},
  month            = sep,
  number           = {2},
  pages            = {703-718},
  volume           = {371},
  archiveprefix    = {arXiv},
  doi              = {10.1111/j.1365-2966.2006.10699.x},
  eprint           = {astro-ph/0607009},
  modificationdate = {2026-01-21T16:27:38},
  primaryclass     = {astro-ph},
  url              = {https://ui.adsabs.harvard.edu/abs/2006MNRAS.371..703S},
}

@Article{Shapley2023,
  author           = {Shapley, Alice E. and Reddy, Naveen A. and Sanders, Ryan L. and Topping, Michael W. and Brammer, Gabriel B.},
  journal          = {\apjl},
  title            = {JWST/NIRSpec Measurements of the Relationships between Nebular Emission-line Ratios and Stellar Mass at z 3-6},
  year             = {2023},
  month            = jun,
  number           = {1},
  pages            = {L1},
  volume           = {950},
  archiveprefix    = {arXiv},
  doi              = {10.3847/2041-8213/acd939},
  eid              = {L1},
  eprint           = {2303.00410},
  modificationdate = {2026-01-23T15:28:15},
  primaryclass     = {astro-ph.GA},
  url              = {https://ui.adsabs.harvard.edu/abs/2023ApJ...950L...1S},
}

@Article{Chisholm2019,
  author           = {Chisholm, J. and Rigby, J.~R. and Bayliss, M. and Berg, D.~A. and Dahle, H. and Gladders, M. and Sharon, K.},
  journal          = {\apj},
  title            = {Constraining the Metallicities, Ages, Star Formation Histories, and Ionizing Continua of Extragalactic Massive Star Populations},
  year             = {2019},
  month            = sep,
  number           = {2},
  pages            = {182},
  volume           = {882},
  archiveprefix    = {arXiv},
  doi              = {10.3847/1538-4357/ab3104},
  eid              = {182},
  eprint           = {1905.04314},
  modificationdate = {2026-01-23T19:52:08},
  primaryclass     = {astro-ph.GA},
  url              = {https://ui.adsabs.harvard.edu/abs/2019ApJ...882..182C},
}

@Article{Plat2019,
  author           = {Plat, A. and Charlot, S. and Bruzual, G. and Feltre, A. and Vidal-Garc{\'\i}a, A. and Morisset, C. and Chevallard, J. and Todt, H.},
  journal          = {\mnras},
  title            = {Constraints on the production and escape of ionizing radiation from the emission-line spectra of metal-poor star-forming galaxies},
  year             = {2019},
  month            = nov,
  number           = {1},
  pages            = {978-1009},
  volume           = {490},
  archiveprefix    = {arXiv},
  doi              = {10.1093/mnras/stz2616},
  eprint           = {1909.07386},
  modificationdate = {2026-01-23T21:06:35},
  primaryclass     = {astro-ph.GA},
  url              = {https://ui.adsabs.harvard.edu/abs/2019MNRAS.490..978P},
}

@Article{Mainali2023,
  author           = {Mainali, Ramesh and Stark, Daniel P. and Jones, Tucker and Ellis, Richard S. and Hezaveh, Yashar D. and Rigby, Jane R.},
  journal          = {\mnras},
  title            = {Spectroscopy of CASSOWARY gravitationally lensed galaxies in SDSS: characterization of an extremely bright reionization-era analogue at z = 1.42},
  year             = {2023},
  month            = apr,
  number           = {3},
  pages            = {4037-4056},
  volume           = {520},
  archiveprefix    = {arXiv},
  doi              = {10.1093/mnras/stad387},
  eprint           = {2301.11264},
  modificationdate = {2026-01-24T16:13:52},
  primaryclass     = {astro-ph.GA},
  url              = {https://ui.adsabs.harvard.edu/abs/2023MNRAS.520.4037M},
}

@Article{Endsley2025,
  author           = {Endsley, Ryan and Shapley, Alice E. and Topping, Michael W. and Stark, Daniel P. and Bouwens, Rychard J. and Rowland, Lucie E. and Sommovigo, Laura and Algera, Hiddo S.~B. and Aravena, Manuel and Bowler, Rebecca A.~A. and da Cunha, Elisabete and de Looze, Ilse and Ferrara, Andrea and Fisher, Rebecca and Gonz{\'a}lez, Valentino and Inami, Hanae and Nanayakkara, Themiya and Schouws, Sander and Tang, Mengtao},
  journal          = {arXiv e-prints},
  title            = {REBELS-MOSFIRE: Weak CIII] Emission is Typical Among Extremely UV-bright, Massive Galaxies at $z\sim7$},
  year             = {2025},
  month            = jun,
  pages            = {arXiv:2506.21674},
  archiveprefix    = {arXiv},
  doi              = {10.48550/arXiv.2506.21674},
  eid              = {arXiv:2506.21674},
  eprint           = {2506.21674},
  modificationdate = {2026-01-24T16:27:50},
  primaryclass     = {astro-ph.GA},
  url              = {https://ui.adsabs.harvard.edu/abs/2025arXiv250621674E},
}

@Article{LeFevre2019,
  author           = {Le F{\`e}vre, O. and Lemaux, B.~C. and Nakajima, K. and Schaerer, D. and Talia, M. and Zamorani, G. and Cassata, P. and Garilli, B. and Maccagni, D. and Pentericci, L. and Tasca, L.~A.~M. and Zucca, E. and Amorin, R. and Bardelli, S. and Cimatti, A. and Giavalisco, M. and Guaita, L. and Hathi, N.~P. and Marchi, F. and Vanzella, E. and Vergani, D. and Dunlop, J.},
  journal          = {\aap},
  title            = {The VIMOS Ultra-Deep Survey: evidence for AGN feedback in galaxies with CIII]-{\ensuremath{\lambda}}1908 {\r{A}} emission 10.8 to 12.5 Gyr ago},
  year             = {2019},
  month            = may,
  pages            = {A51},
  volume           = {625},
  archiveprefix    = {arXiv},
  doi              = {10.1051/0004-6361/201732197},
  eid              = {A51},
  eprint           = {1710.10715},
  modificationdate = {2026-01-24T16:31:18},
  primaryclass     = {astro-ph.GA},
  url              = {https://ui.adsabs.harvard.edu/abs/2019A&A...625A..51L},
}

@Article{Senchyna2024,
  author           = {Senchyna, Peter and Plat, Adele and Stark, Daniel P. and Rudie, Gwen C. and Berg, Danielle and Charlot, Stéphane and James, Bethan L. and Mingozzi, Matilde},
  journal          = {The Astrophysical Journal},
  title            = {GN-z11 in Context: Possible Signatures of Globular Cluster Precursors at Redshift 10},
  year             = {2024},
  issn             = {1538-4357},
  month            = apr,
  number           = {1},
  pages            = {92},
  volume           = {966},
  doi              = {10.3847/1538-4357/ad235e},
  modificationdate = {2026-01-24T16:42:09},
  publisher        = {American Astronomical Society},
}

@Article{Chisholm2015,
  author           = {Chisholm, John and Tremonti, Christy A. and Leitherer, Claus and Chen, Yanmei and Wofford, Aida and Lundgren, Britt},
  journal          = {The Astrophysical Journal},
  title            = {SCALING RELATIONS BETWEEN WARM GALACTIC OUTFLOWS AND THEIR HOST GALAXIES},
  year             = {2015},
  issn             = {1538-4357},
  month            = sep,
  number           = {2},
  pages            = {149},
  volume           = {811},
  doi              = {10.1088/0004-637x/811/2/149},
  modificationdate = {2026-01-24T18:23:20},
  publisher        = {American Astronomical Society},
}

@Article{Glazer2025,
  author           = {Glazer, Kelsey S. and Jones, Tucker and Chen, Yuguang and Sanders, Ryan L. and Bradač, Maruša and Pahl, Anthony J. and Shapley, Alice E. and Ellis, Richard S. and Topping, Michael W. and Reddy, Naveen A.},
  journal          = {The Astrophysical Journal},
  title            = {Stacking PANCAKEZ: Spectroscopic Analysis with NIRSpec Stacks in the Epoch of Reionization. Weak Interstellar Medium Absorption and Implications for Ionizing Photon Escape at z ∼ 7},
  year             = {2025},
  issn             = {1538-4357},
  month            = oct,
  number           = {2},
  pages            = {191},
  volume           = {992},
  doi              = {10.3847/1538-4357/ae0194},
  modificationdate = {2026-01-24T18:44:20},
  publisher        = {American Astronomical Society},
}

@Article{Pahl2020,
  author           = {Pahl, Anthony J. and Shapley, Alice and Faisst, Andreas L. and Capak, Peter L. and Du, Xinnan and Reddy, Naveen A. and Laursen, Peter and Topping, Michael W.},
  journal          = {\mnras},
  title            = {The redshift evolution of rest-UV spectroscopic properties to z {\ensuremath{\sim}} 5},
  year             = {2020},
  month            = apr,
  number           = {3},
  pages            = {3194-3211},
  volume           = {493},
  archiveprefix    = {arXiv},
  doi              = {10.1093/mnras/staa355},
  eprint           = {1910.04179},
  modificationdate = {2026-01-24T18:51:14},
  primaryclass     = {astro-ph.GA},
  url              = {https://ui.adsabs.harvard.edu/abs/2020MNRAS.493.3194P},
}

@Article{G.C.2025,
  author           = {G. C., Keerthi Vasan and Jones, Tucker and Shajib, Anowar J. and Rhoades, Sunny and Chen, Yuguang and Sanders, Ryan L. and Stark, Daniel P. and Ellis, Richard S. and Leethochawalit, Nicha and Kacprzak, Glenn G. and Barone, Tania M. and Glazebrook, Karl and Tran, Kim-Vy H. and Skobe, Hannah and Mortensen, Kris and Barisic, Ivana},
  journal          = {The Astrophysical Journal},
  title            = {Spatially Resolved Galactic Winds at Cosmic Noon: Outflow Kinematics and Mass Loading in a Lensed Star-forming Galaxy at z = 1.87},
  year             = {2025},
  issn             = {1538-4357},
  month            = mar,
  number           = {2},
  pages            = {105},
  volume           = {981},
  doi              = {10.3847/1538-4357/ada95b},
  modificationdate = {2026-01-24T19:38:33},
  publisher        = {American Astronomical Society},
}

@Article{AlvarezMarquez2024,
  author           = {Álvarez-Márquez, J. and Gómez, A. Crespo and Colina, L. and Langeroodi, D. and Marques-Chaves, R. and Prieto-Jiménez, C. and Bik, A. and Alonso-Herrero, A. and Boogaard, L. and Costantin, L. and García-Marín, M. and Gillman, S. and Hjorth, J. and Iani, E. and Jermann, I. and Labiano, A. and Melinder, J. and Meyer, R. and Östlin, G. and Pérez-González, P. G. and Rinaldi, P. and Walter, F. and van der Werf, P. and Wright, G.},
  journal          = {A\&A},
  title            = {Insight into the Starburst Nature of Galaxy GN-z11 with JWST MIRI Spectroscopy},
  year             = {2024},
  issn             = {1432-0746},
  month            = mar,
  pages            = {A250},
  volume           = {695},
  archiveprefix    = {arXiv},
  copyright        = {Creative Commons Attribution Non Commercial No Derivatives 4.0 International},
  date             = {2024-12-17},
  doi              = {10.1051/0004-6361/202451731},
  eprint           = {2412.12826},
  modificationdate = {2026-01-24T22:47:38},
  primaryclass     = {astro-ph.GA},
  publisher        = {EDP Sciences},
}

@Article{Zavala2024,
  author           = {Zavala, Jorge A. and Castellano, Marco and Akins, Hollis B. and Bakx, Tom J. L. C. and Burgarella, Denis and Casey, Caitlin M. and Ortiz, Óscar A. Chávez and Dickinson, Mark and Finkelstein, Steven L. and Mitsuhashi, Ikki and Nakajima, Kimihiko and Pérez-González, Pablo G. and Haro, Pablo Arrabal and Bergamini, Pietro and Buat, Veronique and Backhaus, Bren and Calabrò, Antonello and Cleri, Nikko J. and Fernández-Arenas, David and Fontana, Adriano and Franco, Maximilien and Grillo, Claudio and Giavalisco, Mauro and Grogin, Norman A. and Hathi, Nimish and Hirschmann, Michaela and Ikeda, Ryota and Jung, Intae and Kartaltepe, Jeyhan S. and Koekemoer, Anton M. and Larson, Rebeca L. and McKinney, Jed and Papovich, Casey and Rosati, Piero and Saito, Toshiki and Santini, Paola and Terlevich, Roberto and Terlevich, Elena and Treu, Tommaso and Yung, L. Y. Aaron},
  journal          = {Nature Astronomy},
  title            = {A luminous and young galaxy at z=12.33 revealed by a JWST/MIRI detection of Hα and [OIII]},
  year             = {2024},
  issn             = {2397-3366},
  month            = oct,
  number           = {1},
  pages            = {155--164},
  volume           = {9},
  archiveprefix    = {arXiv},
  copyright        = {arXiv.org perpetual, non-exclusive license},
  date             = {2024-03-15},
  doi              = {10.1038/s41550-024-02397-3},
  eprint           = {2403.10491},
  modificationdate = {2026-01-24T20:52:14},
  primaryclass     = {astro-ph.GA},
  publisher        = {Springer Science and Business Media LLC},
}

@Article{Sanders2021,
  author           = {Sanders, Ryan L. and Shapley, Alice E. and Jones, Tucker and Reddy, Naveen A. and Kriek, Mariska and Siana, Brian and Coil, Alison L. and Mobasher, Bahram and Shivaei, Irene and Dav{\'e}, Romeel and Azadi, Mojegan and Price, Sedona H. and Leung, Gene and Freeman, William R. and Fetherolf, Tara and de Groot, Laura and Zick, Tom and Barro, Guillermo},
  journal          = {\apj},
  title            = {The MOSDEF Survey: The Evolution of the Mass-Metallicity Relation from z = 0 to z 3.3},
  year             = {2021},
  month            = jun,
  number           = {1},
  pages            = {19},
  volume           = {914},
  archiveprefix    = {arXiv},
  doi              = {10.3847/1538-4357/abf4c1},
  eid              = {19},
  eprint           = {2009.07292},
  modificationdate = {2026-01-24T20:57:23},
  primaryclass     = {astro-ph.GA},
  url              = {https://ui.adsabs.harvard.edu/abs/2021ApJ...914...19S},
}

@Article{Hsiao2024a,
  author           = {Hsiao, Tiger Yu-Yang and {\'A}lvarez-M{\'a}rquez, Javier and Coe, Dan and Crespo G{\'o}mez, Alejandro and {Abdurro'uf} and Dayal, Pratika and Larson, Rebecca L. and Bik, Arjan and Blanco-Prieto, Carmen and Colina, Luis and P{\'e}rez-Gonz{\'a}lez, Pablo Guillermo and Costantin, Luca and Prieto-Jim{\'e}nez, Carlota and Adamo, Angela and Bradley, Larry D. and Conselice, Christopher J. and Fujimoto, Seiji and Furtak, Lukas J. and Hutchison, Taylor A. and James, Bethan L. and Jim{\'e}nez-Teja, Yolanda and Jung, Intae and Kokorev, Vasily and Mingozzi, Matilde and Norman, Colin and Ricotti, Massimo and Rigby, Jane R. and Sharon, Keren and Vanzella, Eros and Welch, Brian and Xu, Xinfeng and Zackrisson, Erik and Zitrin, Adi},
  journal          = {\apj},
  title            = {JWST MIRI Detections of H{\ensuremath{\alpha}} and [O III] and a Direct Metallicity Measurement of the z = 10.17 Lensed Galaxy MACS0647‑JD},
  year             = {2024},
  month            = oct,
  number           = {2},
  pages            = {81},
  volume           = {973},
  archiveprefix    = {arXiv},
  doi              = {10.3847/1538-4357/ad6562},
  eid              = {81},
  eprint           = {2404.16200},
  modificationdate = {2026-01-24T22:47:25},
  primaryclass     = {astro-ph.GA},
  url              = {https://ui.adsabs.harvard.edu/abs/2024ApJ...973...81H},
}

@Article{Fanelli1992,
  author           = {Fanelli, Michael N. and O'Connell, Robert W. and Burstein, David and Wu, Chi-Chao},
  journal          = {\apjs},
  title            = {Spectral Synthesis in the Ultraviolet. IV. A Library of Mean Stellar Groups},
  year             = {1992},
  month            = sep,
  pages            = {197},
  volume           = {82},
  doi              = {10.1086/191714},
  modificationdate = {2026-01-25T19:23:35},
  url              = {https://ui.adsabs.harvard.edu/abs/1992ApJS...82..197F},
}

@Article{Gazagnes2018,
  author           = {Gazagnes, S. and Chisholm, J. and Schaerer, D. and Verhamme, A. and Rigby, J.~R. and Bayliss, M.},
  journal          = {\aap},
  title            = {Neutral gas properties of Lyman continuum emitting galaxies: Column densities and covering fractions from UV absorption lines},
  year             = {2018},
  month            = aug,
  pages            = {A29},
  volume           = {616},
  archiveprefix    = {arXiv},
  doi              = {10.1051/0004-6361/201832759},
  eid              = {A29},
  eprint           = {1802.06378},
  modificationdate = {2026-01-25T20:00:05},
  primaryclass     = {astro-ph.GA},
  url              = {https://ui.adsabs.harvard.edu/abs/2018A&A...616A..29G},
}

@Article{RiveraThorsen2017,
  author           = {Rivera-Thorsen, T.~E. and Dahle, H. and Gronke, M. and Bayliss, M. and Rigby, J.~R. and Simcoe, R. and Bordoloi, R. and Turner, M. and Furesz, G.},
  journal          = {\aap},
  title            = {The Sunburst Arc: Direct Lyman {\ensuremath{\alpha}} escape observed in the brightest known lensed galaxy},
  year             = {2017},
  month            = nov,
  pages            = {L4},
  volume           = {608},
  archiveprefix    = {arXiv},
  doi              = {10.1051/0004-6361/201732173},
  eid              = {L4},
  eprint           = {1710.09482},
  modificationdate = {2026-01-26T21:58:39},
  primaryclass     = {astro-ph.GA},
  url              = {https://ui.adsabs.harvard.edu/abs/2017A&A...608L...4R},
}

@Article{McLeod2024,
  author           = {McLeod, D.~J. and Donnan, C.~T. and McLure, R.~J. and Dunlop, J.~S. and Magee, D. and Begley, R. and Carnall, A.~C. and Cullen, F. and Ellis, R.~S. and Hamadouche, M.~L. and Stanton, T.~M.},
  journal          = {\mnras},
  title            = {The galaxy UV luminosity function at z ≃ 11 from a suite of public JWST ERS, ERO, and Cycle-1 programs},
  year             = {2024},
  month            = jan,
  number           = {3},
  pages            = {5004-5022},
  volume           = {527},
  archiveprefix    = {arXiv},
  doi              = {10.1093/mnras/stad3471},
  eprint           = {2304.14469},
  modificationdate = {2026-01-26T22:10:09},
  primaryclass     = {astro-ph.GA},
  url              = {https://ui.adsabs.harvard.edu/abs/2024MNRAS.527.5004M},
}

@Article{Weibel2025a,
  author           = {Weibel, Andrea and Oesch, Pascal A. and Williams, Christina C. and Jespersen, Christian Kragh and Shuntov, Marko and Whitaker, Katherine E. and Atek, Hakim and Bezanson, Rachel and Brammer, Gabriel and Chemerynska, Iryna and Cloonan, Aidan P. and Dayal, Pratika and Furtak, Lukas J. and Hutter, Anne and Ji, Zhiyuan and Maseda, Michael V. and Xiao, Mengyuan},
  journal          = {arXiv e-prints},
  title            = {Exploring Cosmic Dawn with PANORAMIC I: The Bright End of the UVLF at $z\sim9 -17$},
  year             = {2025},
  month            = jul,
  pages            = {arXiv:2507.06292},
  archiveprefix    = {arXiv},
  doi              = {10.48550/arXiv.2507.06292},
  eid              = {arXiv:2507.06292},
  eprint           = {2507.06292},
  modificationdate = {2026-01-26T22:11:54},
  primaryclass     = {astro-ph.GA},
  url              = {https://ui.adsabs.harvard.edu/abs/2025arXiv250706292W},
}

@Article{Whitler2025,
  author           = {Whitler, Lily and Stark, Daniel P. and Topping, Michael W. and Robertson, Brant and Rieke, Marcia and Hainline, Kevin N. and Endsley, Ryan and Chen, Zuyi and Baker, William M. and Bhatawdekar, Rachana and Bunker, Andrew J. and Carniani, Stefano and Charlot, St{\'e}phane and Chevallard, Jacopo and Curtis-Lake, Emma and Egami, Eiichi and Eisenstein, Daniel J. and Helton, Jakob M. and Ji, Zhiyuan and Johnson, Benjamin D. and P{\'e}rez-Gonz{\'a}lez, Pablo G. and Rinaldi, Pierluigi and Tacchella, Sandro and Williams, Christina C. and Willmer, Christopher N.~A. and Willott, Chris and Witstok, Joris},
  journal          = {\apj},
  title            = {The z {\ensuremath{\gtrsim}} 9 Galaxy UV Luminosity Function from the JWST Advanced Deep Extragalactic Survey: Insights into Early Galaxy Evolution and Reionization},
  year             = {2025},
  month            = oct,
  number           = {1},
  pages            = {63},
  volume           = {992},
  archiveprefix    = {arXiv},
  doi              = {10.3847/1538-4357/adfddc},
  eid              = {63},
  eprint           = {2501.00984},
  modificationdate = {2026-01-26T22:12:07},
  primaryclass     = {astro-ph.GA},
  url              = {https://ui.adsabs.harvard.edu/abs/2025ApJ...992...63W},
}

@Article{Dekel2023,
  author           = {Dekel, Avishai and Sarkar, Kartick C. and Birnboim, Yuval and Mandelker, Nir and Li, Zhaozhou},
  journal          = {\mnras},
  title            = {Efficient formation of massive galaxies at cosmic dawn by feedback-free starbursts},
  year             = {2023},
  month            = aug,
  number           = {3},
  pages            = {3201-3218},
  volume           = {523},
  archiveprefix    = {arXiv},
  doi              = {10.1093/mnras/stad1557},
  eprint           = {2303.04827},
  modificationdate = {2026-01-26T22:13:46},
  primaryclass     = {astro-ph.GA},
  url              = {https://ui.adsabs.harvard.edu/abs/2023MNRAS.523.3201D},
}

@Article{Li2024,
  author           = {Li, Zhaozhou and Dekel, Avishai and Sarkar, Kartick C. and Aung, Han and Giavalisco, Mauro and Mandelker, Nir and Tacchella, Sandro},
  journal          = {\aap},
  title            = {Feedback-free starbursts at cosmic dawn: Observable predictions for JWST},
  year             = {2024},
  month            = oct,
  pages            = {A108},
  volume           = {690},
  archiveprefix    = {arXiv},
  doi              = {10.1051/0004-6361/202348727},
  eid              = {A108},
  eprint           = {2311.14662},
  modificationdate = {2026-01-26T22:14:16},
  primaryclass     = {astro-ph.GA},
  url              = {https://ui.adsabs.harvard.edu/abs/2024A&A...690A.108L},
}

@Article{Ceverino2024,
  author           = {Ceverino, D. and Nakazato, Y. and Yoshida, N. and Klessen, R.~S. and Glover, S.~C.~O.},
  journal          = {\aap},
  title            = {Redshift-dependent galaxy formation efficiency at z = 5 ‑ 13 in the FirstLight Simulations},
  year             = {2024},
  month            = sep,
  pages            = {A244},
  volume           = {689},
  archiveprefix    = {arXiv},
  doi              = {10.1051/0004-6361/202450224},
  eid              = {A244},
  eprint           = {2404.02537},
  modificationdate = {2026-01-26T22:14:38},
  primaryclass     = {astro-ph.GA},
  url              = {https://ui.adsabs.harvard.edu/abs/2024A&A...689A.244C},
}

@Article{Feldmann2025,
  author           = {Feldmann, Robert and Boylan-Kolchin, Michael and Bullock, James S. and {\c{C}}atmabacak, Onur and Faucher-Gigu{\`e}re, Claude-Andr{\'e} and Hayward, Christopher C. and Kere{\v{s}}, Du{\v{s}}an and Lazar, Alexandres and Liang, Lichen and Moreno, Jorge and Oesch, Pascal A. and Quataert, Eliot and Shen, Xuejian and Sun, Guochao},
  journal          = {\mnras},
  title            = {Elevated UV luminosity density at Cosmic Dawn explained by non-evolving, weakly mass-dependent star formation efficiency},
  year             = {2025},
  month            = jan,
  number           = {1},
  pages            = {988-1016},
  volume           = {536},
  archiveprefix    = {arXiv},
  doi              = {10.1093/mnras/stae2633},
  eprint           = {2407.02674},
  modificationdate = {2026-01-26T22:14:58},
  primaryclass     = {astro-ph.CO},
  url              = {https://ui.adsabs.harvard.edu/abs/2025MNRAS.536..988F},
}

@Article{Gelli2024,
  author           = {Gelli, Viola and Mason, Charlotte and Hayward, Christopher C.},
  journal          = {\apj},
  title            = {The Impact of Mass-dependent Stochasticity at Cosmic Dawn},
  year             = {2024},
  month            = nov,
  number           = {2},
  pages            = {192},
  volume           = {975},
  archiveprefix    = {arXiv},
  doi              = {10.3847/1538-4357/ad7b36},
  eid              = {192},
  eprint           = {2405.13108},
  modificationdate = {2026-01-26T22:16:04},
  primaryclass     = {astro-ph.GA},
  url              = {https://ui.adsabs.harvard.edu/abs/2024ApJ...975..192G},
}

@Article{Mason2023,
  author           = {Mason, Charlotte A. and Trenti, Michele and Treu, Tommaso},
  journal          = {\mnras},
  title            = {The brightest galaxies at cosmic dawn},
  year             = {2023},
  month            = may,
  number           = {1},
  pages            = {497-503},
  volume           = {521},
  archiveprefix    = {arXiv},
  doi              = {10.1093/mnras/stad035},
  eprint           = {2207.14808},
  modificationdate = {2026-01-26T22:16:30},
  primaryclass     = {astro-ph.GA},
  url              = {https://ui.adsabs.harvard.edu/abs/2023MNRAS.521..497M},
}

@Article{Mirocha2023,
  author           = {Mirocha, Jordan and Furlanetto, Steven R.},
  journal          = {\mnras},
  title            = {Balancing the efficiency and stochasticity of star formation with dust extinction in z {\ensuremath{\gtrsim}} 10 galaxies observed by JWST},
  year             = {2023},
  month            = feb,
  number           = {1},
  pages            = {843-853},
  volume           = {519},
  archiveprefix    = {arXiv},
  doi              = {10.1093/mnras/stac3578},
  eprint           = {2208.12826},
  modificationdate = {2026-01-26T22:16:46},
  primaryclass     = {astro-ph.GA},
  url              = {https://ui.adsabs.harvard.edu/abs/2023MNRAS.519..843M},
}

@Article{Shen2023,
  author           = {Shen, Xuejian and Vogelsberger, Mark and Boylan-Kolchin, Michael and Tacchella, Sandro and Kannan, Rahul},
  journal          = {\mnras},
  title            = {The impact of UV variability on the abundance of bright galaxies at z {\ensuremath{\geq}} 9},
  year             = {2023},
  month            = nov,
  number           = {3},
  pages            = {3254-3261},
  volume           = {525},
  archiveprefix    = {arXiv},
  doi              = {10.1093/mnras/stad2508},
  eprint           = {2305.05679},
  modificationdate = {2026-01-26T22:17:03},
  primaryclass     = {astro-ph.GA},
  url              = {https://ui.adsabs.harvard.edu/abs/2023MNRAS.525.3254S},
}

@Article{Sun2023,
  author           = {Sun, Guochao and Faucher-Gigu{\`e}re, Claude-Andr{\'e} and Hayward, Christopher C. and Shen, Xuejian and Wetzel, Andrew and Cochrane, Rachel K.},
  journal          = {\apjl},
  title            = {Bursty Star Formation Naturally Explains the Abundance of Bright Galaxies at Cosmic Dawn},
  year             = {2023},
  month            = oct,
  number           = {2},
  pages            = {L35},
  volume           = {955},
  archiveprefix    = {arXiv},
  doi              = {10.3847/2041-8213/acf85a},
  eid              = {L35},
  eprint           = {2307.15305},
  modificationdate = {2026-01-26T22:17:34},
  primaryclass     = {astro-ph.GA},
  url              = {https://ui.adsabs.harvard.edu/abs/2023ApJ...955L..35S},
}

@Article{Kravtsov2024,
  author           = {Kravtsov, Andrey and Belokurov, Vasily},
  journal          = {arXiv e-prints},
  title            = {Stochastic star formation and the abundance of $z>10$ UV-bright galaxies},
  year             = {2024},
  month            = may,
  pages            = {arXiv:2405.04578},
  archiveprefix    = {arXiv},
  doi              = {10.48550/arXiv.2405.04578},
  eid              = {arXiv:2405.04578},
  eprint           = {2405.04578},
  modificationdate = {2026-01-26T22:18:02},
  primaryclass     = {astro-ph.GA},
  url              = {https://ui.adsabs.harvard.edu/abs/2024arXiv240504578K},
}

@Article{Reddy2022,
  author           = {Reddy, Naveen A. and Topping, Michael W. and Shapley, Alice E. and Steidel, Charles C. and Sanders, Ryan L. and Du, Xinnan and Coil, Alison L. and Mobasher, Bahram and Price, Sedona H. and Shivaei, Irene},
  journal          = {\apj},
  title            = {The Effects of Stellar Population and Gas Covering Fraction on the Emergent Ly{\ensuremath{\alpha}} Emission of High-redshift Galaxies},
  year             = {2022},
  month            = feb,
  number           = {1},
  pages            = {31},
  volume           = {926},
  archiveprefix    = {arXiv},
  doi              = {10.3847/1538-4357/ac3b4c},
  eid              = {31},
  eprint           = {2108.05363},
  modificationdate = {2026-01-27T13:31:06},
  primaryclass     = {astro-ph.GA},
  url              = {https://ui.adsabs.harvard.edu/abs/2022ApJ...926...31R},
}

@Article{Topping2020,
  author           = {Topping, Michael W. and Shapley, Alice E. and Reddy, Naveen A. and Sanders, Ryan L. and Coil, Alison L. and Kriek, Mariska and Mobasher, Bahram and Siana, Brian},
  journal          = {\mnras},
  title            = {The MOSDEF-LRIS Survey: the interplay between massive stars and ionized gas in high-redshift star-forming galaxies},
  year             = {2020},
  month            = jul,
  number           = {4},
  pages            = {4430-4444},
  volume           = {495},
  archiveprefix    = {arXiv},
  doi              = {10.1093/mnras/staa1410},
  eprint           = {1912.10243},
  modificationdate = {2026-01-30T21:04:09},
  primaryclass     = {astro-ph.GA},
  url              = {https://ui.adsabs.harvard.edu/abs/2020MNRAS.495.4430T},
}

@Article{Runco2021,
  author           = {Runco, Jordan N and Shapley, Alice E and Sanders, Ryan L and Topping, Michael W and Kriek, Mariska and Reddy, Naveen A and Coil, Alison L and Mobasher, Bahram and Siana, Brian and Freeman, William R and Shivaei, Irene and Azadi, Mojegan and Price, Sedona H and Leung, Gene C K and Fetherolf, Tara and de Groot, Laura and Zick, Tom and Fornasini, Francesca M and Barro, Guillermo},
  journal          = {Monthly Notices of the Royal Astronomical Society},
  title            = {The MOSDEF survey: a comprehensive analysis of the rest-optical emission-line properties of z ∼ 2.3 star-forming galaxies},
  year             = {2021},
  issn             = {1365-2966},
  month            = jan,
  number           = {2},
  pages            = {2600--2614},
  volume           = {502},
  doi              = {10.1093/mnras/stab119},
  modificationdate = {2026-01-30T21:05:22},
  publisher        = {Oxford University Press (OUP)},
}

@Article{Harikane2026,
  author           = {Harikane, Yuichi and Perez-Gonzalez, Pablo G. and Alvarez-Marquez, Javier and Ouchi, Masami and Nakazato, Yurina and Ono, Yoshiaki and Nakajima, Kimihiko and Umeda, Hiroya and Isobe, Yuki and Xu, Yi and Zhang, Yechi},
  journal          = {arXiv e-prints},
  title            = {A UV-Luminous Galaxy at z=11 with Surprisingly Weak Star Formation Activity},
  year             = {2026},
  month            = jan,
  pages            = {arXiv:2601.21833},
  archiveprefix    = {arXiv},
  doi              = {10.48550/arXiv.2601.21833},
  eid              = {arXiv:2601.21833},
  eprint           = {2601.21833},
  modificationdate = {2026-02-01T15:59:23},
  primaryclass     = {astro-ph.GA},
  url              = {https://ui.adsabs.harvard.edu/abs/2026arXiv260121833H},
}

@Article{CurtisLake2023,
  author           = {Curtis-Lake, Emma and Carniani, Stefano and Cameron, Alex and Charlot, Stephane and Jakobsen, Peter and Maiolino, Roberto and Bunker, Andrew and Witstok, Joris and Smit, Renske and Chevallard, Jacopo and Willott, Chris and Ferruit, Pierre and Arribas, Santiago and Bonaventura, Nina and Curti, Mirko and D’Eugenio, Francesco and Franx, Marijn and Giardino, Giovanna and Looser, Tobias J. and Lützgendorf, Nora and Maseda, Michael V. and Rawle, Tim and Rix, Hans-Walter and Rodríguez del Pino, Bruno and Übler, Hannah and Sirianni, Marco and Dressler, Alan and Egami, Eiichi and Eisenstein, Daniel J. and Endsley, Ryan and Hainline, Kevin and Hausen, Ryan and Johnson, Benjamin D. and Rieke, Marcia and Robertson, Brant and Shivaei, Irene and Stark, Daniel P. and Tacchella, Sandro and Williams, Christina C. and Willmer, Christopher N. A. and Bhatawdekar, Rachana and Bowler, Rebecca and Boyett, Kristan and Chen, Zuyi and de Graaff, Anna and Helton, Jakob M. and Hviding, Raphael E. and Jones, Gareth C. and Kumari, Nimisha and Lyu, Jianwei and Nelson, Erica and Perna, Michele and Sandles, Lester and Saxena, Aayush and Suess, Katherine A. and Sun, Fengwu and Topping, Michael W. and Wallace, Imaan E. B. and Whitler, Lily},
  journal          = {Nature Astronomy},
  title            = {Spectroscopic confirmation of four metal-poor galaxies at z = 10.3–13.2},
  year             = {2023},
  issn             = {2397-3366},
  month            = apr,
  number           = {5},
  pages            = {622--632},
  volume           = {7},
  doi              = {10.1038/s41550-023-01918-w},
  modificationdate = {2026-02-03T19:39:09},
  publisher        = {Springer Science and Business Media LLC},
}

@Article{MarquesChaves2026,
  author           = {Marques-Chaves, R. and Álvarez-Márquez, J. and Colina, L. and Kendrew, S. and {Abdurro'uf} and Blanco-Prieto, C. and Boogaard, L. A. and Castellano, M. and Caputi, K. I. and Crespo-Gomez, A. and Fontana, A. and Fudamoto, Y. and Fujimoto, S. and García-Marín, M. and Harikane, Y. and Harish, S. and Hashimoto, T. and Hsiao, T. and Iani, E. and Inoue, A. K. and Langeroodi, D. and Lin, R. and Melinder, J. and Napolitano, L. and Ostlin, G. and Pérez-González, P. G. and Prieto-Jiménez, C. and Rinaldi, P. and Del Pino, B. Rodríguez and Santini, P. and Sugahara, Y. and Varo-O'ferral, A. and Wright, G. and Zavala, J.},
  title            = {PRISMS. U37126, a very blue, ISM-naked starburst at z=10.255 with nearly 100% Lyman continuum escape fraction},
  year             = {2026},
  month            = feb,
  archiveprefix    = {arXiv},
  copyright        = {Creative Commons Attribution 4.0 International},
  doi              = {10.48550/ARXIV.2602.02322},
  eprint           = {2602.02322},
  modificationdate = {2026-02-03T19:42:06},
  primaryclass     = {astro-ph.GA},
  publisher        = {arXiv},
}

@Article{AlvarezMarquez2026,
  author           = {Álvarez-Márquez, J. and Colina, L. and Crespo-Gomez, A. and Kendrew, S. and Zavala, J. and Marques-Chaves, R. and Prieto-Jiménez, C. and {Abdurro'uf} and Blanco-Prieto, C. and Boogaard, L. A. and Castellano, M. and Fontana, A. and Fudamoto, Y. and Fujimoto, S. and García-Marín, M. and Harikane, Y. and Harish, S. and Hashimoto, T. and Hsiao, T. and Iani, E. and Inoue, A. K. and Langeroodi, D. and Lin, R. and Melinder, J. and Napolitano, L. and Ostlin, G. and Pérez-González, P. G. and Rinaldi, P. and Del Pino, B. Rodríguez and Santini, P. and Sugahara, Y. and Treu, T. and Varo-O'ferral, A. and Wright, G.},
  title            = {PRISMS. UNCOVER-26185, a metal-poor SFG at z=10.05 with no evidence for a X-ray-luminous AGN},
  year             = {2026},
  month            = feb,
  archiveprefix    = {arXiv},
  copyright        = {Creative Commons Attribution 4.0 International},
  doi              = {10.48550/ARXIV.2602.02323},
  eprint           = {2602.02323},
  modificationdate = {2026-02-03T19:42:22},
  primaryclass     = {astro-ph.GA},
  publisher        = {arXiv},
}

@Article{Munoz2026,
  author           = {Muñoz, Julian B. and Chisholm, John and Sun, Guochao and Samuel, Jenna and Mirocha, Jordan and Bregou, Emily and Venditti, Alessandra and Qezlou, Mahdi and Simmonds, Charlotte and Endsley, Ryan},
  title            = {Relatively Fast and Reasonably Furious: Evidence for Increased Burstiness in Smaller Halos at Cosmic Dawn},
  year             = {2026},
  month            = jan,
  archiveprefix    = {arXiv},
  copyright        = {arXiv.org perpetual, non-exclusive license},
  doi              = {10.48550/ARXIV.2601.07912},
  eprint           = {2601.07912},
  modificationdate = {2026-02-05T16:06:15},
  primaryclass     = {astro-ph.GA},
  publisher        = {arXiv},
}

@Article{Helton2025_MIRIphot,
  author           = {Helton, Jakob M. and Rieke, George H. and Alberts, Stacey and Wu, Zihao and Eisenstein, Daniel J. and Hainline, Kevin N. and Carniani, Stefano and Ji, Zhiyuan and Baker, William M. and Bhatawdekar, Rachana and Bunker, Andrew J. and Cargile, Phillip A. and Charlot, St{\'e}phane and Chevallard, Jacopo and D'Eugenio, Francesco and Egami, Eiichi and Johnson, Benjamin D. and Jones, Gareth C. and Lyu, Jianwei and Maiolino, Roberto and P{\'e}rez-Gonz{\'a}lez, Pablo G. and Rieke, Marcia J. and Robertson, Brant and Saxena, Aayush and Scholtz, Jan and Shivaei, Irene and Sun, Fengwu and Tacchella, Sandro and Whitler, Lily and Williams, Christina C. and Willmer, Christopher N.~A. and Willott, Chris and Witstok, Joris and Zhu, Yongda},
  journal          = {Nature Astronomy},
  title            = {Photometric detection at 7.7 {\ensuremath{\mu}}m of a galaxy beyond redshift 14 with JWST/MIRI},
  year             = {2025},
  month            = may,
  pages            = {729-740},
  volume           = {9},
  archiveprefix    = {arXiv},
  doi              = {10.1038/s41550-025-02503-z},
  eprint           = {2405.18462},
  modificationdate = {2026-02-07T19:31:28},
  primaryclass     = {astro-ph.GA},
  url              = {https://ui.adsabs.harvard.edu/abs/2025NatAs...9..729H},
}

@Article{Shajib2025_NRSresol,
  author           = {Shajib, Anowar J. and Treu, Tommaso and Melo, Alejandra and Roberts-Borsani, Guido and Knabel, Shawn and Cappellari, Michele and Frieman, Joshua A.},
  journal          = {A\&A},
  title            = {An accurate measurement of the spectral resolution of the JWST Near Infrared Spectrograph},
  year             = {2025},
  issn             = {1432-0746},
  month            = oct,
  pages            = {L12},
  volume           = {702},
  archiveprefix    = {arXiv},
  copyright        = {Creative Commons Attribution 4.0 International},
  date             = {2025-07-04},
  doi              = {10.1051/0004-6361/202556281},
  eprint           = {2507.03746},
  modificationdate = {2026-02-10T21:42:38},
  primaryclass     = {astro-ph.IM},
  publisher        = {EDP Sciences},
}

@Article{Tacchella2023_GNz11,
  author           = {Tacchella, Sandro and Eisenstein, Daniel J. and Hainline, Kevin and Johnson, Benjamin D. and Baker, William M. and Helton, Jakob M. and Robertson, Brant and Suess, Katherine A. and Chen, Zuyi and Nelson, Erica and Pusk{\'a}s, D{\'a}vid and Sun, Fengwu and Alberts, Stacey and Egami, Eiichi and Hausen, Ryan and Rieke, George and Rieke, Marcia and Shivaei, Irene and Williams, Christina C. and Willmer, Christopher N.~A. and Bunker, Andrew and Cameron, Alex J. and Carniani, Stefano and Charlot, Stephane and Curti, Mirko and Curtis-Lake, Emma and Looser, Tobias J. and Maiolino, Roberto and Maseda, Michael V. and Rawle, Tim and Rix, Hans-Walter and Smit, Renske and {\"U}bler, Hannah and Willott, Chris and Witstok, Joris and Baum, Stefi and Bhatawdekar, Rachana and Boyett, Kristan and Danhaive, A. Lola and de Graaff, Anna and Endsley, Ryan and Ji, Zhiyuan and Lyu, Jianwei and Sandles, Lester and Saxena, Aayush and Scholtz, Jan and Topping, Michael W. and Whitler, Lily},
  journal          = {\apj},
  title            = {JADES Imaging of GN-z11: Revealing the Morphology and Environment of a Luminous Galaxy 430 Myr after the Big Bang},
  year             = {2023},
  month            = jul,
  number           = {1},
  pages            = {74},
  volume           = {952},
  archiveprefix    = {arXiv},
  doi              = {10.3847/1538-4357/acdbc6},
  eid              = {74},
  eprint           = {2302.07234},
  modificationdate = {2026-02-09T19:10:53},
  primaryclass     = {astro-ph.GA},
  url              = {https://ui.adsabs.harvard.edu/abs/2023ApJ...952...74T},
}

@Article{Powell2011,
  author           = {Powell, Leila C. and Slyz, Adrianne and Devriendt, Julien},
  journal          = {\mnras},
  title            = {The impact of supernova-driven winds on stream-fed protogalaxies},
  year             = {2011},
  month            = jul,
  number           = {4},
  pages            = {3671-3689},
  volume           = {414},
  archiveprefix    = {arXiv},
  doi              = {10.1111/j.1365-2966.2011.18668.x},
  eprint           = {1012.2839},
  modificationdate = {2026-02-11T20:32:41},
  primaryclass     = {astro-ph.CO},
  url              = {https://ui.adsabs.harvard.edu/abs/2011MNRAS.414.3671P},
}

@Article{Ortiz2025,
  author           = {Ortiz, Oscar A. Chavez and Finkelstein, Steven L. and Plat, Adele and Silcock, Maddie and Lake, Emma Curtis and Gupta, Ansh R. and Napolitano, Lorenzo and Castellano, Marco and Bromm, Volker and Mitsuhashi, Ikki and Charlot, Stephane and Fontana, Adriano and Zavala, Jorge A. and Chevallard, Jacopo and Burgarella, Denis and Hirschmann, Michaela and Bakx, Tom and Vidal-Garcia, Alba and Calabrò, Antonello and Feltre, Anna},
  title            = {Significant Evidence of an AGN Contribution in GHZ2 at z = 12.34},
  year             = {2025},
  month            = nov,
  archiveprefix    = {arXiv},
  copyright        = {Creative Commons Attribution 4.0 International},
  doi              = {10.48550/ARXIV.2511.03035},
  eprint           = {2511.03035},
  modificationdate = {2026-02-12T15:32:35},
  primaryclass     = {astro-ph.GA},
  publisher        = {arXiv},
}

@Article{Castellano2025,
  author           = {Castellano, M. and Napolitano, L. and Moreschini, B. and Calabrò, A. and Christensen, L. and Llerena, M. and Bakx, T. J. L. C. and Belfiore, F. and Bevacqua, D. and Dickinson, M. and Fontana, A. and Gandolfi, G. and Gasparetto, T. and Marconi, A. and Mascia, S. and Merlin, E. and Morishita, T. and Nanayakkara, T. and Paris, D. and Pentericci, L. and Pérez-Díaz, B. and Roberts-Borsani, G. and Ruiz, S. Rojas and Santini, P. and Treu, T. and Vanzella, E. and Vulcani, B. and Wang, X. and Yoon, I. and Zavala, J.},
  title            = {Investigating ionising sources and the complex interstellar medium of GHZ2 at $z=12.3$},
  year             = {2025},
  month            = dec,
  archiveprefix    = {arXiv},
  copyright        = {Creative Commons Attribution 4.0 International},
  doi              = {10.48550/ARXIV.2512.08490},
  eprint           = {2512.08490},
  modificationdate = {2026-02-12T15:34:28},
  primaryclass     = {astro-ph.GA},
  publisher        = {arXiv},
}

@Article{Endsley2025_bursty,
  author           = {Endsley, Ryan and Chisholm, John and Stark, Daniel P. and Topping, Michael W. and Whitler, Lily},
  journal          = {\apj},
  title            = {The Burstiness of Star Formation at z {\ensuremath{\sim}} 6: A Huge Diversity in the Recent Star Formation Histories of Very UV-faint Galaxies},
  year             = {2025},
  month            = jul,
  number           = {2},
  pages            = {189},
  volume           = {987},
  archiveprefix    = {arXiv},
  doi              = {10.3847/1538-4357/addc74},
  eid              = {189},
  eprint           = {2410.01905},
  modificationdate = {2026-02-15T13:54:30},
  primaryclass     = {astro-ph.GA},
  url              = {https://ui.adsabs.harvard.edu/abs/2025ApJ...987..189E},
}

@Article{Austin2025,
  author           = {Austin, Duncan and Conselice, Christopher J. and Adams, Nathan J. and Harvey, Thomas and Duan, Qiao and Trussler, James and Li, Qiong and Juod{\v{z}}balis, Ignas and Ormerod, Katherine and Ferreira, Leonardo and Westcott, Lewi and Harris, Honor and Wilkins, Stephen M. and Bhatawdekar, Rachana and Caruana, Joseph and Coe, Dan and Cohen, Seth H. and Driver, Simon P. and D'Silva, Jordan C.~J. and Frye, Brenda and Furtak, Lukas J. and Grogin, Norman A. and Hathi, Nimish P. and Holwerda, Benne W. and Jansen, Rolf A. and Koekemoer, Anton M. and Marshall, Madeline A. and Nonino, Mario and Ortiz, III, Rafael and Pirzkal, Nor and Robotham, Aaron and Ryan, Jr., Russell E. and Summers, Jake and Willmer, Christopher N.~A. and Windhorst, Rogier A. and Yan, Haojing and Zackrisson, Erik},
  journal          = {\apj},
  title            = {EPOCHS. III. Unbiased UV Continuum Slopes at 6.5 < z < 13 from Combined PEARLS GTO and Public JWST/NIRCam Imaging},
  year             = {2025},
  month            = dec,
  number           = {1},
  pages            = {43},
  volume           = {995},
  archiveprefix    = {arXiv},
  doi              = {10.3847/1538-4357/ae07db},
  eid              = {43},
  eprint           = {2404.10751},
  modificationdate = {2026-02-15T14:01:55},
  primaryclass     = {astro-ph.GA},
  url              = {https://ui.adsabs.harvard.edu/abs/2025ApJ...995...43A},
}

@Article{Boyett2024,
  author           = {Boyett, Kit and Bunker, Andrew J. and Curtis-Lake, Emma and Chevallard, Jacopo and Cameron, Alex J. and Jones, Gareth C. and Saxena, Aayush and Charlot, St{\'e}phane and Curti, Mirko and Wallace, Imaan E.~B. and Arribas, Santiago and Carniani, Stefano and Willott, Chris and Alberts, Stacey and Eisenstein, Daniel J. and Hainline, Kevin and Hausen, Ryan and Johnson, Benjamin D. and Rieke, Marcia and Robertson, Brant and Stark, Daniel P. and Tacchella, Sandro and Williams, Christina C. and Chen, Zuyi and Egami, Eiichi and Endsley, Ryan and Kumari, Nimisha and Laseter, Isaac and Looser, Tobias J. and Maseda, Michael V. and Scholtz, Jan and Shivaei, Irene and Simmonds, Charlotte and Smit, Renske and {\"U}bler, Hannah and Witstok, Joris},
  journal          = {\mnras},
  title            = {Extreme emission line galaxies detected in JADES JWST/NIRSpec - I. Inferred galaxy properties},
  year             = {2024},
  month            = dec,
  number           = {2},
  pages            = {1796-1828},
  volume           = {535},
  archiveprefix    = {arXiv},
  doi              = {10.1093/mnras/stae2430},
  eprint           = {2401.16934},
  modificationdate = {2026-02-15T14:02:57},
  primaryclass     = {astro-ph.GA},
  url              = {https://ui.adsabs.harvard.edu/abs/2024MNRAS.535.1796B},
}

@Article{Fujimoto2024_uncover,
  author           = {Fujimoto, Seiji and Wang, Bingjie and Weaver, John R. and Kokorev, Vasily and Atek, Hakim and Bezanson, Rachel and Labbe, Ivo and Brammer, Gabriel and Greene, Jenny E. and Chemerynska, Iryna and Dayal, Pratika and de Graaff, Anna and Furtak, Lukas J. and Oesch, Pascal A. and Setton, David J. and Price, Sedona H. and Miller, Tim B. and Williams, Christina C. and Whitaker, Katherine E. and Zitrin, Adi and Cutler, Sam E. and Leja, Joel and Pan, Richard and Coe, Dan and van Dokkum, Pieter and Feldmann, Robert and Fudamoto, Yoshinobu and Goulding, Andy D. and Khullar, Gourav and Marchesini, Danilo and Maseda, Michael and Nanayakkara, Themiya and Nelson, Erica J. and Smit, Renske and Stefanon, Mauro and Weibel, Andrea},
  journal          = {\apj},
  title            = {UNCOVER: A NIRSpec Census of Lensed Galaxies at z = 8.50{\textendash}13.08 Probing a High-AGN Fraction and Ionized Bubbles in the Shadow},
  year             = {2024},
  month            = dec,
  number           = {2},
  pages            = {250},
  volume           = {977},
  archiveprefix    = {arXiv},
  doi              = {10.3847/1538-4357/ad9027},
  eid              = {250},
  eprint           = {2308.11609},
  modificationdate = {2026-02-15T14:06:48},
  primaryclass     = {astro-ph.GA},
  url              = {https://ui.adsabs.harvard.edu/abs/2024ApJ...977..250F},
}

@Article{Kageura2025,
  author           = {Kageura, Yuta and Ouchi, Masami and Nakane, Minami and Umeda, Hiroya and Harikane, Yuichi and Yoshiura, Shintaro and Nakajima, Kimihiko and Yajima, Hidenobu and Thai, Tran Thi},
  journal          = {\apjs},
  title            = {Census of Ly{\ensuremath{\alpha}} Emission from {\ensuremath{\sim}}600 Galaxies at z = 5{\textendash}14: Evolution of the Ly{\ensuremath{\alpha}} Luminosity Function and a Late Sharp Cosmic Reionization},
  year             = {2025},
  month            = jun,
  number           = {2},
  pages            = {33},
  volume           = {278},
  archiveprefix    = {arXiv},
  doi              = {10.3847/1538-4365/adc690},
  eid              = {33},
  eprint           = {2501.05834},
  modificationdate = {2026-02-15T14:07:47},
  primaryclass     = {astro-ph.GA},
  url              = {https://ui.adsabs.harvard.edu/abs/2025ApJS..278...33K},
}

@Article{Ji2026,
  author           = {Ji, Xihan and Belokurov, Vasily and Maiolino, Roberto and Monty, Stephanie and Isobe, Yuki and Kravtsov, Andrey and McClymont, William and {\"U}bler, Hannah},
  journal          = {\mnras},
  title            = {Connecting JWST discovered N/O-enhanced galaxies to globular clusters: evidence from chemical imprints},
  year             = {2026},
  month            = jan,
  number           = {3},
  pages            = {staf2110},
  volume           = {545},
  archiveprefix    = {arXiv},
  doi              = {10.1093/mnras/staf2110},
  eid              = {staf2110},
  eprint           = {2505.12505},
  modificationdate = {2026-02-15T14:08:14},
  primaryclass     = {astro-ph.GA},
  url              = {https://ui.adsabs.harvard.edu/abs/2026MNRAS.545f2110J},
}

@Article{Maiolino2024_GNz11,
  author           = {Maiolino, Roberto and Scholtz, Jan and Witstok, Joris and Carniani, Stefano and D'Eugenio, Francesco and de Graaff, Anna and {\"U}bler, Hannah and Tacchella, Sandro and Curtis-Lake, Emma and Arribas, Santiago and Bunker, Andrew and Charlot, St{\'e}phane and Chevallard, Jacopo and Curti, Mirko and Looser, Tobias J. and Maseda, Michael V. and Rawle, Timothy D. and Rodr{\'\i}guez del Pino, Bruno and Willott, Chris J. and Egami, Eiichi and Eisenstein, Daniel J. and Hainline, Kevin N. and Robertson, Brant and Williams, Christina C. and Willmer, Christopher N.~A. and Baker, William M. and Boyett, Kristan and DeCoursey, Christa and Fabian, Andrew C. and Helton, Jakob M. and Ji, Zhiyuan and Jones, Gareth C. and Kumari, Nimisha and Laporte, Nicolas and Nelson, Erica J. and Perna, Michele and Sandles, Lester and Shivaei, Irene and Sun, Fengwu},
  journal          = {\nat},
  title            = {A small and vigorous black hole in the early Universe},
  year             = {2024},
  month            = mar,
  number           = {8002},
  pages            = {59-63},
  volume           = {627},
  archiveprefix    = {arXiv},
  doi              = {10.1038/s41586-024-07052-5},
  eprint           = {2305.12492},
  modificationdate = {2026-02-15T14:10:46},
  primaryclass     = {astro-ph.GA},
  url              = {https://ui.adsabs.harvard.edu/abs/2024Natur.627...59M},
}

@Article{Mason2026,
  author           = {Mason, Charlotte A. and Chen, Zuyi and Stark, Daniel P. and Yi Lu, Ting and Topping, Michael and Tang, Mengtao},
  journal          = {\aap},
  title            = {Constraints on the z {\ensuremath{\sim}} 6{\ensuremath{-}}13 intergalactic medium from JWST spectroscopy of Lyman-alpha damping wings in galaxies},
  year             = {2026},
  month            = jan,
  pages            = {A114},
  volume           = {705},
  archiveprefix    = {arXiv},
  doi              = {10.1051/0004-6361/202553820},
  eid              = {A114},
  eprint           = {2501.11702},
  modificationdate = {2026-02-15T14:11:16},
  primaryclass     = {astro-ph.GA},
  url              = {https://ui.adsabs.harvard.edu/abs/2026A&A...705A.114M},
}

@Misc{Ellis2025,
  author           = {Ellis, Richard S},
  title            = {Galaxies and Black Holes in the First Billion Years},
  year             = {2025},
  copyright        = {arXiv.org perpetual, non-exclusive license},
  doi              = {10.48550/ARXIV.2508.16948},
  modificationdate = {2026-04-19T17:38:33},
  publisher        = {arXiv},
}
\bibliographystyle{aasjournal}



\end{document}